\documentclass[12pt]{article}
\usepackage[colorlinks,linkcolor=Cobalt,citecolor=Cobalt,bookmarks,bookmarksnumbered]{hyperref}
\usepackage[scaled=0.85]{helvet}
\usepackage{XoohmE}
\usepackage{graphicx,color}
\usepackage{booktabs,multirow}

\def\bsf{\sffamily\bfseries}
\def\cA{\mathcal{A}}
\def\bB{{\boldsymbol B}}

\def\rD{{\rm D}}
\def\Db{\rule[1.75ex]{6pt}{.67pt}\mkern-13mu{D}}
\def\BD{\boldsymbol{D}}

\def\ddt{\partial_\tau}
\def\cF{{\cal F}}
\def\Bf{{\boldsymbol f}}
\def\bF{{\boldsymbol F}}
\def\BF{{\boldsymbol\F}}
\def\bBF{{\boldsymbol{\mit\Phi}}}

\def\BJ{{\boldsymbol\J}}
\def\bJ{{\boldsymbol{\mit\J}}}

\def\hP#1{\skew4\widehat{P}^{#1}}

\def\BR{\boldsymbol{R}}
\def\BS{\boldsymbol{\mit\Sigma}}
\def\bS{\boldsymbol\Sigma}
\def\cS{\mathcal{S}}
\def\mS#1#2{{\skew1\widehat\sigma}_{\scriptscriptstyle#1}{}^{\scriptscriptstyle#2}}
\def\TS#1{\mathop{\textrm{Tr}}\nolimits_{_{#1}}\,[\,\widehat{\boldsymbol\sigma}\,]}
\def\BX{{\boldsymbol\X}}
\def\bX{{\boldsymbol{\mit\Xi}}}

\def\Spin{\mathop{\textsl{Spin}}}
\def\SU{\mathop{\textsl{SU}}}
\def\spin{\mathop{\textrm{spin}}}

\def\pp{{\vphantom{+}\smash{\mathchar'75\mkern-9mu|\mkern5mu}}}
\def\mm{{=\,}}
\def\4{\text{\bsf-}}

\def\Dt#1{\accentset{\hbox{\LARGE.}}{#1}}	
\def\ad{{\accentset{\hbox{\large.}}{\a}}}
\def\bd{{\accentset{\hbox{\large.}}{\b}}}

\def\fc#1#2{\relax\ifmmode{\scriptstyle\frac{#1}{#2}} 
                    \else$\scriptstyle\frac{#1}{#2}$\fi}    

\def\vC#1{\vcenter{\hbox{\hss#1\hss}}}
\def\sm#1{\begin{smallmatrix}#1\end{smallmatrix}}

\def\Bm#1{\begin{bmatrix}#1\end{bmatrix}}

\def\Lx#1{\makebox[0pt][l]{#1}}
\def\Cx#1{\makebox[0pt][c]{#1}}
\def\Rx#1{\makebox[0pt][r]{#1}}

\def\Am{\textsl{Adinkramat}}

\definecolor{Red}    {rgb}{1.00,0.00,0.00} 
\definecolor{Green}  {rgb}{0.00,0.75,0.00} 
\definecolor{Blue}   {rgb}{0.00,0.00,1.00} 
\definecolor{Orange} {rgb}{1.00,0.67,0.00} 
\definecolor{Purple} {rgb}{0.50,0.00,0.50} 
\definecolor{Gold}   {rgb}{1.00,0.90,0.00} 
\definecolor{Magenta}{rgb}{1.00,0.00,1.00} 
\definecolor{Turque} {rgb}{0.00,0.90,0.90} 
\definecolor{Seaweed}{rgb}{0.00,0.25,0.00} 
\definecolor{Brown}  {rgb}{0.50,0.13,0.00} 
\definecolor{Cobalt} {rgb}{0.00,0.00,0.50} 
\definecolor{Sage}   {rgb}{0.00,0.50,0.38} 
\definecolor{grey1}  {rgb}{0.20,0.20,0.20} 
\definecolor{grey2}  {rgb}{0.40,0.40,0.40} 
\definecolor{grey3}  {rgb}{0.60,0.60,0.60} 
\definecolor{grey4}  {rgb}{0.80,0.80,0.80} 
\definecolor{grey5}  {rgb}{0.90,0.90,0.90} 
\def\C#1#2{{\ifcase#1\or
             \color{Red}\or\color{Green}\or\color{Blue}\or
              \color{Orange}\or\color{Purple}\or\color{Gold}\or
             \color{Magenta}\or\color{Turque}\or\color{Seaweed}\or
               \color{Brown}\or\color{Cobalt}\or\color{Sage}\or
                 \color{grey1}\or\color{grey2}\or\color{grey3}\or
                 \color{grey4}\else\color{grey5}\fi#2}}
\definecolor{gray}{rgb}{.7,.7,.7}
\def\XXX{\colorbox{yellow}{\color{red}\bf X\kern-4pt{\Large$\bs*$}\kern-4.125ptX}}
\def\cb#1#2{\setlength\fboxsep{1pt}\colorbox{#1}{\color{#1}\fbox{\color{black}#2}}}
\def\cB#1{\hbox to0pt{\setlength\fboxsep{0pt}\hss\color{gray}\fbox{\cb{white}{#1}}\hss}}

 \font\rOpe=cmsy10                        
 \def\ktl{{\hbox{\rOpe\char'170}}}        
 \def\kbl{{\hbox{\rOpe\char'170}}}        
 \def\kcr{{\reflectbox{\rOpe\char'170}}}        
 \def\ktr{{\reflectbox{\rOpe\char'170}}}        
 \def\kbr{{\reflectbox{\rOpe\char'170}}}        
 \def\Border{\vbox{\hsize0pt
        \setlength{\unitlength}{1mm}
        \newcount\xco
        \newcount\yco
        \xco=-21
        \yco=12
        \begin{picture}(0,0)(-7.5,0)
        \put(\xco,\yco){$\ktl$}
        \advance\yco by-1
        {\loop
        \put(\xco,\yco){$\kcr$}
        \advance\yco by-2
        \ifnum\yco>-240
        \repeat
        \put(\xco,\yco){$\kbl$}}
        \xco=170
        \yco=12
        \put(\xco,\yco){$\ktr$}
        \advance\yco by-1
        {\loop
        \put(\xco,\yco){$\kcr$}
        \advance\yco by-2
        \ifnum\yco>-240
        \repeat
        \put(\xco,\yco){$\kbr$}}
        \put(-19.5,13){\scalebox{.6067}{%
         University of Maryland Center for String and Particle  Theory \&\ Physics Department%
        |University of Maryland Center for String and Particle  Theory \&\ Physics Department}}
        \put(-19.5,-241.5){\scalebox{.58385}{%
         Howard University Department of Physics and Astronomy%
        |Howard University Department of Physics and Astronomy%
        |Howard University Department of Physics and Astronomy}}
        \end{picture}
        \par\vskip-8mm}}
\definecolor{UMred}{rgb}{.9,.05,.2}
\definecolor{HUblue}{rgb}{.0,.3,.7}
 \def\UMbanner{\vbox{\hsize0pt
        \setlength{\unitlength}{.4mm}
        \thicklines\color{UMred}
        \begin{picture}(0,0)(-30,-10)
        \put(165,16){\line(1,0){4}}
        \put(170,16){\line(1,0){4}}
        \put(180,16){\line(1,0){4}}
        \put(175,0){\line(1,0){4}}
        \put(180,0){\line(1,0){4}}
        \put(185,0){\line(1,0){4}}
        \put(169,0){\line(0,1){16}}
        \put(170,0){\line(0,1){16}}
        \put(179,0){\line(0,1){16}}
        \put(180,0){\line(0,1){16}}
        \put(184,0){\line(0,1){16}}
        \put(185,0){\line(0,1){16}}
        \put(169,16){\oval(8,32)[bl]}
        \put(170,16){\oval(8,32)[br]}
        \put(179,0){\oval(8,32)[tl]}
        \put(185,0){\oval(8,32)[tr]}
        \color{HUblue}
        \put(167.75,-2){\line(1,0){4}}
        \put(172.75,-2){\line(1,0){4}}
        \put(177.75,-2){\line(1,0){4}}
        \put(182.75,-2){\line(1,0){4}}
        \put(167.75,-2){\line(0,-1){16}}
        \put(171.75,-2){\line(0,-1){16}}
        \put(172.75,-2){\line(0,-1){16}}
        \put(176.75,-2){\line(0,-1){16}}
        \put(181.75,-2){\line(0,-1){16}}
        \put(182.75,-2){\line(0,-1){16}}
        \put(181.75,-2){\oval(8,32)[bl]}
        \put(182.75,-2){\oval(8,32)[br]}
        \put(167.75,-18){\line(1,0){4}}
        \put(172.75,-18){\line(1,0){4}}
        \end{picture}
        \par\vskip-6.5mm
        \thicklines}}

 \SfTitles 
\makeatletter
\def\@begintheorem#1#2{\trivlist\leftskip=2pc\rightskip=.25pc%
                        \item[\hskip\labelsep{\bsf\slshape#1\ #2}]%
                         \let\em=\itshape\slshape}
\def\@opargbegintheorem#1#2#3{\trivlist\leftskip=2pc\rightskip=.25pc%
                        \item[\hskip\labelsep{\bsf\slshape#1\ #2\ (#3)}]%
                         \let\em=\itshape\slshape}
\def\@endtheorem{\endtrivlist}
\def\paragraph{\@startsection{paragraph}{4}{\z@}
           {.75ex \@plus.5ex \@minus.2ex}{-2mm}{\sf\bfseries\boldmath}}
\def\Remk{\par\noindent\addtocounter{remark}{1}%
           \def\@currentlabel{{\thesection.\arabic{remark}}}%
           {\bsf Remark~\thesection.\arabic{remark}}:~\ignorespaces\relax}
\long\def\@makecaption#1#2{\vskip\abovecaptionskip\sbox\@tempboxa{\small{\bsf#1}: #2}%
          \ifdim\wd\@tempboxa>\hsize
          {\addtolength{\leftskip}{1pc}\addtolength{\rightskip}{1pc}
            \small\baselineskip12pt{\bsf#1}: #2\par}%
           \else\global\@minipagefalse\hb@xt@\hsize{\hfil\box\small\@tempboxa\hfil}%
            \fi\vskip\belowcaptionskip}
\renewenvironment{proof}{{\noindent\bsf Proof:}}
               {\QED\par}
\makeatother

 \allowdisplaybreaks
\begin{document}
\thispagestyle{empty}
\vbox{\Border\UMbanner}
 \noindent
 \today
  \hfill\smash{\parbox[t]{50mm}{\raggedleft\small
                           UMDEPP-11 005\\[-1mm] 
                           MIT-CTP  4232
 }}
 \vglue0mm
 \begin{center}
{\LARGE\bsf\boldmath
  On Dimensional Extension of Supersymmetry:\\*[1mm]
  From Worldlines to Worldsheets
 }\\*[5mm]
{\large\bsf 
      S.J.\,Gates, Jr.$^*$ and
      T.\,H\"{u}bsch$^\dag$
}\\*[2mm]
\parbox[t]{80mm}{\small\centering\it
      $^*$Center for String and Particle Theory\\[-1mm]
Department of Physics, University of Maryland\\[-1mm]
College Park, MD 20742-4111 USA
  \\[-4pt] {\tt  gatess@wam.umd.edu}}\hfil
\parbox[t]{80mm}{\small\centering\it
      $^\dag$Department of Physics \&\ Astronomy,\\[-1mm]
      Howard University, Washington, DC 20059
  \\[-4pt] {\tt  thubsch@howard.edu}}
\\[5mm]
{\sf\bfseries ABSTRACT}\\[3mm]
\parbox{154mm}{\addtolength{\baselineskip}{-2pt}\parindent=2pc\noindent
There exist myriads of off-shell worldline supermultiplets for $(N\,{\leq}\,32)$-extended supersymmetry in which every supercharge maps a component field to precisely one other component field or its derivative. A subset of these extends to off-shell worldsheet $(p,q)$-supersymmetry and is characterized by the twin theorems~\ref{T:bowT} and~\ref{T:SSR} in this note. The evasion of the {\em\/obstruction\/} defined in these theorems is conjectured to be sufficient for a worldline supermultiplet to extend to worldsheet supersymmetry; it is also a necessary filter for dimensional extension to higher-dimensional spacetime. We show explicitly how to ``re-engineer'' an Adinkra---if permitted by the twin theorems~\ref{T:bowT} and~\ref{T:SSR}---so as to depict an off-shell supermultiplet of worldsheet $(p,q)$-supersymmetry.
}
\end{center}
\vspace{5mm}
\noindent
\parbox[t]{60mm}{PACS: 11.30.Pb, 12.60.Jv}\hfill
\parbox[t]{100mm}{\raggedleft\small\baselineskip=12pt\sl
            When eating an elephant, take a bite at a time.\\[-1pt]
            \dots and keep an eye on the elephant.\\[-1pt]
            |\,Anonymous\hphantom{.}}
\section{Introduction, Results and Summary}
Supersymmetry has been studied for almost four decades in physics and more than that in mathematics, yet there is still no complete theory of off-shell representations. That is, the complete off-shell structure of supermultiplets is known only for a low enough total number of supercharges, counting independent components of spinors separately\cite{r1001,rPW,rWB,rBK,rES-SuGra}. To remedy this, Ref.\cite{rGR0} proposed to dimensionally reduce to $1d$ (worldline) supersymmetric Quantum Mechanics, obtain a complete off-shell representation theory, then dimensionally {\em\/extend\/}\ft{In a bout of chemical inspiration, Ref.\cite{rGR0} used the term ``oxidization'' as the reverse of dimensional reduction. Subsequently, ``enhance'' was used in Refs.\cite{rFIL,rFL}. Herein, ``extend'' and ``extension'' will be used instead, in their standard group-theoretic, representation-theoretic and geometric sense.} back to the spacetime of desired dimensionality, employing the geometric fact that all higher-dimensional spacetimes include continua of worldlines.

In this spirit, Refs.\cite{rA,r6-1,r6--1,r6-3,r6-3.2,r6-1.2} developed a detailed classification of a huge class (${\sim}\,10^{12}$ for no more than 32 supersymmetries) of worldline supermultiplets wherein each supercharge maps each component field to precisely one other component field or its derivative, and which are faithfully represented by graphs called {\em\/Adinkras\/}; see also\cite{rPT,rT01,rT01a,rCRT,rKRT,rKT07,rGKT10}. The subsequently intended dimensional extension has been addressed recently\cite{rFIL,rFL}, and the purpose of the present note is to complement this effort and identify an easily verifiable obstruction to dimensional extension.

To this end, we focus on the worldline to worldsheet extension, being that all higher-dimensional spacetimes include worldsheets, which in turn include worldlines. Worldsheet dimensional extension is thus a stepping stone towards dimensional extension to higher-dimensional spacetimes. Of course, worldsheet supersymmetry is also important in its own right\cite{rGSW1,rGSW2,rJPS,rBBS} and affords comparison with numerous known results; see Refs.\cite{rTwSJG0,rGHR,rUDSS01,rPW,rES-SuGra,rDGR,rSChSF0,rHSS,rSChSF,rGSS,rHP1}, to name but a few.

The paper is organized as follows:
 Requisite definitions and notation are provided in the remainder of this introduction, whereupon Section~\ref{s:WSR} presents the main result (twin theorems~\ref{T:bowT} and~\ref{T:SSR}, and Corollary~\ref{C:ObSuSy}), a criterion that all worldsheet supermultiplets must satisfy.
 Section~\ref{s:XMpls} illustrates the use of this criterion by presenting the case of dimensional extension from worldline $N\,{=}\,4$ supersymmetry without central charges to worldsheet $(2,2)$-supersymmetry without central charges.
 Section~\ref{s:XMpls+} illustrates the ease of use of this criterion in $(4k,4k)$-supersymmetric examples.
 Our conclusions are summed up in section~\ref{e:coda}, and technically more involved (and explicit) details are deferred to the appendix.

\paragraph{Notation and Definitions:}
We study off-shell and on-the-half-shell\cite{rHP1}
 linear and finite-dimensional representations of the centrally unextended $(1,1|p,q)$-super\-sym\-metry, \ie, the worldsheet $(p,q)$-extended super-Poincar\'e symmetry generated by $p$ Majorana-Weyl (real, 1-component), left-handed superderivatives\ft{While not strictly necessary to use superdifferential operators to study supersymmetry, we find it simpler to do so, and there is no loss of generality: supersymmetry implies superspace\cite{rHTSSp08}.} $\rD_{\a+}$, $q$ Majorana-Weyl right-handed ones, $\rD_{\bd-}$, and the light-cone worldsheet derivatives $\vd_\pp$ and $\vd_\mm$. Amongst these,
\begin{equation}
 \big\{\,\rD_{\a+}\,,\,\rD_{\b+}\,\big\}=2i\,\d_{\a\b}\,\vd_\pp,\qquad
 \big\{\,\rD_{\ad-}\,,\,\rD_{\bd-}\,\big\}=2i\,\d_{\ad\bd}\,\vd_\mm,
 \label{e:pqSuSy}
\end{equation}
are the only nonzero supercommutators.
 Being abelian, worldsheet Lorentz symmetry $\Spin(1,1)\simeq\textsl{GL}(1;\IR)\simeq\IR^\times$ (the multiplicative group of nonzero real numbers, \ie, the non-compact cousin of $U(1)$) has only 1-dimensional irreducible representations, upon each of which it acts by a multiplicative real number\cite{rWyb,rFRH}. Eigenvalues of the Lorentz generator are called {\bsf spin} for simplicity:
\begin{equation}
 \spin(\rD_{\a+})=+\inv2=-\spin(\rD_{\ad-}),\qquad
 \spin(\vd_\pp)=+1=-\spin(\vd_\mm),
\end{equation}
where the ``$\pm$'' subscripts count spin in units of $\pm\frac12\hbar$; superscripts count oppositely. We emphasize that the $\a$ and $\Dt\a$ indices count ``internal'' (not spacetime) degrees of freedom.
 In addition to spin, all objects also have an {\bsf engineering (mass-) dimension}, defined by
\begin{equation}
  [\rD_{\a+}] = \inv2 = [\rD_{\ad-}],\qquad
  [\vd_\pp] = 1 = [\vd_\mm].
\end{equation}
 The operators in\eq{e:pqSuSy} are first order differential operators in $(1,1|p,q)$-dimensional superspace, and act on {\bsf superfields} $\BF,\BJ$, \etc\
 Component fields
\begin{equation}
  \f\Defl\BF|,\quad
  \j^-_\a\Defl i\rD_{\a+}\BF|,\quad
  \j^+_\ad\Defl i\rD_{\ad-}\BF|,\quad\cdots\quad
  F^{\,\mm}_{\a\b}\Defl \frc i2[\rD_{\a+},\rD_{\b+}]\BF|,\quad\etc,
 \label{e:CompF}
\end{equation}
are|up to numerical factors chosen for convenience|defined by projecting the
\begin{equation}
  \rD^{\bf a,b}\Defl \rD_{1+}^{~a_1}\wedge\cdots\wedge\rD_{p+}^{~a_p}
                       \rD_{1-}^{~b_1}\wedge\cdots\wedge\rD_{q-}^{~b_q},\qquad
  a_\a,b_\ad\in\{0,1\},
 \label{e:MonD}
\end{equation}
superderivatives of the superfields to the $(1,1|0,0)$-dimen\-si\-o\-nal (purely bosonic) worldsheet.
 A worldsheet superfield is {\bsf off-shell} if it is subject to no worldsheet differential equation (one involving $\vd_\pp$ and/or $\vd_\mm$, but neither $\rD_{\a+}$ nor $\rD_{\ad-}$).
 If it is subject to only {\bsf unidextrous} worldsheet differential equations\cite{rUDSS01,rHSS} (involving {\em\/either\/} $\vd_\pp$ {\em\/or\/} $\vd_\mm$ but not both),
 it is said to be {\bsf on the half-shell}\cite{rHP1}; such superfields are not off-shell in the standard field-theoretic sense on the worldsheet, but {\em\/are\/} off-shell on a continuum of unidextrously embedded worldlines and can provide for dynamics not describable otherwise\cite{rHT-UCS}.
 Calling a superfield, operator, expression, equation or another construct thereof {\bsf ambidextrous} emphasizes that it is not unidextrous.

\paragraph{Adinkras:}
{\em Adinkraic\/} supermultiplets admit mutually compatible bases of component fields and supersymmetry generators, such that each supersymmetry generator maps each component field to precisely one other component field or its spacetime derivative. All such worldline supermultiplets are faithfully depicted by Adinkras (see Table~\ref{t:A}),
\begin{table}[ht]
  \centering
  \begin{tabular}{@{} cc|cc @{}}
    \makebox[15mm]{\bsf Adinkra} & \makebox[40mm]{\bsf Supersymmetry Action} 
  & \makebox[15mm]{\bsf Adinkra} & \makebox[40mm]{\bsf Supersymmetry Action} \\ 
    \hline
    \begin{picture}(5,9)(0,5)
     \put(0,0){\includegraphics[height=11mm]{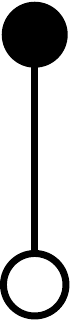}}
     \put(3,0){\scriptsize$A$}
     \put(3,9){\scriptsize$B$}
     \put(-1,4){\scriptsize$I$}
    \end{picture}\vrule depth4mm width0mm
     & $\rD_I\begin{bmatrix}\BJ_B\\\BF_A\end{bmatrix}
           =\begin{bmatrix}\Dt\BF_A\\i\BJ_B\end{bmatrix}$
  & \begin{picture}(5,9)(0,5)
     \put(0,0){\includegraphics[height=11mm]{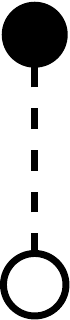}}
     \put(3,0){\scriptsize$A$}
     \put(3,9){\scriptsize$B$}
     \put(-1,4){\scriptsize$I$}
    \end{picture}\vrule depth4mm width0mm
     & $\rD_I\begin{bmatrix}\BJ_B\\\BF_A\end{bmatrix}
           =\begin{bmatrix}-\Dt\BF_A\\-i\BJ_B\end{bmatrix}$ \\[5mm]
    \hline
    \begin{picture}(5,9)(0,5)
     \put(0,0){\includegraphics[height=11mm]{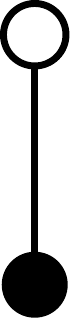}}
     \put(3,0){\scriptsize$B$}
     \put(3,9){\scriptsize$A$}
     \put(-1,4){\scriptsize$I$}
    \end{picture}\vrule depth4mm width0mm
     &  $\rD_I\begin{bmatrix}\BF_A\\\BJ_B\end{bmatrix}
           =\begin{bmatrix}i\dot\BJ_B\\\BF_A\end{bmatrix}$
  & \begin{picture}(5,9)(0,5)
     \put(0,0){\includegraphics[height=11mm]{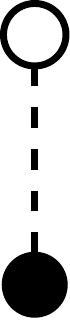}}
     \put(3,0){\scriptsize$B$}
     \put(3,9){\scriptsize$A$}
     \put(-1,4){\scriptsize$I$}
    \end{picture}\vrule depth4mm width0mm
     &  $\rD_I\begin{bmatrix}\BF_A\\\BJ_B\end{bmatrix}
           =\begin{bmatrix}-i\Dt\BJ_B\\-\BF_A\end{bmatrix}$ \\[5mm]
    \hline
  \multicolumn{4}{l}{\vrule height3.0ex width0pt\parbox{120mm}{\small\baselineskip12pt
   The edges may labeled by $I$ or drawn in the $I^{\text{th}}$ color.}}
  \end{tabular}
  \caption{Adinkras depict supermultiplets\eq{eSM=SF} by assigning:
    (white/black) vertices\,$\iff$\,(boson/fermion) component fields;
    edge color/index\,$\iff$\,$\rD_I$;
    edge dashed\,$\iff$\,$c=-1$;
    nodes are placed at heights equal to the engineering dimension of the corresponding component field, thus determining $\l$ in Eqs.\eq{eSM=SF}.}
  \label{t:A}
\end{table}
 which are far more compact and comprehensible than the often very large systems of supersymmetry transformation rules that they depict. The present note explores adopting this graphical tool for worldsheet supermultiplets.
 As done e.g.\ in\cite{r1001,rGSS,rFGH}, we introduce a collection of otherwise {\bsf intact} (\ie, unconstrained, ungauged, unprojected\dots) component superfields, and correspond the supersymmetry transformation with superderivative constraint equations\ft{The correspondence\eq{eSM=SF} derives from the superspace relation $Q=i\rD+2\q{\cdot}\Sl{\nabla}$ between supercharges $Q$ and superderivatives, and the fact that if the $\rD$ act from the left then the $Q$ act from the right\cite{r1001,rBK}.}
\begin{equation}
 \left. \begin{array}{r@{\>}l}
          \rD_I \,\BF_A   &= i(\IL_I)_A{}^{\ha B}\,(\ddt^{1-\l}\BJ_{\ha B})\\
          \rD_I \,\BJ_{\ha B} &= (\IL^{-1}_I)_{\ha B}{}^A\,(\ddt^\l\BF_A)
        \end{array}\right\}
 \quad\Iff\quad
 \left\{\begin{array}{r@{\>}lr@{\>}l}
          Q_I\,\f_A &=-(\IL_I)_A{}^{\ha B}\,(\ddt^{1-\l}\j_{\ha B}),\quad
           & \f_A   &:=\BF_A|,\\
          Q_I\,\j_{\ha B} &=-i(\IL^{-1}_I)_{\ha B}{}^A\,(\ddt^\l\f_A),\quad
           & \j_{\ha B} &:=\BJ_{\ha B}|,
        \end{array}\right.
 \label{eSM=SF}
\end{equation}
where the exponent $\l=0,1$ depends on $I,A,\ha{B}$, and the matrices $\IL_I$ have exactly one entry, $\pm1$, in every row and in every column. This type of (adinkraic) supersymmetry action is then depicted using the ``dictionary'' provided in Table~\ref{t:A}.
 For example,
\begin{subequations}\label{e:N2ES}
 \begin{align}
 \rD_1\,\BF   &=i\,\bs\J_1,    & \C1{\rD_2}\,\BF   &=i\,\BJ_2,\label{Q.f1}\\*
 \rD_1\,\BJ_1 &=\Dt\BF,        & \C1{\rD_2}\,\BJ_1 &=-\bF,\label{Q.j1}\\*
 \rD_1\,\BJ_2 &=\bF,           & \C1{\rD_2}\,\BJ_2 &=\Dt\BF,\label{Q.j2}\\*
 \rD_1\,\bF   &=i\Dt\BJ_2,
  \begin{picture}(30,0)(-12,0)
  \put(10,-2){\includegraphics[height=25mm]{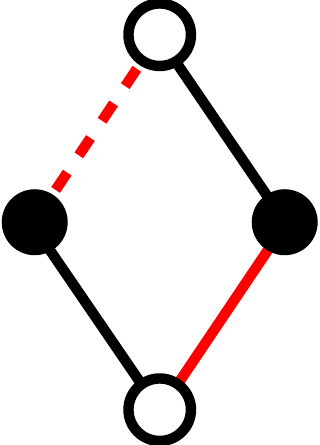}}
  \put(22,-1){\small$\bs\F$}
  \put(4,10){\small$\bs\J_1$}
  \put(29,10){\small$\bs\J_2$}
  \put(22,20){\small$\bs{F}$}
 \end{picture}\quad
                               & \C1{\rD_2}\,\bF     &=-i\,\Dt\BJ_1,\label{Q.f2}
 \end{align}
\end{subequations}
and
\begin{subequations}\label{e:N2IS}
 \begin{align}
 \rD_1\,\bB_1 &=i\,\BX_1,     & \C1{\rD_2}\,\bB_1 &=i\,\BX_2,\label{Qf1}\\*
 \rD_1\,\bB_2 &=i\,\BX_2,     & \C1{\rD_2}\,\bB_2 &=-i\,\BX_1,\label{Qf2}\\*
 \rD_1\,\BX_1 &=\Dt\bB_1,     & \C1{\rD_2}\,\BX_1 &=-\Dt\bB_2,\label{Qj1}\\*
 \rD_1\,\BX_2 &=\Dt\bB_2,
  \begin{picture}(30,0)(-12,0)
  \put(10,0){\includegraphics[height=20mm]{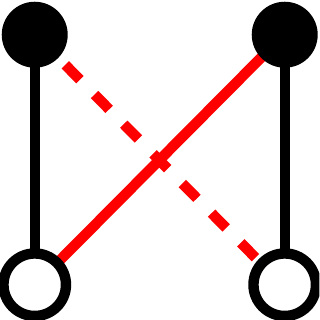}}
  \put(5,0){\small$\bB_1$}
  \put(5,18){\small$\BX_1$}
  \put(31,0){\small$\bB_2$}
  \put(31,18){\small$\BX_2$}
 \end{picture}\quad
                              & \C1{\rD_2}\,\BX_2&=\Dt\bB_1,\label{Qj2}
 \end{align}
\end{subequations}
define two clearly distinct {\em\/worldline\/} $N{=}2$ supermultiplets. Since all nodes of an Adinkra are always placed at heights proportional to the engineering dimensions of the component fields that they represent, we may use `height' and `engineering dimension' interchangeably.
In the superdifferential systems\eqs{e:N2ES}{e:N2IS}, all superfields $\BF,\BJ_i,\bF,\bB_i,\BX_i$ may be chosen to be real, as seen by writing the superderivative action in terms of supercommutators, so that
\begin{subequations}
\begin{align}
  (\rD_I\BF)\Defl[\rD_I,\BF],\quad&\To\quad
  (\BJ_I)^\dag=[(-i\rD_I),\BF]^\dag=[\BF^\dag,(-i\rD_I)^\dag]=-[i\rD_I,\BF]
               =\BJ_I,\\
  (\rD_1\BJ_2)\Defl\{\rD_1,\BJ_2\},\quad&\To\quad
  (\bF)^\dag=\{\rD_1,\BJ_2\}^\dag=\{\BJ_2^\dag,\rD_1^\dag\}
   =+\{\rD_1,\BJ_2\}=\bF,\quad\etc
\end{align}
\end{subequations}
Given the comparative brevity and ease of comprehension, supersymmetry transformation rules such as\eqs{e:N2ES}{e:N2IS} will subsequently be depicted by Adinkras rather than written out explicitly; see the appendix for examples of the relation Adinkra\,$\iff$\,explicit equations. This formulation affords writing supersymmetric Lagrangians in the manifestly supersymmetric fashion, in superspace\cite{r1001,rBK}.

Given the obvious distinction between the Adinkras\eq{e:N2ES} and\eq{e:N2IS} we refer to nodes drawn at the same height as being on the same {\bsf level}; the number of levels then counts the number of distinct engineering dimensions of the component fields in a supermultiplet\cite{rCRT,r6-1}.

\section{An Obstruction for Extension to Worldsheet Supersymmetry}
\label{s:WSR}
Worldlines and worldsheets|unlike higher-dimensional spacetimes|have in common the abelian nature of the Lorentz groups, $\Spin(1,0)\simeq\ZZ_2$ and $\Spin(1,1)\simeq\IR^\times$ respectively, whereupon all physical quantities may be parametrized in terms of independent real 1-component variables.
 In addition, the operator $\ddt$ transforms as the trivial representation of the worldline Lorentz group $\Spin(1,0)$, which underlies the classification efforts of Refs.\cite{r6-1,r6-3,r6-3.2,r6-1.2}.
 However, the operators $\vd_\pp$ and $\vd_\mm$ do not transform trivially under the worldsheet Lorentz group $\Spin(1,1)$, having spin $+1$ and $-1$, respectively, and this induces the key differences between any classification of worldsheet supermultiplets and classifications of representations of worldline supersymmetry; see also the ``bi-filtration'' of Ref.\cite{r6--1}.
 
Herein, this feature is employed as a {\em\/filter\/} to distinguish those worldline off-shell supermultiplets, say from the huge\ft{The sheer number, ${\gtrsim}\,\,10^{47}$, of distinct real worldline supermultiplets|which come in ${\gtrsim10}\,^{12}$ equivalence classes|is daunting\cite{r6-3,r6-3.2,r6-1.2}. This is further multiplied by a combinatorially growing abundance of height assignments, as well as added structures, such as complexification and group actions, which further diversify the possible interactions; see for example\cite{rDGR,rEW-ADHM} for direct consequences of this fact.} collection of Refs.\cite{r6-3,r6-3.2,r6-1.2} that do extend to worldsheet off-shell supermultiplets, some of which would possibly further extend to higher-dimensional off-shell supermultiplets.

In particular, the combinatorial explosion of worldline supermultiplets\cite{r6-3,r6-3.2,r6-1.2} owes also to the fact that replacing a worldline field with its derivative, $\f\mapsto\Dt\f=(\ddt\f)$, only changes the engineering dimension of the field and produces only minor, though important changes in the supersymmetry relations\cite{rGR-1}.
 By contrast, replacing a worldsheet field $\f$ with either $(\vd_\pp\f)$ or $(\vd_\mm\f)$ changes both the engineering dimension and the spin of the relevant field:
\begin{equation}
 \textrm{spin}(\vd_\pp\f)=\text{spin}(\f)+1,
  \qquad\text{and}\qquad
 \textrm{spin}(\vd_\mm\f)=\text{spin}(\f)-1.
 \label{e:2dR/L}
\end{equation}
Replacements such as $\f\mapsto(\vd_\pp\f)$ and $\f\mapsto(\vd_\mm\f)$ will then|in general|{\em\/obstruct\/} the supersymmetry relations! For example, consider the $(p,q)=(1,1)$ case\eq{e:N2ES}:
\begin{equation}
 \vC{\begin{picture}(150,35)\footnotesize
   \put(0,0){\includegraphics[width=150mm]{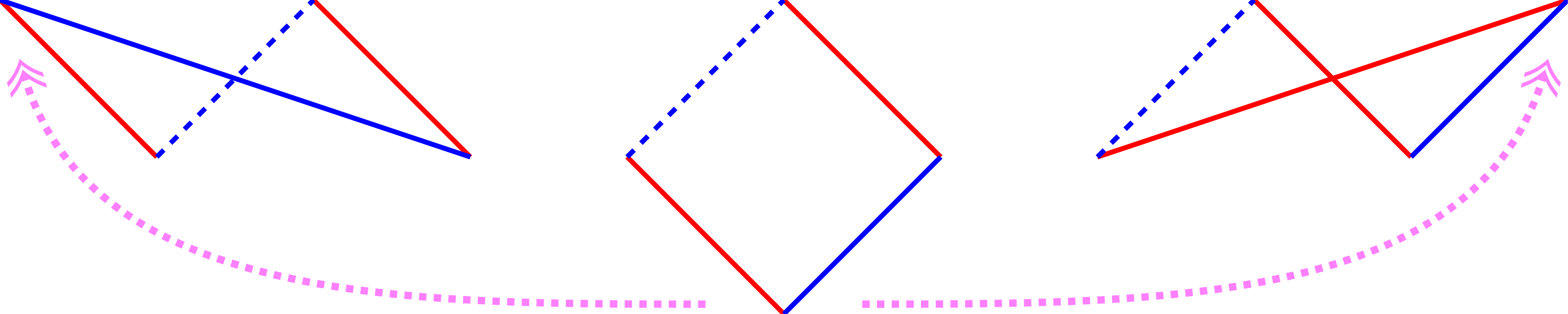}}
    \put(15,14){\cB{$\BJ_+$}}
    \put(45,14){\cB{$\BJ_-$}}
    \put(30,28){\cB{\bf F}}
    \put(0,28){\cB{$(i\vd_\pp\BF)$}}
    \put(29.5,18.5){\large\bf?}
    \put(75,0){\cB{$\BF$}}
    \put(60,14){\cB{$\BJ_+$}}
    \put(90,14){\cB{$\BJ_-$}}
    \put(75,28){\cB{\bf F}}
    \put(150,28){\cB{$(i\vd_\mm\BF)$}}
    \put(105,14){\cB{$\BJ_+$}}
    \put(135,14){\cB{$\BJ_-$}}
    \put(120,28){\cB{\bf F}}
    \put(118.5,18.5){\large\bf?}
 \end{picture}}
 \label{e:11n}
\end{equation}
Let us discuss the meaning of this diagram with some care. If the central Adinkra is taken to depict a worldline $N = 2$ supermultiplet, the node $\bs\F$ could be ``lifted'' to obtain the Adinkras to the right or left, wherein both $\vd_\pp$ and $\vd_\mm$ would be $\ddt$ and the Adinkra on to the right would be identical with the one to the left.

In the context of a representation of worldsheet $(1,1)$-supersymmetry, either $\vd_\pp$ {\em or} $\vd_{\mm}$ can be used as shown in the Adinkras to the left and to the right,  respectively.  The use of distinct partial derivatives then leads to the distinct diagrams as shown. It is easy to see that in these diagrams all the edges|except for the ones flagged by a question-mark|do depict the action by either \C1{$\rD_+$ (red)} or \C3{$\rD_-$ (blue)}, such as:
\begin{equation}
 \big(\BJ_+\Defl\rD_+\BF) \tooo{-\rD_-} \big(\BF_\pm\Defl\rD_+\rD_-\BF\big),\quad
 \big(\BJ_+\Defl\rD_+\BF) \tooo{\rD_+} (i\vd_\pp\BF),\quad
 \textit{etc.}
\end{equation}
However, the edges flagged by the question-marks
\begin{equation}
 \big(\BJ_-\Defl\rD_-\BF) \tooo{~\bs?~} (i\vd_\pp\BF),\qquad\text{and}\qquad
 \big(\BJ_+\Defl\rD_+\BF) \tooo{~\bs?~} (i\vd_\mm\BF)
 \label{e:s3/2}
\end{equation}
evidently require a spin-$(\pm\frc32)$ operator\ft{The component superfields in\eq{e:11n} were drawn with relative horizontal position proportional to spin.} of engineering dimension $+\frc12$, of which there are none within the supersymmetry algebra\eq{e:pqSuSy}; tracing along the Haag-{\Lv}opusa\'nski-Sohnius theorem\cite{rHLS-SuSy} shows that no such local operator exists.
 Thus, in adinkraic worldsheet $(p,q)$-supermultiplets for $p,q\neq0$, height (\ie, engineering dimension) rearrangements are much more restricted than they are in the worldline case\cite{rA,r6-1,r6-3.2}.

More to the point, the inability to perform individual ``node-raising''\eq{e:11n} leads to:
\begin{thrm}\label{T:bowT}
Adinkras depicting off-shell supermultiplets of worldsheet $(p,q)$-supersymmetry contain no ``ambidextrous 2-color bow-ties'':
\begin{equation}
 \vC{\begin{picture}(50,30)
   \put(12,3){\includegraphics[height=24mm]{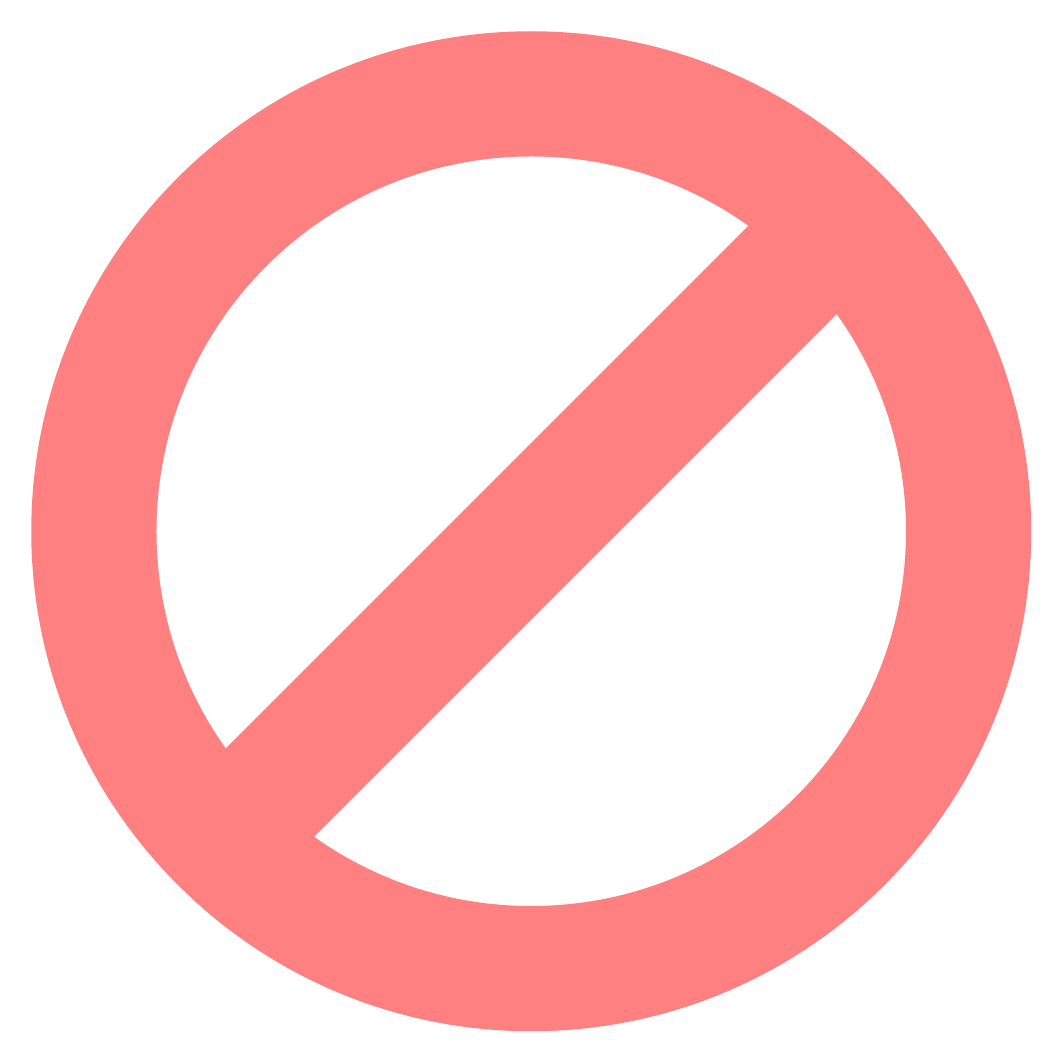}}
   \put(0,3){\includegraphics[height=24mm]{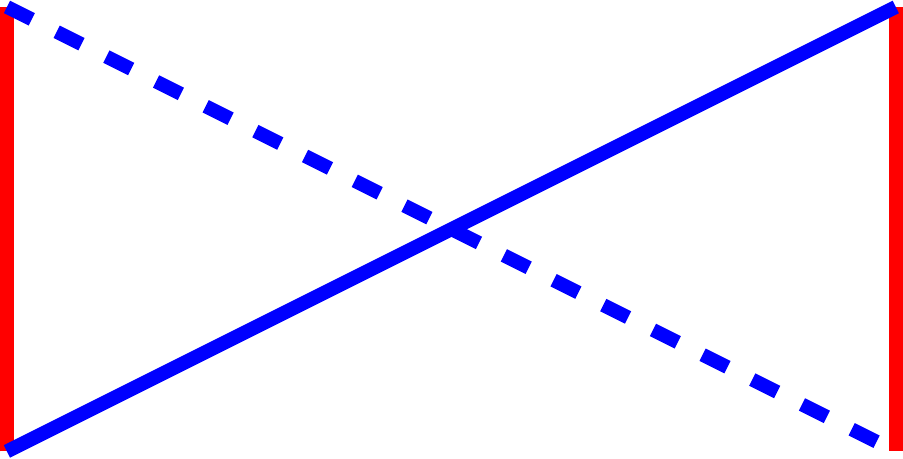}}
    \put(-6,14){\C1{$\rD_+$}}
    \put(48,14){\C1{$\rD_+$}}
    \put(3,9){\C3{$\rD_-$}}
    \put(36,9){\C3{$-\rD_-$}}
    \put(1,2){\cB{$\BF$}}
    \put(47,2){\cB{{\bf X}}}
    \put(1,25){\cB{$\BJ_+$}}
    \put(47,25){\cB{$\BJ_-$}}
 \end{picture}}
 \label{e:AmBow22}
\end{equation}
\end{thrm}
\begin{proof}
Following the edges from $\BF$ to $\bf X$ in two different ways, we conclude:
\begin{alignat}7
  \rD_+\BF &=\BJ_+&&=-\rD_-{\bf X},\quad&&\To&\quad
   \spin(\BF){+}1&=\spin(\BJ_+)=\spin({\bf X}){-}1,\nn\\
   &&&&&\To&\quad\spin({\bf X})&=\spin(\BF)+2;\label{e:F+2}\\[2mm]
  \rD_-\BF &=\BJ_-&&=\rD_+{\bf X},\quad&&\To&\quad
   \spin(\BF){-}1&=\spin(\BJ_-)=\spin({\bf X}){+}1,\nn\\
   &&&&&\To&\quad\spin({\bf X})&=\spin(\BF)-2.\label{e:F-2}
\end{alignat}
Clearly, \Eq{e:F+2} and\eq{e:F-2} cannot both be true, \ie, no {\bf X} can exist to complete the (sub)super\-mul\-ti\-plet as depicted in\eq{e:AmBow22}.
\end{proof}
 This result may be reformulated in the following useful form:
\begin{thrm}[Spin Sum Rule]\label{T:SSR}
 Within any Adinkra depicting an off-shell supermultiplet of worldsheet $(p,q)$-supersymmetry, to every edge depicting the transformation $\rD_I:\bF\!_A\to\bF\!_B$ (up to worldsheet derivatives and multiplicative constants of convenience), assign the height-weighted spin:
\begin{equation}
  \mS{I\,B}A\Defl\spin(\rD_I)\big([\bF\!_B]-[\bF\!_A]\big),\qquad
  I=(\a+),(\ad-).
 \label{e:DefSiBA}
\end{equation}
 The sum of $\mS{I\,B}A$ around any 2-colored closed quadrangle must vanish:
\begin{equation}
  \TS{IJ}\Defl
   \mS{J\,A}D+\mS{I\,D}C+\mS{J\,C}B+\mS{I\,B}A,\qquad
  \TS{IJ}\Is0,
 \label{e:SumS=0}
\end{equation}
with no sum on $I,J$, indicating two edge-colors, \ie, two supersymmetry transformations.
\end{thrm}
For example, in the putative (sub-)Adinkra\eq{e:AmBow22}, if we start from {\bf X} of engineering dimension $-\inv2$, and follow the edges: \C1{$\rD_+$} (straight up), \C3{$\rD_-$} (down left), \C1{$\rD_+$} (straight up), \C3{$-\rD_-$} (down right) through the ``ambidextrous 2-colored bow-tie,'' the sum of $\mS{I\,B}A$'s is:
\begin{equation}
 \begin{aligned}
  &(+\inv2)\big[(0)-(-\inv2)\big]
  +(-\inv2)\big[(-\inv2)-(0)\big]
  +(+\inv2)\big[(0)-(-\inv2)\big]
  +(-\inv2)\big[(-\inv2)-(0)\big]\\
  &=+\inv4+\inv4+\inv4+\inv4=+1\neq0,\quad
  \simeq
   \spin\big((\C3{-\rD_-})^{-1}\circ\C1{\rD_+}\circ(\C3{\rD_-})^{-1}\circ\C1{\rD_+}\big),
 \end{aligned}
\end{equation}
where the inverses denote that the action (and the spin) of the operator is being reversed, going from a higher to a lower node. By way of contrast, the same computation for the diamond-shaped quadrangle in the middle of\eq{e:11n}, starting from the bottom node and following counter-clockwise gives:
\begin{equation}
 \begin{aligned}
  &(+\inv2)\big[(0)-(-\inv2)\big]
  +(-\inv2)\big[(0)-(-\inv2)\big]
  +(+\inv2)\big[(-\inv2)-(0)\big]
  +(-\inv2)\big[(-\inv2)-(0)\big]\\
  &=+\inv4-\inv4-\inv4+\inv4=0,\quad
  \simeq
   \spin\big((\C3{-\rD_-})^{-1}\circ(\C1{\rD_+})^{-1}\circ\C3{\rD_-}\circ\C1{\rD_+}\big).
 \end{aligned}
\end{equation}
The next section will demonstrate the filtering efficiency of theorems~\ref{T:bowT} and~\ref{T:SSR}, \ie, Corollary~\ref{C:ObSuSy} on the particular case of extending worldline $N\,{=}\,4$ supersymmetry without central charges into worldsheet $(2,2)$-supersymmetry without central charges.

In view of their r\^ole in theorems~\ref{T:bowT} and~\ref{T:SSR},
\begin{corl}[extension obstruction]\label{C:ObSuSy}
The ambidextrous 2-color bow-tie sub-Adinkra\eq{e:AmBow22} and its numerical counterpoint, $\TS{IJ}$, is an obstruction for dimensional extension of worldline off-shell $N$-extended supersymmetry without central charge into worldsheet off-shell $(p,q)$-supersymmetry without central charges for $p,q\neq0$.
\end{corl}

However, before continuing with explicit examples, few general remarks are in order:
\begin{enumerate}\itemsep=-3pt\vspace{-2mm}

 \item\label{i:1}
  The ``spin sum rule'' is independent of the ``dashing rule''\cite{rA,r6-1} whereby the number of dashed edges in any quadrangle within any Adinkra must be odd, and which stems from the anticommutivity of the $\rD$'s; see below, at the end of this section. In fact, the above ``spin sum rule'' is unaffected by changes in the solid/dashing assignments.

 \item\label{i:2}
  Since spins are eigenvalues of the Lorentz generator, their engineering dimension weighted sum along concatenated edges is the net {\em\/height-weighted spin\/} of the corresponding D-monomial.
 Each closed path depicts a {\em\/closed $\rD$-orbit\/}\ft{An orbit generated by a sequential application of superderivatives is considered closed it if returns to a constant multiple of a $\vd^{\,m}_\pp\vd^{\,n}_\mm$-derivative of the initial component field for some non-negative integers $m,n$.}, so that the sum of height-weighted spins along a closed path is the trace of the height-weighted $\Spin(1,1)$-action over the given closed D-orbit.

 \item\label{i:3}
  This height-weighted ``spin sum rule'' generalizes straightforwardly to all higher dimensional spacetimes, since every higher-dimensional Lorentz group:
  \begin{enumerate}\itemsep=-3pt\vspace{-2mm}
   \item contains a continuum of $\Spin(1,1)$ subgroups,
   \item acts on supercharges ($Q$) and superderivatives ($\rD$) admitting the $\inv2\ZZ$-grading defined by the engineering dimensions $[\rD]=+\inv2=[Q]$, with dimensionless Lorentz generators.
 \end{enumerate}\vspace{-2mm}
The ``spin sum rule'' must hold for all $\Spin(1,1)\subset\Spin(1,d{-}1)$ subgroups. The ``remaining'' $\Spin(1,d{-}1)/\Spin(1,1)$ Lorentz symmetry evidently must also act consistently, which poses additional obstructions|presumably generating the criteria of Refs.\cite{rFIL,rFL}.
 
 \item\label{i:4}
  In turn, note that both
  {\em\/2-colored unidextrous\/} and
  {\em\/4-colored ambidextrous\/} bow-ties:
\begin{equation}
 \vC{\begin{picture}(65,28)(0,-2)
   \put(0,0){\includegraphics[height=24mm]{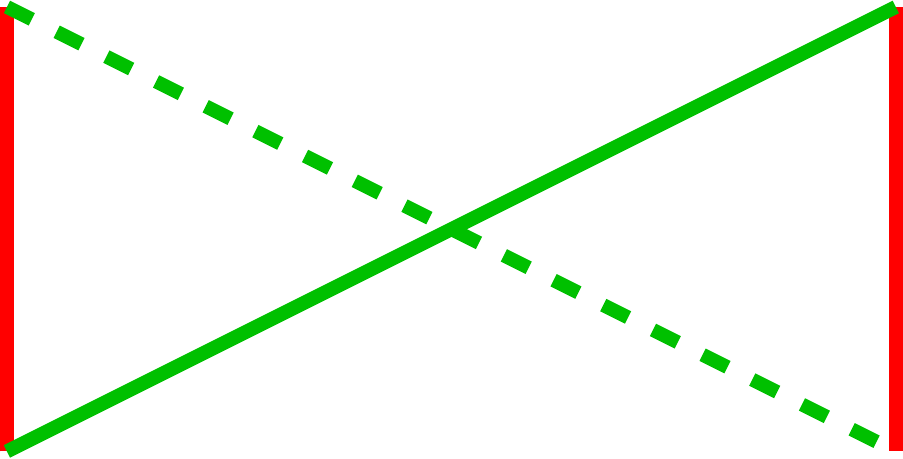}}
    \put(0,0){\cB{$\bs\F_1$}}
    \put(47,0){\cB{$\bs\F_2$}}
    \put(0,22){\cB{$\bs\J_{1+}$}}
    \put(47,22){\cB{$\bs\J_{2+}$}}
    \put(-1,13){\Rx{\C1{$\rD_{1+}$}}}
    \put(13,8){\Rx{\C2{$\rD_{2+}$}}}
    \put(34,8){\Lx{\C2{$-\rD_{2+}$}}}
    \put(49,13){\Lx{\C1{$\rD_{1+}$}}}
 \end{picture}}
 \qquad
 \vC{\begin{picture}(55,28)(0,-2)
   \put(0,0){\includegraphics[height=24mm]{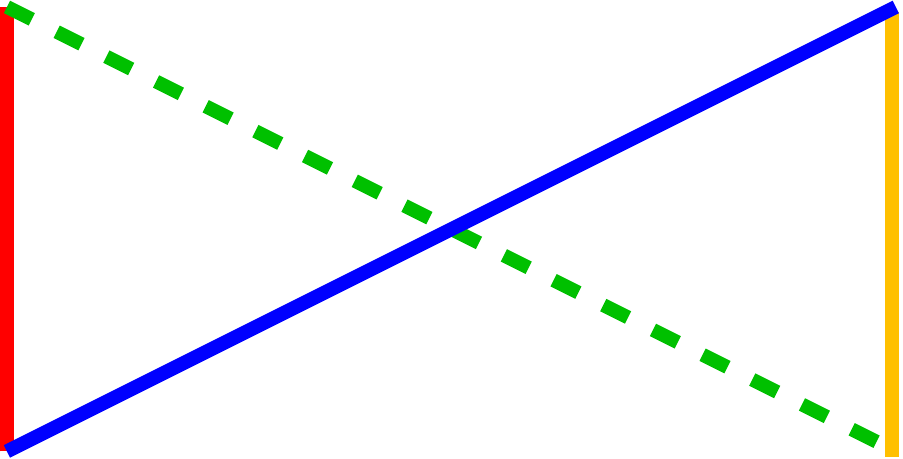}}
    \put(0,0){\cB{$\bs\F_1$}}
    \put(47,0){\cB{$\bs\F_2$}}
    \put(0,22){\cB{$\bs\J_+$}}
    \put(47,22){\cB{$\bs\J_-$}}
    \put(-1,13){\Rx{\C1{$\rD_{1+}$}}}
    \put(13,8){\Rx{\C3{$\rD_{1-}$}}}
    \put(34,8){\Lx{\C2{$-\rD_{2+}$}}}
    \put(49,13){\Lx{\C4{$\rD_{2-}$}}}
 \end{picture}}
\end{equation}
satisfy theorems~\ref{T:bowT} and~\ref{T:SSR}, and indeed are perfectly consistent (sub-)Adinkras.
\end{enumerate}
We than immediately have:
\begin{corl}[valises]\label{C:v}
Off-shell adinkraic 2-level supermultiplets (so-called {\em\/valises\/}\cite{r6-3}; also called ``short multiplets'' in Ref.\cite{rPT} and the default member of the ``root superfield representations'' of Ref.\cite{rGLP}) with no gauge equivalence condition can only exist in 1-dimensional (no space, one time) models and for unidextrous supersymmetry, which is known to exist only in spacetimes of signatures $(t,s)$ with $t{-}s=0\bmod8$: only these cases admit models with Majorana-Weyl (real chiral) fermions of only one helicity\cite{rSohn}, and reality (Hermiticity) is required for lagrangians of all physically acceptable models.
\end{corl}
For example, the four off-shell supermultiplets of worldsheet $(4,0)$-supersymmetry examined in Ref.\cite{rDGR} are all {\em\/valises\/}; they conform to our criterion straightforwardly, since these models exhibit unidextrous (chiral) supersymmetry, wherein no ``ambidextrous 2-color bow-tie'' obstruction\eq{e:AmBow22} can exist. Evidently, this implies:
\begin{corl}[unidextrous extension, off-shell]\label{C:N0}
All off-shell supermultiplets of $N$-extended worldline supersymmetry without central charges extend to off-shell supermultiplets of worldsheet $(N,0)$-supersymmetry without central charges through the {\em\/unidextrous\/} identifications
\begin{equation}
  \{\,\rD_I,\ddt\,\} \mapsto \{\,\rD_{I+},\vd_\pp\,\},
\end{equation}
where the $\vd_\mm$-action on component (super)fields remains unrestricted.
The analogous holds for parity-mirrored extension to worldsheet $(0,N)$-supersymmetry.
\end{corl}
It is also possible to use the unidextrous extension $\{\,\rD_I,\ddt\,\} \mapsto \{\,\rD_{I+},\vd_\pp\,\}$ even within a supersymmetric theory that does include nontrivial $\rD_{\ad-}$-transformations, but these superderivatives|and then also $\vd_\mm$|must annihilate the supermultiplet:
\begin{corl}[unidextrous extension, on the half-shell]\label{C:Nq}
All off-shell supermultiplets of $N$-ex\-ten\-ded worldline supersymmetry without central charges extend to supermultiplets of worldsheet $(N,q$)-super\-sym\-metry, for arbitrary $q$, but such supermultiplets are annihilated by\/ $\rD_{\ad-}$ and consequently also by $\vd_\mm$, so that they are on-the-half-shell\cite{rHP1}.
The analogous holds for parity-mirrored extension to worldsheet $(p,N)$-supersymmetry, for arbitrary $p$.
\end{corl}
Finally, owing to the sweeping extensions in Corollaries~\ref{C:N0} and~\ref{C:Nq} and the fact that the only essential difference between the worldline and the worldsheet Poincar\'e groups is spin, which has been accounted for by the twin theorems~\ref{T:bowT} and~\ref{T:SSR} and Corollary~\ref{C:ObSuSy}, we conclude:
\begin{corl}
The obstruction identified in the twin theorems~\ref{T:bowT} and~\ref{T:SSR} and Corollary~\ref{C:ObSuSy} is the only obstruction for extending worldline $N$-extended supersymmetry without central charge into worldsheet $(p,q)$-super\-sym\-metry without central charges, for all $p,q$.
\end{corl}

\paragraph{Combinatorial Complexity}
The dashing rule for turning a ``one-hooked hypercube'' (where one of the vertices is ``hooked'' while the rest extend freely upward) into an Adinkra\cite{r6-1} is partially responsible for the existence of the myriad of Adinkras.
 The number of ways a hypercube can be turned into such an Adinkra was recently calculated by Yan Zhang\cite{rYZpc2011} and found to be:
\begin{equation}
  \#(\text{``one-hooked Adinkras''}) = 2 \cdot N! \cdot 2^{2^N - 1 }
 \label{e:Yan}
\end{equation}
so that it can be seen a double exponential drives some of the growth with $N$.

In turn, there is also a fast-growing group of equivalences: An $N$-cube has $2^N$ vertices, each of which corresponds to a component field. The component field redefinition of changing the sign of a component field also flips the solid/dashed assignment of all the edges incident with that vertex. This yields $2^{2^N}$ distinct sign choices for the component fields. Also, the horizontal rearrangements of the nodes within the one-hooked Adinkra, generates another (sub)group of evident equivalences, and has $\prod_{k=0}^N {N \choose k}!$
elements. Thus, the group of equivalences has at least $2^{2^N}\cdot\prod_{k=0}^N {N \choose k}!$ elements. Finally, note that this huge group of equivalences does not act transitively on the set of all variously dashed one-hooked hypercubes, so the number of inequivalently dashed one-hooked hypercubes is not simply the ratio of\eq{e:Yan} by $\big(2^{2^N}{\cdot}\prod_{k=0}^N {N \choose k}!\big)$. In fact, all one-hooked N-cubical Adinkras are equivalent to each other.

Adinkras with more complicated height-arrangements depict supermultiplets with more complicated choices of relative engineering dimension assignments for the component fields. The numbers of such inequivalent Adinkras then grows combinatorially with $N$; the ``node choice group'' of Ref.\cite{r6-3.2} was defined to encode the symmetries in arbitrary Adinkras.

It is then gratifying to note that the severe restrictions on the combinatorial variety of Adinkras placed by the twin theorems~\ref{T:bowT} and~\ref{T:SSR} markedly reduce the number of Adinkras that may depict off-shell worldsheet $(p,q)$-supermultiplets for any fixed $p,q$. Clearly, a computer-aided classification of worldsheet supermultiplets akin to the classification of Refs.\cite{r6-1,r6-3,r6-3.2,rRLM-Codes} would highly desirable, and hopefully can be implemented by appropriate encoding the results presented herein.

\section{Examples: Worldsheet (2,2)-Supersymmetry}
\label{s:XMpls}
Ref.\cite{r6-3.2} depict all adinkraic representations of $(N\,{=}\,4)$-extended worldline supersymmetry by listing 28 Adinkras, but without showing
 ({\bsf1})~the solid/dashed edge distinction,
 ({\bsf2})~the boson\,$\iff$\,fermion flipped versions, 
 ({\bsf3})~the {\em\/twisted\/}\ft{Twisting flips the solid/dashed parity of edges of an odd number of colors in an Adinkra\cite{r6-1}.} versions of the 4 ``half-sized'' supermultiplets. This gives a total of $2{\cdot}(24+2{\cdot}4)=64$ distinct Adinkras, still not counting {\em\/nodal permutations\/}\ft{These are permutations of white and separately black nodes across different heights; they leave the node-per-height count unchanged and horizontal permutations are inconsequential.}. Of these, only:
\begin{equation}
 \vC{\begin{picture}(160,23)
   \put(0,0){\includegraphics[width=160mm]{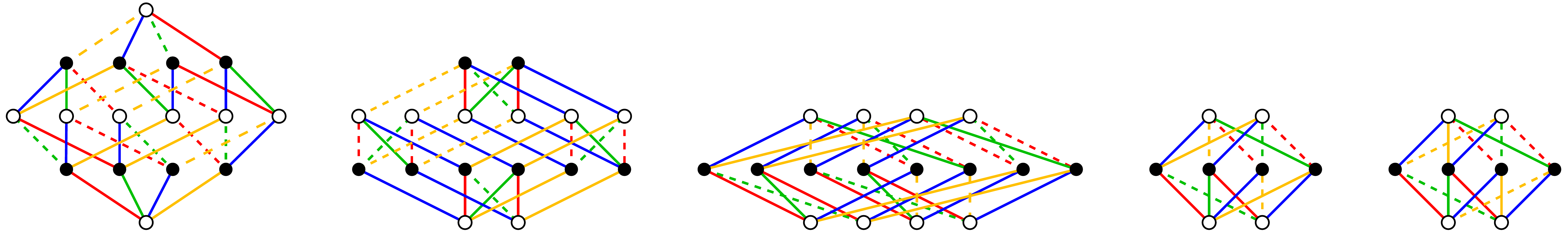}}
    \put(5,1){\bsf A}
     \put(-2,21){\footnotesize(intact)}
    \put(37,1){\bsf B}
     \put(41,21){\footnotesize(semi-chiral)}
    \put(74,1){\bsf C}
     \put(73,15){\footnotesize(decomposes $\to$ {\bsf D}+{\bsf E})}
    \put(117,1){\bsf D}
     \put(120.5,15){\footnotesize(chiral)}
    \put(141,1){\bsf E}
     \put(142,15){\footnotesize(tw.-chiral)}
 \end{picture}}
 \label{e:Five2.2}
\end{equation}
and their upside-down and boson\,$\iff$\,fermion flipped versions have no ambidextrous 2-colored bow-ties, and so satisfy theorems~\ref{T:bowT} and~\ref{T:SSR}, respectively. For example, one of the ``half-sized'' supermultiplets from the tables of Ref.\cite{r6-3.2} that does not pass this requirement is
\begin{equation}
 \vC{\begin{picture}(120,22)(0,1)
   \put(11.8,10.5){\includegraphics[width=10mm]{No.pdf}}
   \put(79,10.5){\includegraphics[width=10mm]{No.pdf}}
   \put(0,0){\includegraphics[width=120mm]{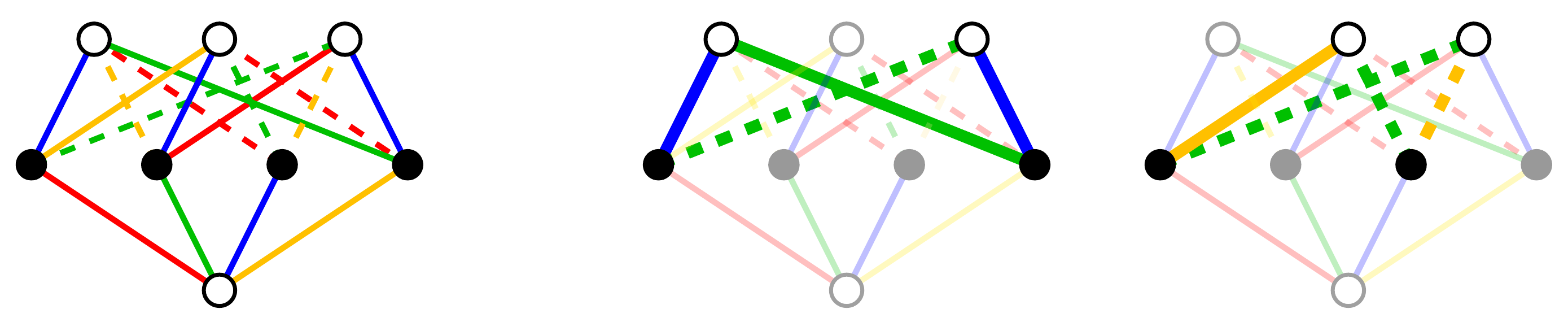}}
    \put(7,4){\scriptsize\C1{$\rD_{1+}$}}
    \put(8,9){\scriptsize\C2{$\rD_{2+}$}}
    \put(21,9){\scriptsize\C3{$\rD_{1-}$}}
    \put(23.5,4){\scriptsize\C4{$\rD_{2-}$}}
 \end{picture}}
 \label{e:143bows}
\end{equation}
which contains two ambidextrous 2-colored bow-ties, as highlighted in the two copies to the right, where the edges forming the extension-obstructing ``bow-ties'' have exaggerated thickness and the remaining graph elements are rendered in paler hues.

We now examine the five Adinkras\eq{e:Five2.2}, of which {\bsf A}--{\bsf C} have the chromotopology of a 4-cube, while the chromotopologies of {\bsf D} and {\bsf E} are two inequivalent $\ZZ_2$ quotients thereof; see below. Suffice it here to say that the first three of these are 1-color-decomposable: they disconnect upon deleting all edges of a single color; the last two are 2-color-decomposable.

\paragraph{Adinkra A:}  Up to flipping the sign of the four ``inner'' four component bosons in the middle row\ft{Flipping the sign of a component field depicted by the node $n$ also flips the solid/dashed assignment of each edge incident to $n$; edges connecting two sign-flipped nodes remain unchanged.}, the nodes in this Adinkra depict the superderivatives used to project component fields\cite{rHSS,r6-1}:
\begin{equation}
 \vC{\unitlength=1.067mm
     \begin{picture}(140,40)(1,0)
      \put(-5,.5){\includegraphics[width=160mm,height=40mm]{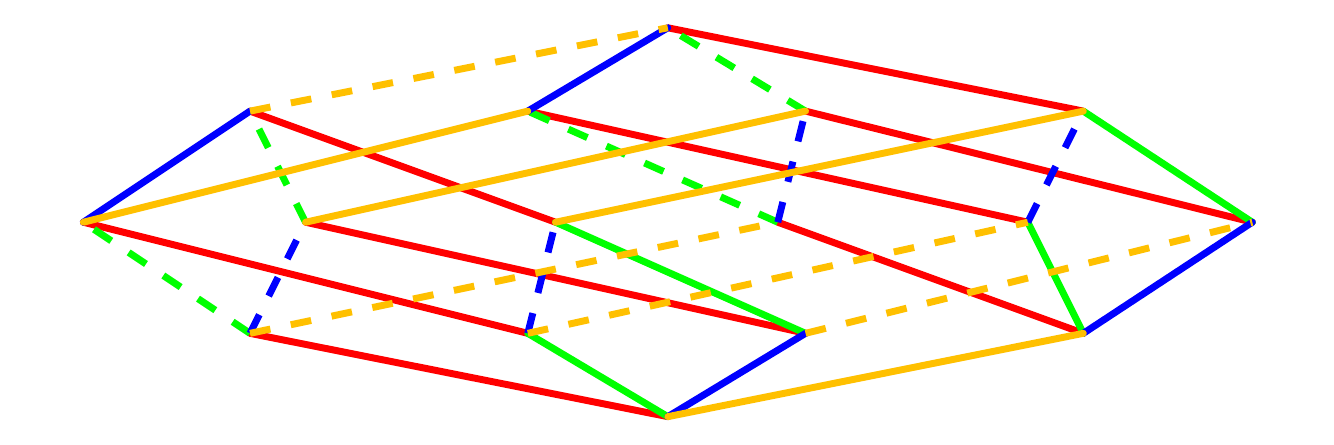}}
      \footnotesize
      \put(70,36){\cB{$\inv4[\rD_{1+},\rD_{2+}][\rD_{1-},\rD_{2-}]$}}
      \put(22,27.5){\cB{$\inv2[\rD_{1+},\rD_{2+}]\rD_{1-}$}}
      \put(54,27.5){\cB{$\inv2[\rD_{1+},\rD_{2+}]\rD_{2-}$}}
      \put(86.5,27.5){\cB{$\inv2[\rD_{1-},\rD_{2-}]\rD_{1+}$}}
      \put(118,27.5){\cB{$\inv2[\rD_{1-},\rD_{2-}]\rD_{2+}$}}
      \put(4.75,18){\cB{$\frc i2[\rD_{1+},\rD_{2+}]$}}
      \put(31,18){\cB{$i\rD_{1+}\rD_{1-}$}}
      \put(57.5,18){\cB{$i\rD_{2+}\rD_{1-}$}}
      \put(83.5,18){\cB{$i\rD_{1+}\rD_{2-}$}}
      \put(110,18){\cB{$i\rD_{2+}\rD_{2-}$}}
      \put(135.75,18){\cB{$\frc i2[\rD_{1-},\rD_{2-}]$}}
      \put(22,9){\cB{$i\rD_{1+}$}}
      \put(53.5,9){\cB{$i\rD_{2+}$}}
      \put(86.5,9){\cB{$i\rD_{1-}$}}
      \put(118,9){\cB{$i\rD_{2-}$}}
      \put(70,2){\cB{$\Ione$}}
     \end{picture}}
 \label{e:Ds}
\end{equation}
Herein, the edges (each associated with a superderivative: $\C1{\rD_{1+}\iff\text{red}}$, $\C2{\rD_{2+}\iff\text{green}}$, $\C3{\rD_{1-}\iff\text{blue}}$, $\C4{\rD_{2-}\iff\text{orange}}$) connect those superderivatives\eq{e:MonD} which differ in precisely that one $\rD$. The factor $i^{\sss[\![{\bf a,b}]\!]}$ is included in\eq{e:Ds} to insure that the component fields\eq{e:CompF} projected with the operators\eq{e:MonD} are real:
\begin{equation}
  [\![\bf{a{,}b}]\!]\Defl{\ttt\binom{|{\bf a,b}|+1}2},\quad
  |{\bf a,b}|\Defl|{\bf a}|+|{\bf b}|,\quad
  |{\bf a}|\Defl\sum_{\a=1}^p a_\a,\quad
  |{\bf b}|\Defl\sum_{\ad=1}^q b_\ad.
\end{equation}
 Both the operators in\eq{e:MonD} and the corresponding component fields of the supermultiplet are stacked in the order of increasing engineering dimension, $\frc12\big(|{\bf a,b}|\big)$. Dashed edges indicate the application (from left) of the negative of the superderivative associated to such a dashed edge. For example, $i\rD_{1+}$ is connected to $\frc{i}2[\rD_{1+},\rD_{2+}]$ by a dashed \C2{green} $(\C2{\rD_{2+}})$ edge, indicating that
\begin{equation}
  i\rD_{1+} ~\C2{\too{~(-\rD_{2+})\cdot~}}~ -i\rD_{2+}\rD_{1+}
  = \frc{i}2(\rD_{1+}\rD_{2+}-\rD_{2+}\rD_{1+}) = \frc{i}2[\rD_{1+},\rD_{2+}].
\end{equation}

By applying this tesseract of superderivatives\eq{e:Ds} to a single, intact superfield {\em\/\`a la\/} Salam and Strathdee\cite{rSSSS1,rSSSS5} and projecting the result to the worldsheet (\ie, setting the fermionic coordinates of the super-worldsheet to zero), we obtain the component fields of this familiar supermultiplet. The edges in the Adinkra\eq{e:Ds} then depict the action of the supersymmetry transformations within the supermultiplet defined by the Salam-Strathdee superfield.
 Alternatively, we may introduce a superfield in place of each node of the tesseract\eq{e:Ds}:
\begin{equation}
 \vC{\unitlength=1.067mm
     \begin{picture}(140,35)(0,1)
      \put(-5,.5){\includegraphics[width=160mm,height=40mm]{Spindle.pdf}}
      \footnotesize
      \put(70,35){\cB{\pmb{\cF~}}}
      \put(22,27.5){\cB{$\BX^-_1$}}
      \put(54,27.5){\cB{$\BX^-_2$}}
      \put(86.5,27.5){\cB{$\BX^+_1$}}
      \put(118,27.5){\cB{$\BX^+_2$}}
      \put(4.75,18){\cB{${\bf f}^\mm$}}
      \put(30,18){\cB{${\bf f}_{11}$}}
      \put(57.5,18){\cB{${\bf f}_{21}$}}
      \put(83,18){\cB{${\bf f}_{12}$}}
      \put(110.5,18){\cB{${\bf f}_{22}$}}
      \put(135.75,18){\cB{${\bf f}^\pp$}}
      \put(22,9){\cB{$\BJ^-_1$}}
      \put(53.5,9){\cB{$\BJ^-_2$}}
      \put(86.5,9){\cB{$\BJ^+_1$}}
      \put(118,9){\cB{$\BJ^+_2$}}
      \put(70,1.5){\cB{$\BF$}}
     \end{picture}}
 \label{e:BF}
\end{equation}
whereupon the edges depict the superdifferential relations generalizing\eqs{e:N2ES}{e:N2IS}. This results in the same supermultiplet as defined by the single Salam-Strathdee superfield; the lowest component fields in the sixteen component superfields\eq{e:BF} are (up to multiplicative constants for convenience) identical with the component fields defined by projection using the tesseract of superderivatives\eq{e:Ds}, applied to a single intact superfield.

\paragraph{Adinkra B:} This Adinkra represents a $\rD_{\a+}\iff\rD_{\ad-}$ mirror {\em\/pair\/} of distinct $(2,2)$-supermultiplets. The Adinkra itself is a tensor product of the two Adinkras\eq{e:N2ES} and\eq{e:N2IS}:
\begin{equation}
 \vC{\begin{picture}(120,30)
   \put(0,0){\includegraphics[height=32mm]{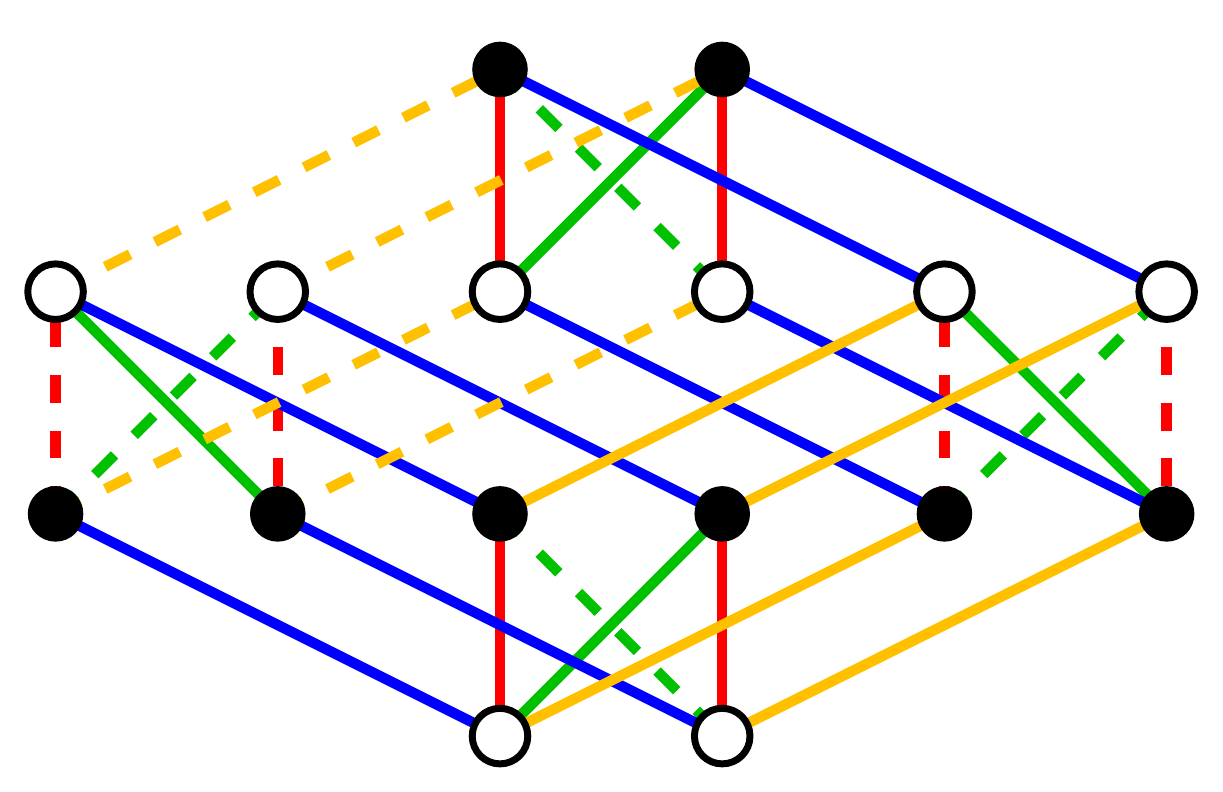}}
   \put(54,10){\LARGE$=$}
   \put(65,0){\includegraphics[height=22.5mm]{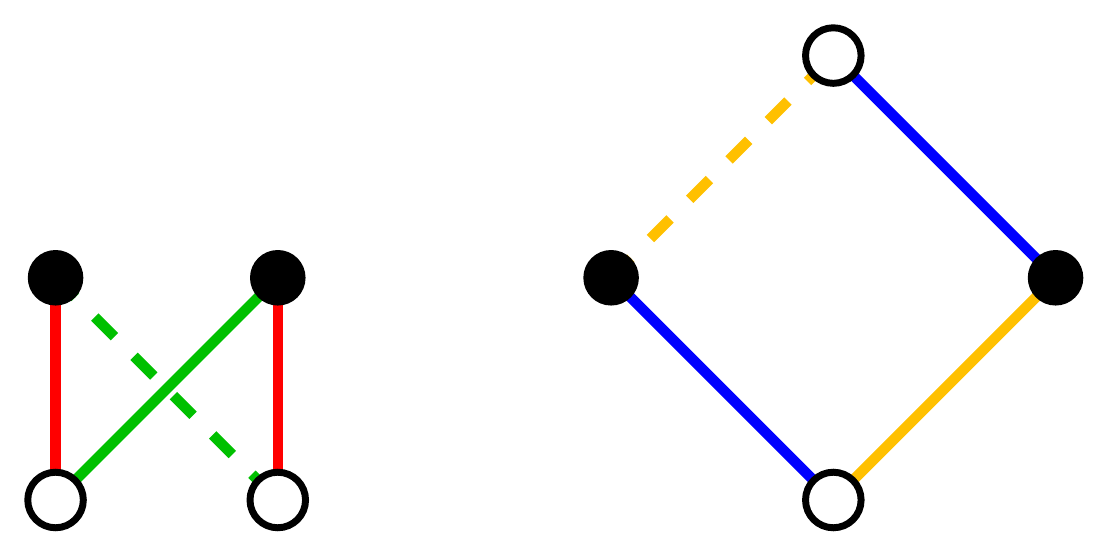}}
   \put(80,10){\LARGE$\otimes$}
 \end{picture}}
 \label{e:B2662}
\end{equation}
and it is possible to identify the first factor as depicting the $\rD_{1+},\rD_{2+}$ action and the second one the $\rD_{1-},\rD_{2-}$ action---or the other way around. In this case, boson\,$\iff$\,fermion flipping coincides with upside-down flipping after some additional judicious component field sign-changes.

 Notice the left-right asymmetry between the $\rD_{\a+}$-action (\C1{red} and \C2{green} edges) and the $\rD_{\ad-}$-action (\C3{blue} and \C4{orange} edges). As shown in the appendix, this implies that the real component (super)fields depicted by the nodes of the Adinkra may be complexified {\em\/simultaneously\/} with the two real components of $\rD_{\a+}$.
 The appendix also proves that the supermultiplet depicted by\eq{e:B2662} is one of the two semi-chiral supermultiplets\cite{rSChSF0,rSChSF}, the other one obtained by swapping the assignment to the edges $\rD_{\a+}\iff\rD_{\ad-}$. The conjugate supermultiplets are of course obtained by swapping $i\iff-i$ in the complex combinations of the component fields and the superderivatives assigned to the edges of the left-hand side factor in the tensor product\eq{e:B2662}; the real Adinkra stays the same.

\paragraph{Adinkra C:} A field redefinition shows this Adinkra to be a direct sum of the Adinkras~{\bsf D} and {\bsf E}; see Ref.\cite{r6-3,r6-3.2} for general criteria. This owes to a $\ZZ_2$ symmetry that commutes with supersymmetry, which is made manifest by rearranging the nodes of this Adinkra horizontally:
\begin{equation}
 \vC{\begin{picture}(160,15)
   \put(0,0){\includegraphics[width=160mm]{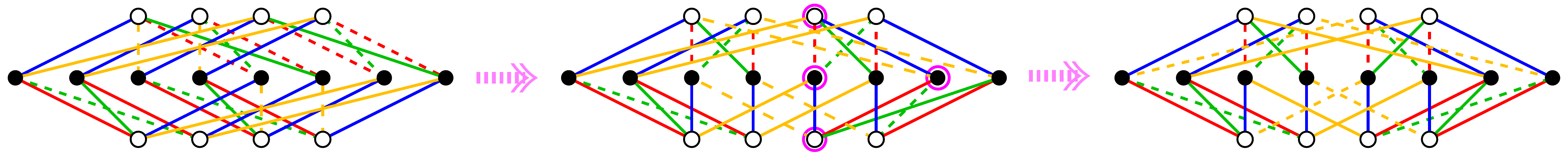}}
 \end{picture}}
 \label{e:484}
\end{equation}
then flipping the signs of the component fields represented by the encircled four nodes. This produces the right-hand side rendition of this Adinkra, which has a perfect literal left-right symmetry, so that its right-hand and left-hand halves may be identified|node-by-node and edge-by-edge:
\begin{equation}
 \vC{\begin{picture}(120,28)
   \put(0,0){\includegraphics[height=30mm]{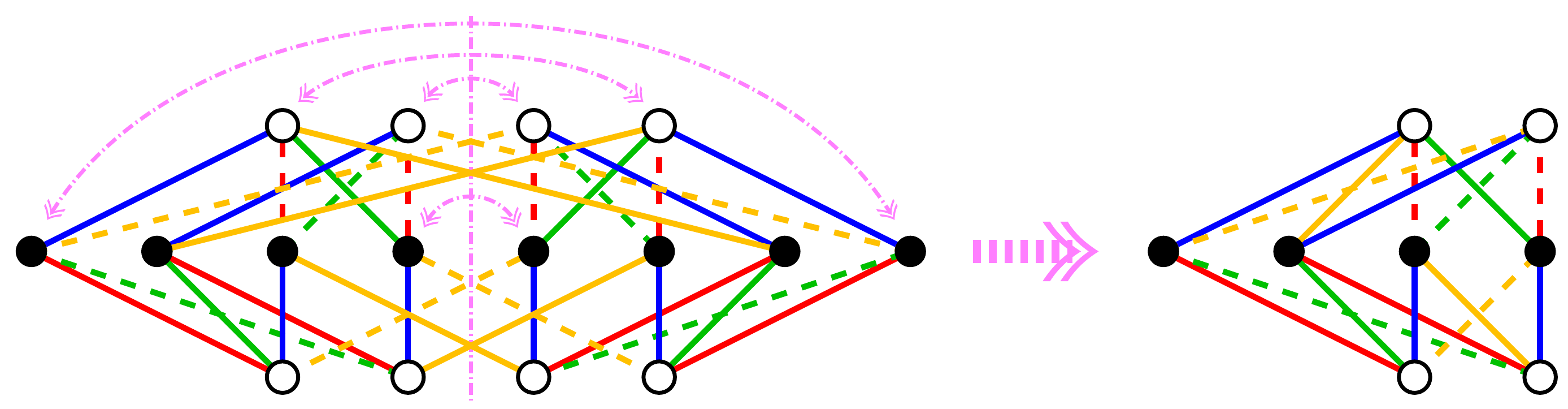}}
 \end{picture}}
 \label{e:B484>242}
\end{equation}
resulting in a half-sized Adinkra, identical to {\bsf E} in\eq{e:Five2.2}. Identifying instead the {\em\/negative\/} of the right-hand nodes with those in the left-hand half flips the solid/dashed parity of only the golden edges, yielding Adinkra~{\bsf D} in\eq{e:Five2.2}.  Such projections are fully explored and catalogued in Refs.\cite{r6-3,r6-3.2}.

\paragraph{Adinkra D and E:} After a complex combination of the component fields, these Adinkras depict the chiral and twisted-chiral worldsheet supermultiplets\cite{rGHR}, respectively. Note that the complex combination of the component fields is consistent with simultaneously combining
\begin{subequations}
\begin{alignat}9
  \BD^c_+&\Defl\C1{\rD_{1+}}+i\C2{\rD_{2+}}&\quad
   &\text{with \C1{red} and \C2{green} edges,}\\
  \BD^c_-&\Defl\C3{\rD_{1-}}+i\C4{\rD_{2-}}&\quad
   &\text{with \C3{blue} and \C4{orange} edges,}
\end{alignat}
\end{subequations}
or the other way around, which however turns out to be equivalent by a simple sign-change in the two (auxiliary) component fields corresponding to the two top nodes.

\paragraph{Remark:} The real supermultiplets discussed here may well be {\em\/endowed\/} with additional structures:
\begin{enumerate}\itemsep=-3pt\vspace{-2mm}
 \item A complex structure indicates the ability to combine the real component fields into complex combinations in a way that is consistent with supersymmetry, as done in the example worked out in the appendix. If possible, the conjugate version is automatically also possible.

 \item A group $G$-action indicates the property that the component fields and the superderivatives may be combined into representations of $G$: $\BR_B$ for the bosons, $\BR_F$ for the fermions and $\BR_S$ for the superderivatives. It then must be the case that:
\begin{equation}
   \BR_S \otimes \BR_B \supset \BR_F\qquad\text{and}\qquad
   \BR_S \otimes \BR_F \supset \BR_B.
\end{equation}
Ref.\cite{rFGH} uses this approach to construct worldline lagrangians for various types of so-called ultra-multiplets, but the approach is equally applicable for worldsheet models, and also for models in higher-dimensional spacetime.
\end{enumerate}
Thus, two supermultiplets may well have identical Adinkras and so be identical as far as supersymmetry transformations are concerned, but differ by such additional structures, which are also applied to the supersymmetry generators. The simplest such difference is between a complex supermultiplet and its complex conjugate: depicted by the same Adinkra, they differ in having the complex structure $i$ \textit{vs}.\ $-i$.

\section{Examples: Worldsheet (4{\slshape k},4{\slshape k}\,)-Supersymmetry}
\label{s:XMpls+}
In fact, the filtering r\^ole of Theorem~\ref{T:bowT} may be used to ``re-engineer'' a worldline Adinkra by raising/lowering its nodes so that the end result|if possible|faithfully depicts an off-shell supermultiplet of worldsheet $(p,q)$-supersymmetry.

\subsection{$(4,4)$-Supersymmetry}
Start with a {\em\/valise\/} version of the smallest $N=8$ worldline supermultiplet, the so-called ``ultramultiplet''\cite{rGR0}, depicted as
\begin{equation}
 \vC{\includegraphics[width=80mm]{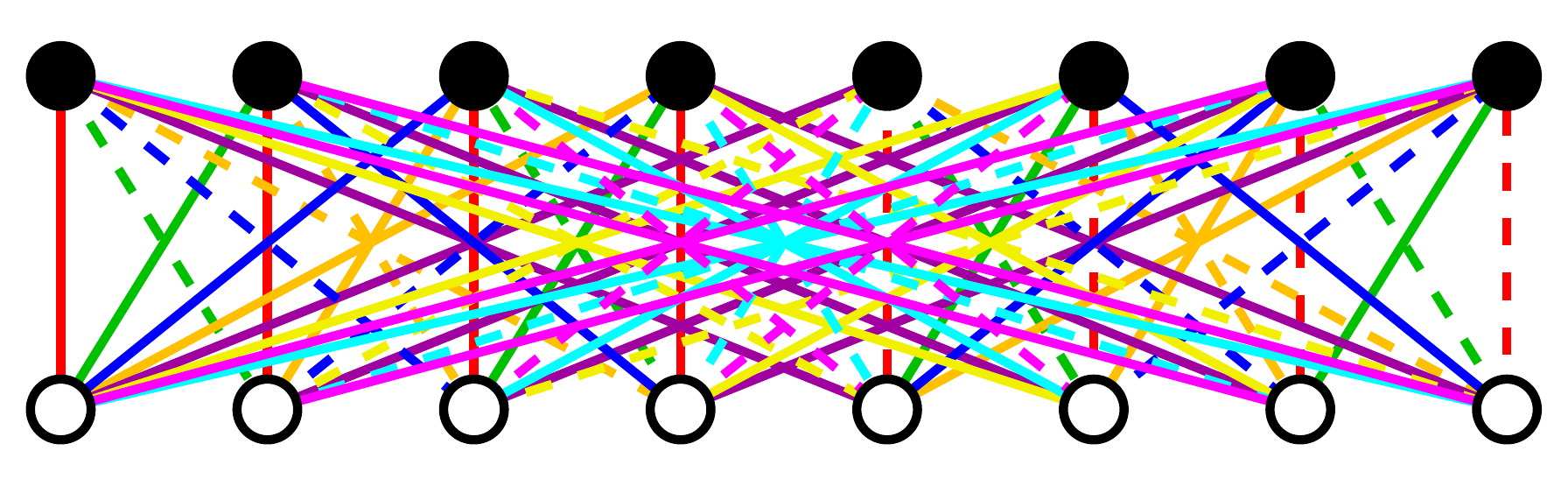}}
 \label{e:E8}
\end{equation}
which was studied extensively in terms of Adinkras in Ref.\cite{rFGH}. As it is, Corollary~\ref{C:v} implies that this Adinkra can represent an off-shell worldsheet supermultiplet only for $(8,0)$- or $(0,8)$-supersymmetry. Aiming for an off-shell supermultiplet of $(4,4)$-supersymmetry, it is clear that the edge-colors must be partitioned into two groups (to be identified with $\rD_{\a+}$- and $\rD_{\ad-}$-action), such that no edges from the first group forms a bow-tie with any of the edges from the second group. 

To this end (although the end result may perhaps appear to be evident), we may start by hiding all but two of the edge-colors, and horizontally rearrange the nodes if necessary so as to exhibit the regular pattern:
\begin{equation}
 \vC{\begin{picture}(70,20)
   \put(0,0){\includegraphics[width=70mm]{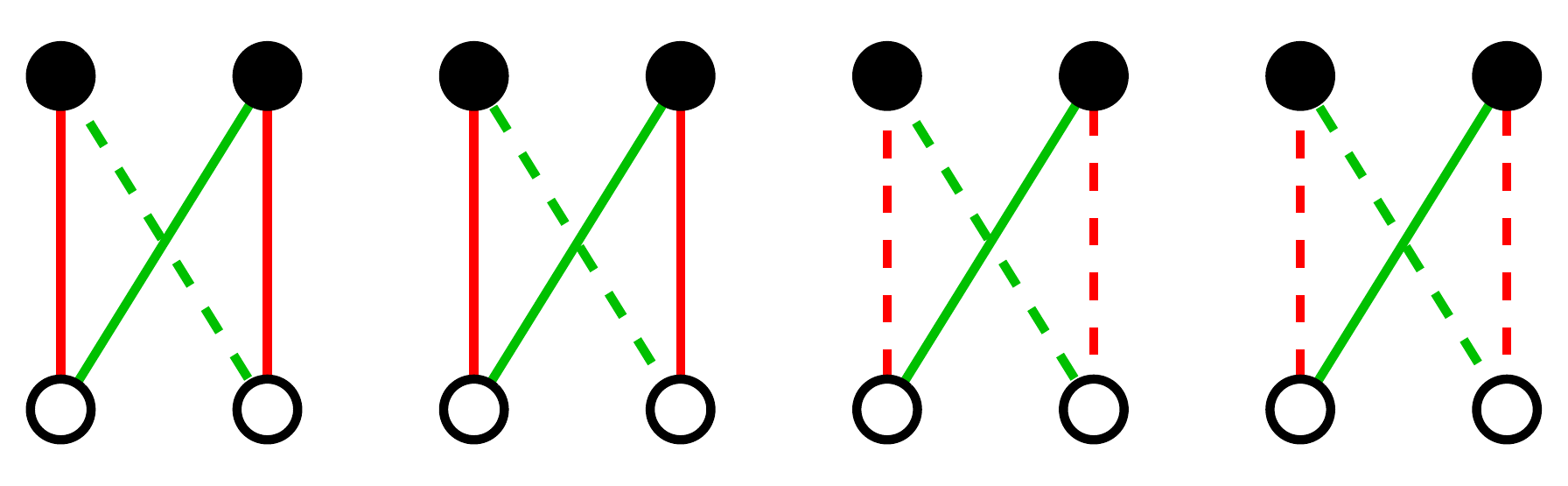}}
 \end{picture}}
\end{equation}
and associate these edges with \C1{$\rD_{1+}$} and \C2{$\rD_{2+}$}.
 We now aim to place edges of two more colors---to be associated with \C3{$\rD_{3+}$} and \C4{$\rD_{4+}$}---in a way that may (and in fact will) form 2-color bow-ties amongst themselves, but will permit adding the remaining 4 colors---to be associated with $\rD_{\ad-}$---without forming ambidextrous 2-color bow-ties. Placing edges in the 3rd color in this fashion|and maintaining an odd number of dashed edges for every quadrangle, we have:
\begin{equation}
 \vC{\begin{picture}(160,35)
   \put(0,0){\includegraphics[width=70mm]{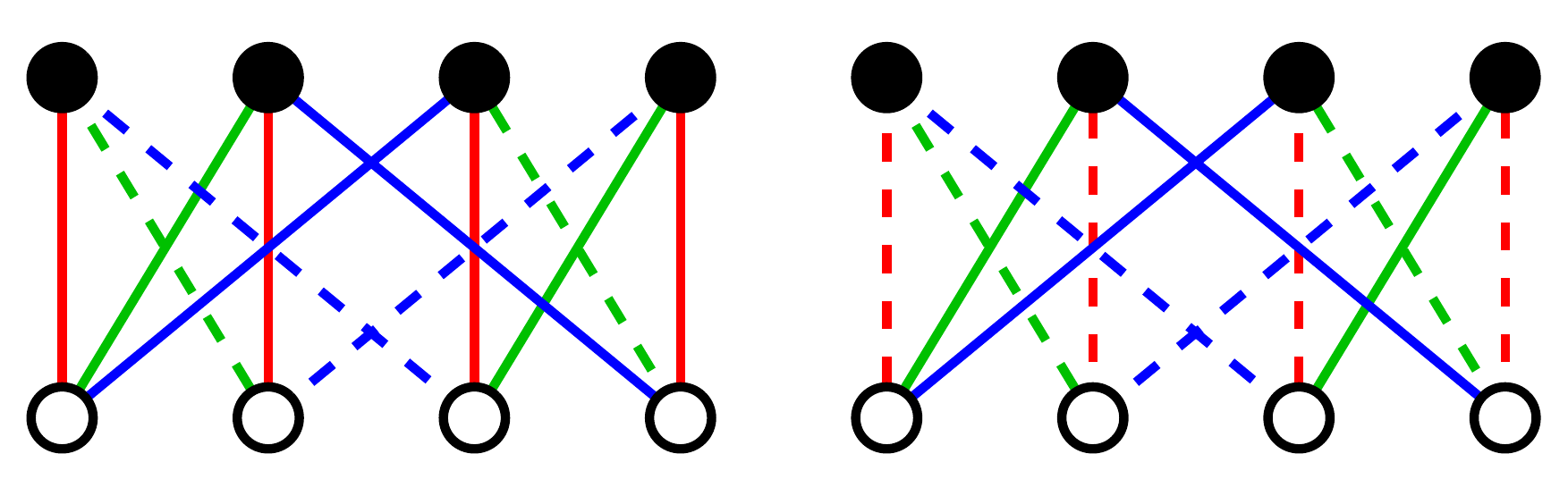}}
    \put(77,14){\LARGE$\to$}
   \put(90,0){\includegraphics[width=70mm]{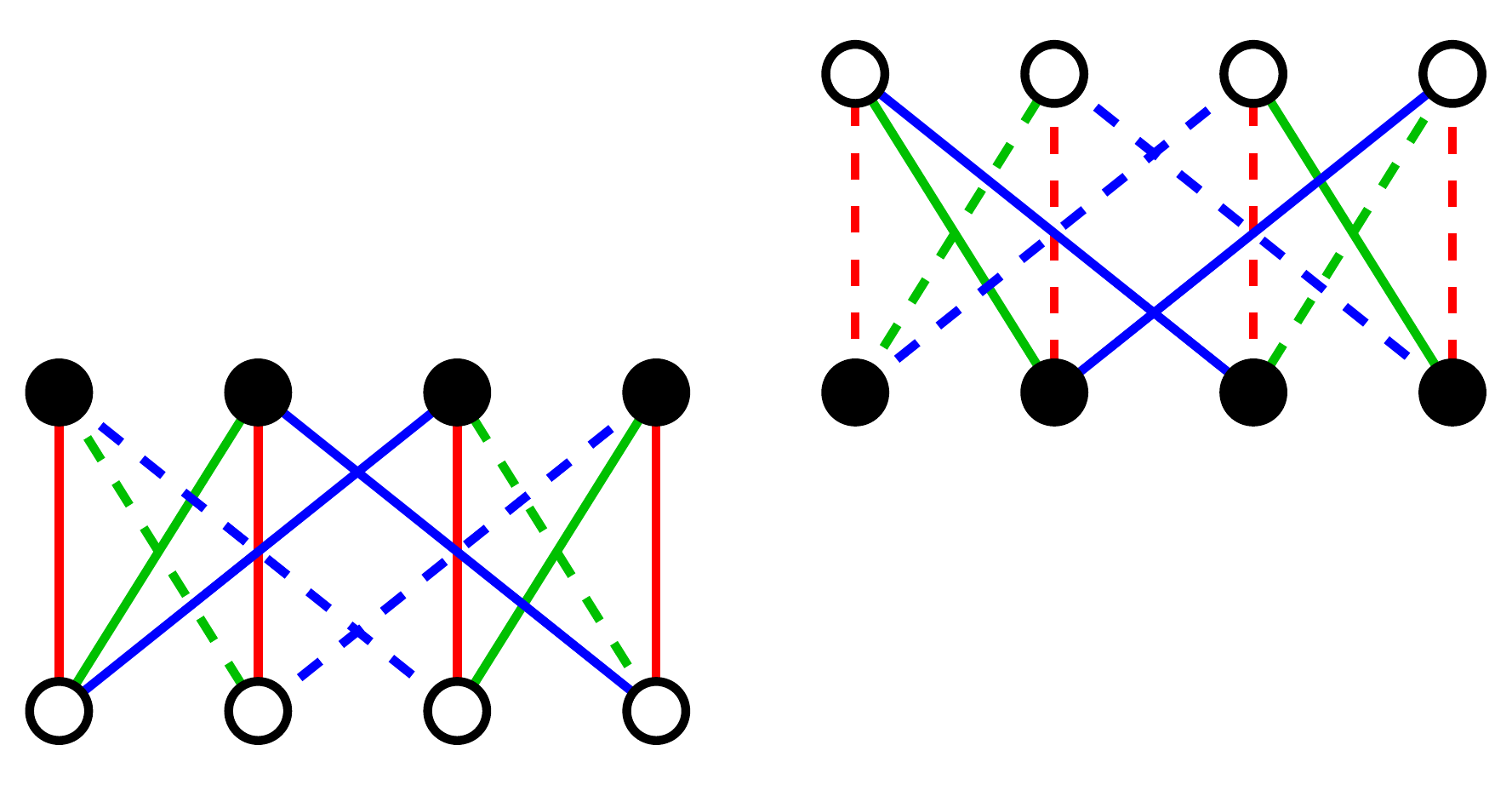}}
 \end{picture}}
\end{equation}
where we have raised the white nodes in the right-hand half, anticipating the second ($\rD_{\ad-}$) group of edges to connect the left-hand half to the right-hand half and so avoid forming ambidextrous 2-color bow-ties. Edges of the fourth color indeed do fit in without connecting the two halves:
\begin{equation}
 \vC{\begin{picture}(70,30)
   \put(0,0){\includegraphics[width=70mm]{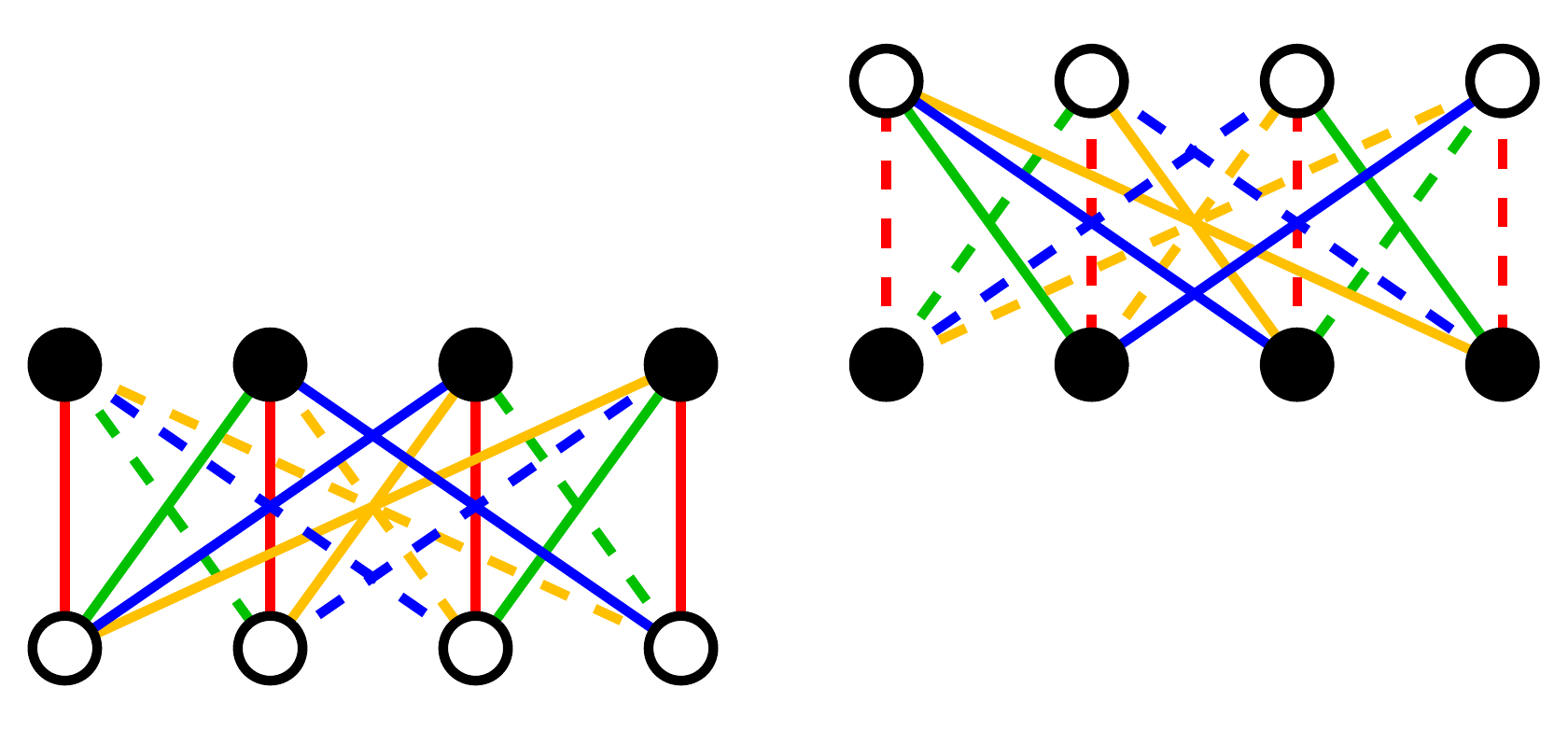}}
    \put(38,3){\parbox[b]{35mm}{\footnotesize\baselineskip=9pt
               (only the first four\hfil\newline\phantom{(}supersymmetries)}}
 \end{picture}}
\end{equation}
The edges of the remaining four colors, to be associated with $\rD_{\ad-}$ may now be added without forming two-colored ambidextrous bow-ties, as shown here pair-wise:
\begin{equation}
 \vC{\begin{picture}(150,30)
   \put(0,0){\includegraphics[width=70mm]{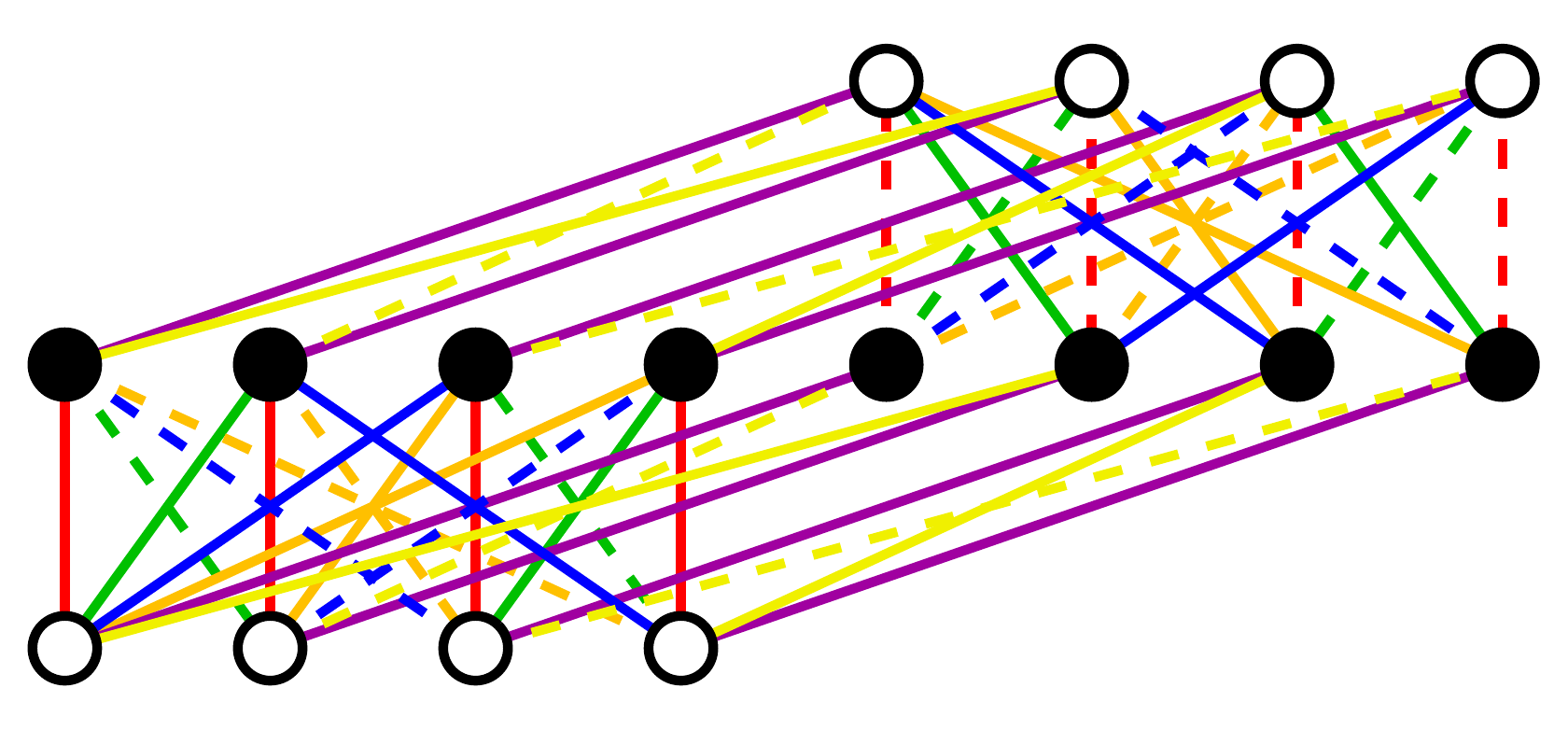}}
   \put(80,0){\includegraphics[width=70mm]{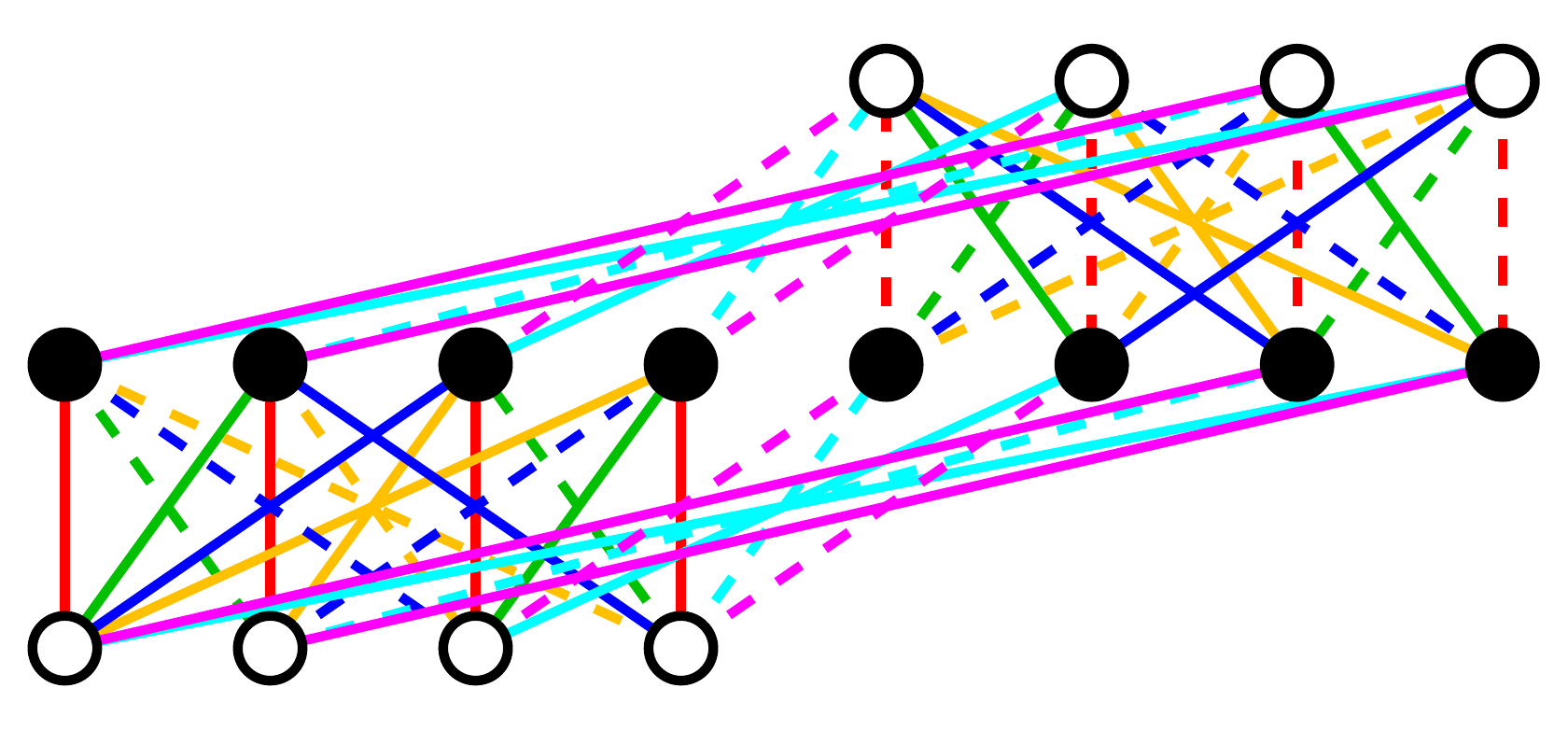}}
 \end{picture}}
\end{equation}
Together, these produce:
\begin{equation}
 \vC{\begin{picture}(160,30)
   \put(0,0){\includegraphics[width=70mm]{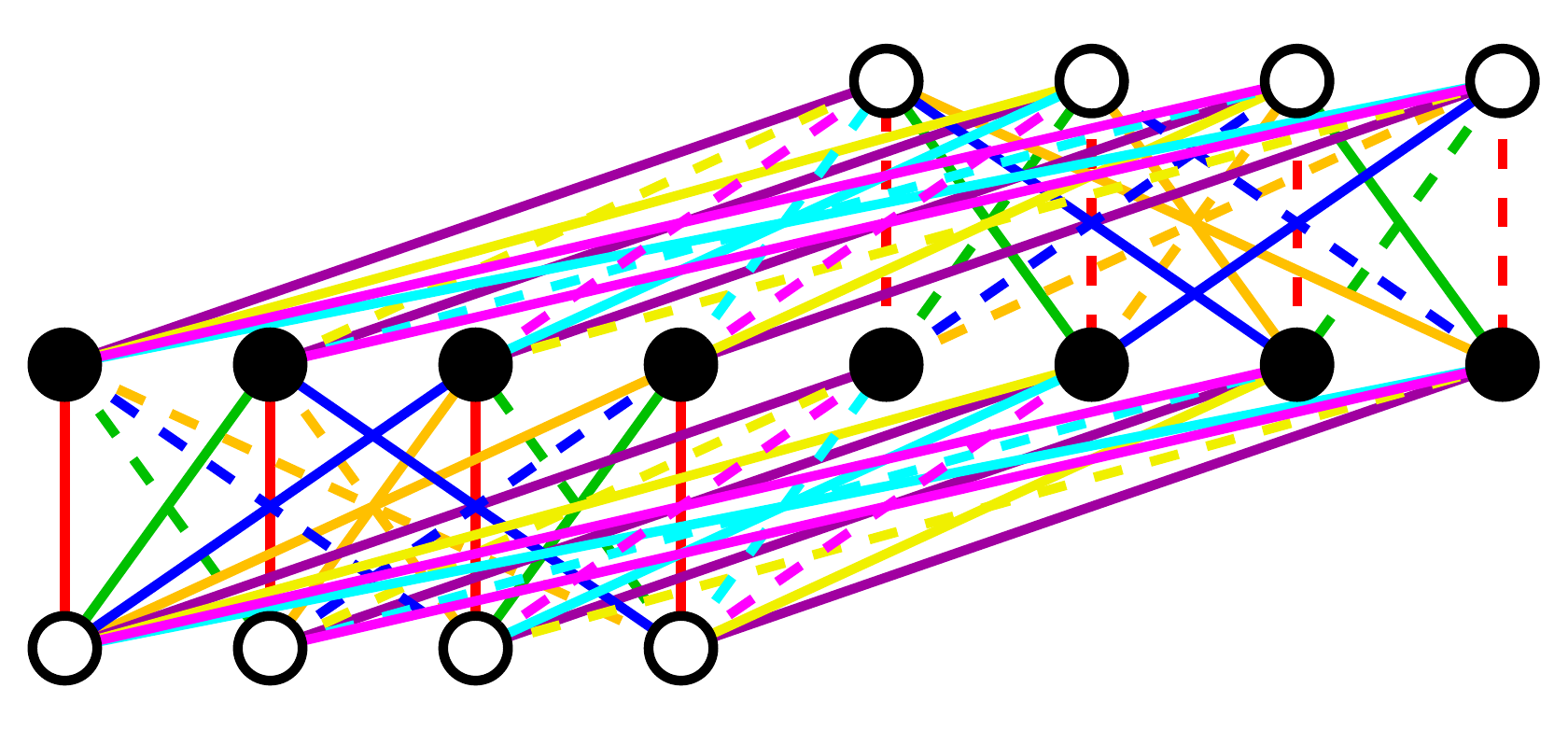}}
    \put(77,14){\LARGE$\to$}
   \put(90,0){\includegraphics[width=70mm]{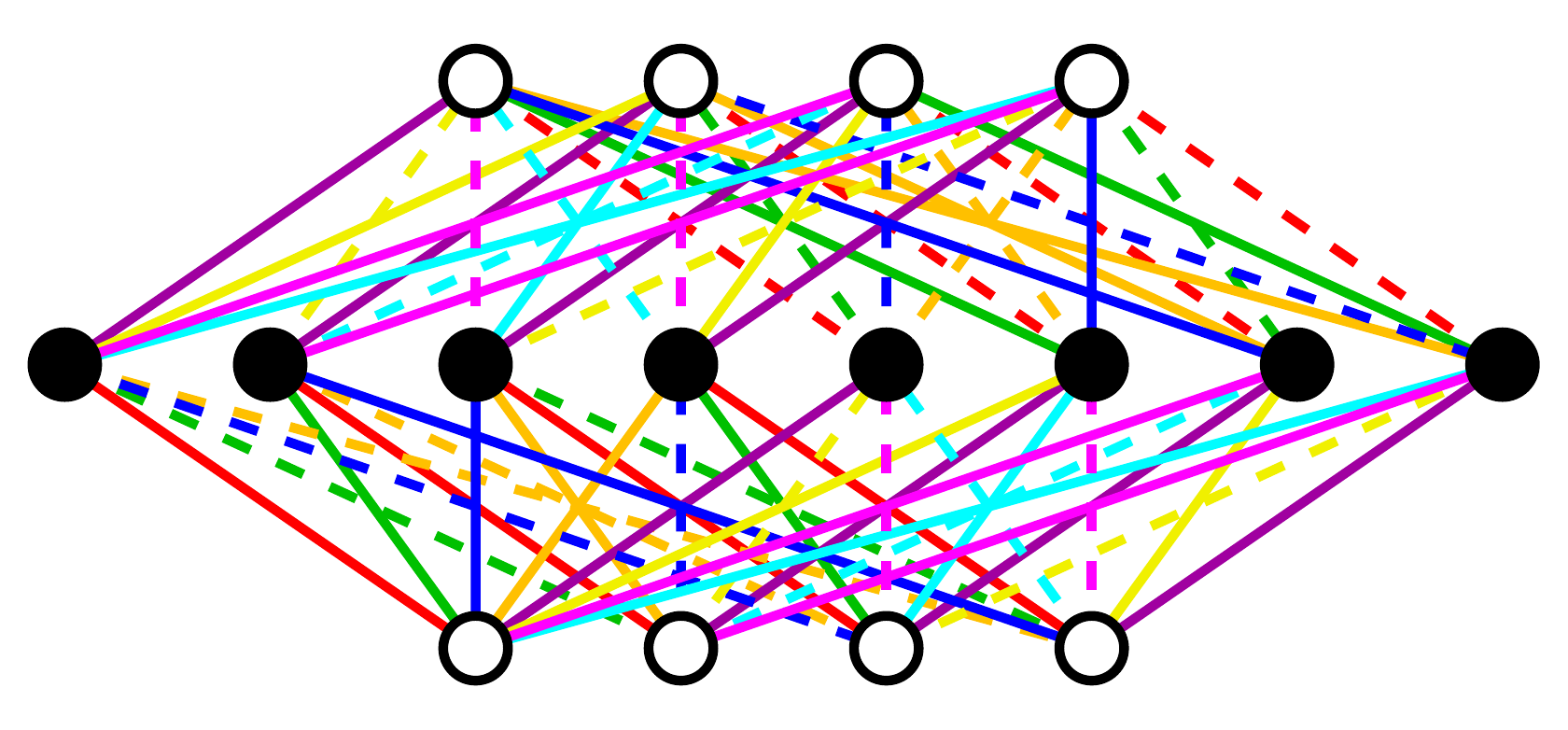}}
 \end{picture}}
 \label{e:484E8}
\end{equation}
where the nodes have been repositioned horizontally simply to highlight the similarity with the left-hand-side Adinkra in\eq{e:484}. Indeed, a careful comparison shows that the Adinkra~{\bsf C} in\eq{e:Five2.2} depicts the a supermultiplet with the same component field content, but with its supersymmetry enhanced from $(2,2)$- to $(4,4)$-supersymmetry in\eq{e:484E8}.

In turn, while Adinkra~{\bsf C} in\eq{e:Five2.2} decomposes as $\text{\bsf D}\oplus\text{\bsf E}$ by virtue of possessing the $\ZZ_2$ symmetry that was made manifest in\eq{e:484}, it is not hard to see that the edges depicting the action of the additional four supersymmetries in\eq{e:484E8} obstruct this symmetry, and thus also the decomposition.

The Adinkra\eq{e:484E8} thus depicts an indecomposable off-shell supermultiplet of worldsheet $(4,4)$-supersymmetry. In fact, it is also irreducible, since there is no smaller $(4,4)$-supermultiplet.
\ping

There currently exists a whole menagerie\cite{rTwSJG0,rGHR,rHST83,rIK84,rHST84,rKLR84,rIK85,rIKL88,rK88,rGI92,rIS94,rLIR94,rI96,rSK-book,rK97,rKUM97,r6-4}
 of  supermultiplets described in the physics literature that all possess the properties of providing a linear realization of off-shell worldsheet $(4,4)$-supersymmetry with:
\begin{enumerate}\itemsep=-3pt\vspace{-3mm}
 \item a finite number of auxiliary fields, and
 \item with {\em\/no\/} central charges.
\end{enumerate}\vspace{-3mm}
There exist an even larger literature for on-shell such supermultiplets and/or supermultiplets
with infinite sets of auxiliary fields and/or central charges.

Since we are only concerned with identifying the supermultiplet described graphically in\eq{e:484E8}, we can restrict our consideration to only the papers\cite{rTwSJG0,rGHR} in the description of the menagerie.
The supermultiplet described graphically in\eq{e:484E8} possess $(8|8)$ bosonic/fermionic degrees
of freedom.  This observation alone informs us that only the works in the first two
cited papers can be relevant as the other supermultiplets possess more off-shell degrees of freedom. We will use the nomenclature of\cite{rGK,rGK98} where these two relevant cases are called the TM-I and TM-II supermultiplets.

The twisted supermultiplet I $(\s,\p,\f|\j_{i\sss A}|G,G_i{}^j)$, denoted TM-I, of\cite{rGHR} was described by supersymmetry transformation laws (${\SSS A,B}=\pm$, $i,j,k,\ell=1,2$ and $\g^1,\g^2,\g^3$ are suitable $2\,{\times}\,2$ Dirac matrices while $C_{ij}$ and $C_{\sss AB}$ are `charge conjugation matrices'\cite{r1001}, chosen here to be equal to the 2nd Pauli matrix)
\begin{subequations}
 \label{e:TM-I}
\begin{gather}
D_{i\sss A} \f = 2\,C_{ij}\,\j^j_{\sss A},\qquad
D_{i\sss A} \s = -i\,\bar\j_{i\sss A},\qquad
D_{i\sss A} \p = (\g^3)_{\sss A}{}^{\sss B}\,{\bar\j}_{i\sss B},\\*[1mm]
D_{i\sss A}\j^{j\sss B}
               = \d_i{}^j \big[\,(\g^c)_{\sss A}{}^{\sss B}\,(\vd_c\s)
                + i(\g^3 \g^c)_{\sss A}{}^{\sss B}\,(\vd_c\p)\,\big]
                + \frc12 \big[\,\d_i{}^j(\g^3)_{\sss A}{}^{\sss B} G
                                   +i\d_{\sss A}{}^{\sss B} G_i{}^j\,\big],\\*[1mm]
\Db_{i\sss A} \j_j^{\sss B}
               = i\,C_{ij}(\g^c)_{\sss A}{}^{\sss B}\,(\vd_c\f),\qquad
D_{i\sss A} G  = -2i\,(\g^3\g^c)_{\sss A}{}^{\sss B}\,(\vd_c{\bar\j}_{i\sss B}),\\*[1mm]
D_{i\sss A} G_j{}^k 
               = 4[\d_j{}^\ell \d_i{}^k - \frc12\d_j{}^k\d_i{}^\ell]
                     (\g^c)_{\sss A}{}^{\sss B}\,(\vd_c\bar\j_{\ell\sss B}).
\end{gather}
\end{subequations}
The fields $\s,\p,G$ are real, $\f$ and ${\j}_{i\sss A}$ (and also the $D_{i\sss A}$) are complex and $G_i{}^j$ form a traceless Hermitian matrix of complex fields:
\begin{equation}
 G_i{}^j = (G_j{}^i)^*,\qquad G_i{}^i=0.
 \label{e:AH}
\end{equation} 
 The invariant component-level action takes the form
\begin{equation}
{\cal S}_\text{TM-I}
  = \int\rd^2\s~
     \Big[\,\frc12\s\Box\s + \frc12\p\Box\p + \frc12\f\Box\bar\f
            ~+~ i\j_{i\sss A} (\g^c)^{\sss AB}(\vd_c\,\bar\j^i{}_{\sss B})
            ~-~\frc12 G^2 - \frc1{16} G_i{}^j\, G_j {}^i \,\Big].
 \label{e:S-I}
\end{equation}

The twisted multipet II $(\vf,\vf_i{}^j|\c_{i\sss A}|S,P,F)$, denoted TM-II,\cite{rIK84,rIK85,rGHR} has the following transformation laws:
\begin{subequations}
 \label{e:TM-II}
\begin{gather}
D_{i\sss A}\vf   = (\g^3)_{\sss A}{}^{\sss B} \c_{i\sss B},\qquad
D_{i\sss A}\vf_j{}^k
                 = i\big[\d_i{}^k\d_j{}^\ell
                         -\frc12\d_j{}^k\d_i{}^\ell \big] \c_{\ell\sss A},\qquad
D_{i\sss A}\c_{j\sss B}
                 = \frc12 C_{ij}C_{\sss AB}\bar{F},\\*[1mm]
\Db^i{}_{\sss A}\c_{j\sss B}
                 = i\d^i{}_j(\g^3\g^a)_{\sss AB}(\vd_a\vf)
                     +2(\g^a)_{\sss AB}(\vd_a\vf_j{}^i)
                     +\frc{i}2 \d^i{}_j C_{\sss AB}\,S
                     +\frc12 \d^i{}_j (\g^3)_{\sss AB}\,P,\\*[1mm]
D_{i\sss A} \bar{F} = 0,\qquad
D_{i\sss A} F = - 4i C_{ij} (\g_a)_{\sss A}{}^{\sss B}(\vd_a\bar\c^j{}_{\sss B}),\\*[1mm]
D_{i\sss A}\,S = - 2(\g_a)_{\sss A}{}^{\sss B}(\vd_a\c_{i\sss B}),\qquad
D_{i\sss A}\,P = - 2i(\g^3\g^a)_{\sss A}{}^{\sss B}(\vd_a\c_{i\sss B}).
\end{gather}
\end{subequations}
where the fields $\vf,S,P$ are real, $F$ and $\c_{i\sss A}$ are complex and $\vf_i {}^j$ form a traceless Hermitian matrix of complex component fields
\begin{equation}
  \vf_i {}^j = (\vf_j{}^i)^*,\qquad
  \vf_i {}^i=0.
\end{equation}
An invariant component-level action for this supermultiplet is
\begin{equation}
 {\cal S}_\text{TM-II}
 =\int\rd^2\s~ \Big[\, \frc12\vf\Box\vf+\vf_j{}^i\Box\vf_i{}^j
                    ~+~i\c_{i\sss A}(\g^c)^{\sss AB}(\vd_c\bar\c^i{}_{\sss B})
                    ~-~\frc18(S^2 + P^2 + F\bar{F}\,) \,\Big].
 \label{e:S-II}
\end{equation}

As first noted in\cite{rIK84,rIK85} and discussed in\cite{rGK,rGK98}, these supermultiplets in two dimensions are easily shown to be {\em\/usefully inequivalent\/} in the sense of Ref.\cite{rChiLin}---\ie, there exist lagrangians that involve combinations of such supermultiplets that cannot be transformed by field redefinition into lagrangians that involve only one type of these  supermultiplets.
 For example, using both the TM-I and TM-II supermultiplets, it is possible write a supersymmetric mass term of the form
\begin{equation}
{\cal S}^\text{(mix)}_\text{mass}
  =\frc12M_0\int\rd^2\s~\Big[\, \s S - \p P - \frc12(\bar\f F +\f\bar{F})
                              - \frc14 \vf_i {}^j G_j{}^i - \vf G
                              + 2(\j_{i\sss A}\,\c^{i\sss A} + \textit{h.c.})\,\Big].
 \label{e:MixMass}
\end{equation}
{\em Via\/} a series of straightforward but involved calculations is can be shown that no such mass term exist using solely TM-I supermultiplets or TM-II supermultiplets.

To compare the transformation laws in\eq{e:TM-I} and\eq{e:TM-II} with those implied
by the graph\eq{e:484E8} we must work in a real basis where light-cone coordinates
of spinors are used. To that end, we switch from the complex 1-component Weyl bases $D_{i\sss A},\j_{i\sss A},$ \etc, to 2-component real (Majorana) bases $\rD_{i\sss A},\J_{i\sss A},$ \etc\
 In doing so, the two components of the Majorana spinor $\J_{i+}$ for each of the two values of the index $i$ provide the four left-handed fermions in\eq{e:484E8} and $\J_{i-}$ provide the right-handed ones. In a similar manner the $\rD_{i\pm}$-operators in these equations are also 2-component real (Majorana) operators, tallying up a total of four left-handed and four right-handed superderivatives, corresponding to $(4,4)$-supersymmetry and depicted by the total of eight edge-colors in\eq{e:484E8}.

In case of\eq{e:TM-I}, we find
\begin{subequations}
 \label{e:TM-Ia}
\begin{alignat}9
\rD_{i\pm}\s &= \J_{i\pm},\quad&\quad
\rD_{i\pm}\p &= \mp i(\Ione_2\otimes\bs\s^2)_i{}^j\,\J_{j\pm}, \\
\rD_{i\pm}\f_{({\sss R})} &=-i(\bs\s^2 \otimes \bs\s^3)_i{}^j\,\J_{j\pm},\quad&\quad
\rD_{i\pm}\f_{({\sss I})} &=+i(\bs\s^2 \otimes \bs\s^1)_i{}^j\,\J_{j\pm}, \\
\rD_{i+}\,G &=-2i(\Ione_2\otimes\bs\s^2)_i{}^j(\vd_\pp\J_{j-}),\quad&\quad 
\rD_{i-}\,G &=-2i(\Ione_2\otimes\bs\s^2)_i{}^j(\vd_\mm\J_{j+}),\\
\rD_{i+}\,G_1{}^1 &=+2(\bs\s^3\otimes\Ione_2)_i{}^j(\vd_\pp\J_{j-}),\quad&\quad 
\rD_{i-}\,G_1{}^1 &=-2(\bs\s^3\otimes\Ione_2)_i{}^j(\vd_\mm\J_{j+}),  \\
\rD_{i+}\,G_1{}^2_{({\sss R})} &=2i(\bs\s^2\otimes\Ione_2)_i{}^j(\vd_\pp\J_{j-}),\quad&\quad 
\rD_{i-}\,G_1{}^2_{({\sss R})} &=-2i(\bs\s^2\otimes\Ione_2)_i{}^j(\vd_\mm\J_{j+}),  \\
\rD_{i+}\,G_1{}^2_{({\sss I})} &=2i(\bs\s^1\otimes\bs\s^2)_i{}^j(\vd_\pp\J_{j-}),\quad&\quad 
\rD_{i-}\,G_1{}^2_{({\sss I})} &=-2i(\bs\s^1\otimes\bs\s^2)_i{}^j(\vd_\mm\J_{j+}),
\end{alignat}
\end{subequations}
where $\f_{({\sss R})}$ and $\f_{({\sss I})}$ respectively denote the real and imaginary parts of the complex spin-0 field $\f$. In the same manner $G_1{}^2_{({\sss R})}$ and $G_1{}^2_{({\sss I})}$ denote the real and imaginary part of the complex spin-0 field $G_1{}^2$.
 We need not give an explicit expressions for $G_2{}^1$ and $G_2{}^2$, since\eq{e:AH} implies  that $G_2{}^2=-G_1{}^1$ and $G_2{}^1=(G_1{}^2)^*$.
 
The key property---of being {\em\/adinkraic\/}\cite{r6-1}---of the supersymmetry transformation rules\eq{e:TM-Ia} is that each supercharge maps each component field to precisely one other component field or its derivative; owing to this feature the supermultiplet\eq{e:TM-I} may be depicted by an Adinkra. Analogous remarks apply to the TM-II supermultiplet\eq{e:TM-II}, and that system of supersymmetry transformation rules also has a real (Majorana) rendition analogous to\eq{e:TM-Ia}.

The fact that both supermultiplets\eq{e:TM-I} and\eq{e:TM-II} may be depicted by the same Adinkra\eq{e:484E8} implies that there is an intimate relation between these two distinct representations of worldsheet $(4,4)$-supersymmetry.  
 The difference between the two supermultiplets evidently owes to the complex and tensor structure (the latter indicated by the indices $i,j,k,\ell$), which may be employed to represent a group action, not unlike those discussed in Ref.\cite{rFGH}, which then serves to further distinguish TM-I supermultiplets from the TM-II ones.

With the appropriate choice of the lagrangian densities\eq{e:S-I} and\eq{e:S-II}, the lower white nodes correspond to propagating physical bosons, whereas the upper white nodes depict non-propagating auxiliary component fields; all the fermions (depicted by black circles) then have Dirac-like 1st order differential equations of motion. The Adinkra\eq{e:484E8} in which the nodes are bundled to reflect the real/complex and tensorial nature of $(\s,\p,\f|\j_{i\sss A}|G,G_i{}^j)$ is then evidently upside-down as compared to the same Adinkra in which the nodes are bundled to reflect the real/complex and tensorial nature of $(\vf,\vf_i{}^j|\c_{i\sss A}|S,P,F)$. This feature then depicts the type of duality between the supermultiplets TM-I\eq{e:TM-I} and TM-II\eq{e:TM-II}, which in turn permits the existence of the mass terms\eq{e:MixMass}. In this duality, the fermions in the two supermultiplets are simply identified, $\j_{i\sss A}\iff\c_{i\sss A}$, but the corresponding identification of the bosons, $(\s,\p,\f|G,G_i{}^j)\iff(\vf,\vf_i{}^j|S,P,F)$, must be non-local owing to the differing engineering (mass) dimensions of the like component fields.
 These and related topics will be explored under a separate cover.

\subsection{$(8,8)$-Supersymmetry}
In a fashion rather similar to the procedure\eqs{e:E8}{e:484E8}, it is straightforward to produce two Adinkras that depict off-shell supermultiplets of worldsheet $(8,8)$-supersymmetry.

We begin with the {\em\/valise\/} version of the $64{+}64$-component $N=16$ worldline supermultiplet with the $E_8\,{\times}\,E_8$ chromotopology\cite{r6-3}, and hide all edges but those corresponding to the first few supersymmetries:
\begin{gather}
 \vC{\begin{picture}(160,17)
      \put(0,0){\includegraphics[width=160mm]{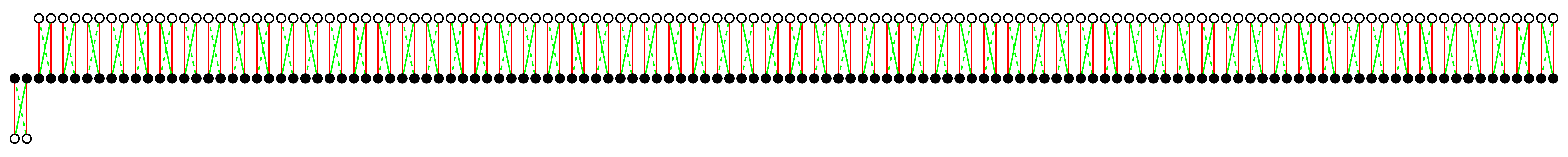}}
       \put(5,3){\footnotesize(first two supersymmetries only)}
     \end{picture}}
 \label{e:E8xE80}\\
 \vC{\begin{picture}(160,17)
      \put(0,0){\includegraphics[width=160mm]{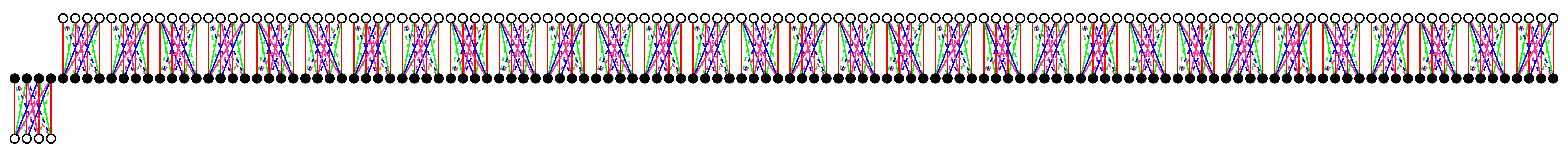}}
       \put(7,3){\footnotesize(first four supersymmetries only)}
     \end{picture}}\\
 \vC{\begin{picture}(160,17)
      \put(0,0){\includegraphics[width=160mm]{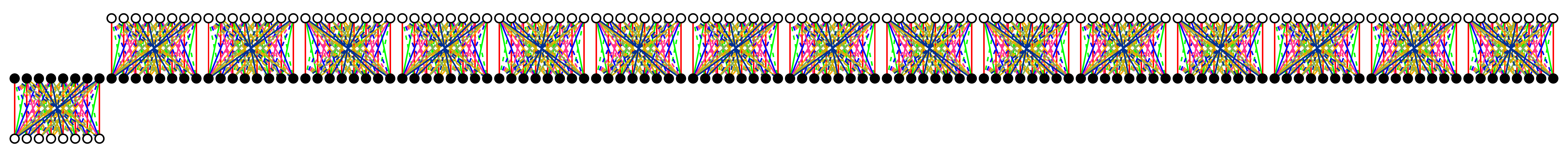}}
       \put(12,3){\footnotesize(first eight supersymmetries only)}
     \end{picture}}
 \label{e:1st8}
\end{gather}
where the 8+8-node blocks have been patterned according to\eq{e:E8} and so use up the edges corresponding to the eight left-handed supersymmetries. Next we start adding the edges corresponding to the right-handed supersymmetries, and lower block after block of 8 bosons, so as to avoid forming ambidextrous 2-color bow-ties:
\begin{equation}
 \vC{\begin{picture}(160,30)
      \put(0,0){\includegraphics[width=160mm]{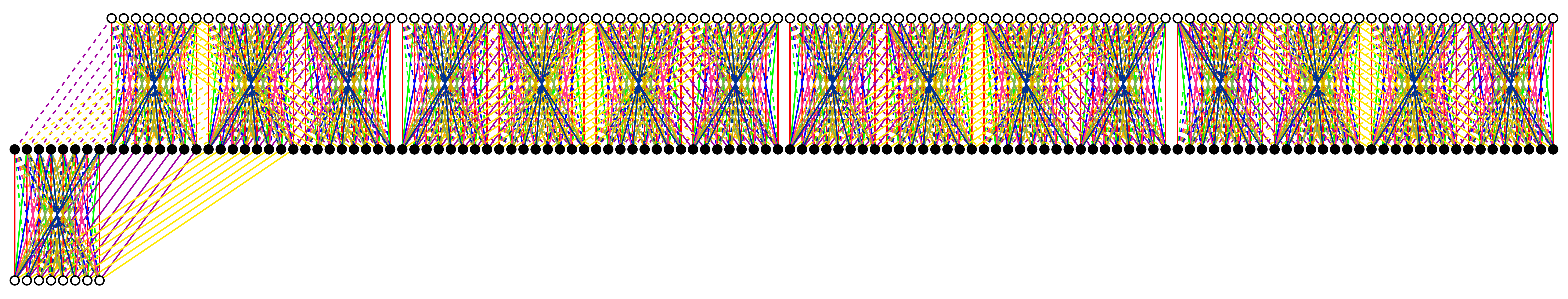}}
       \put(20,4){\footnotesize(first ten supersymmetries only)}
       \put(31,15){$\underbrace{\rule{128mm}{0mm}}
                   _{\text{some $8{+}8$ blocks remain to be judiciously `lowered'}}$}
     \end{picture}}
\end{equation}
\begin{equation}
 \vC{\begin{picture}(160,30)
      \put(0,0){\includegraphics[width=160mm]{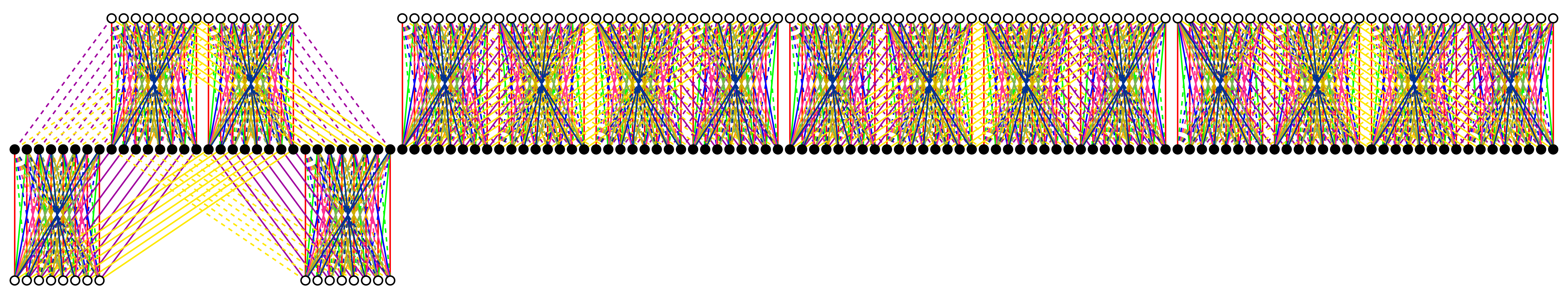}}
       \put(42,4){\footnotesize(first ten supersymmetries only)}
       \put(41,15){$\underbrace{\rule{118mm}{0mm}}
                   _{\text{some $8{+}8$ blocks remain to be judiciously `lowered'}}$}
     \end{picture}}
\end{equation}
Thus, the 9th and 10th color edges indicate the relative positioning of the first four $8{+}8$ blocks. Hiding these edges for clarity and revealing the edges corresponding to the 11th and 12th supersymmetry indicates which of the next $8$-tuples of bosonic nodes need to be `lowered':
\begin{equation}
 \vC{\begin{picture}(160,30)
      \put(0,0){\includegraphics[width=160mm]{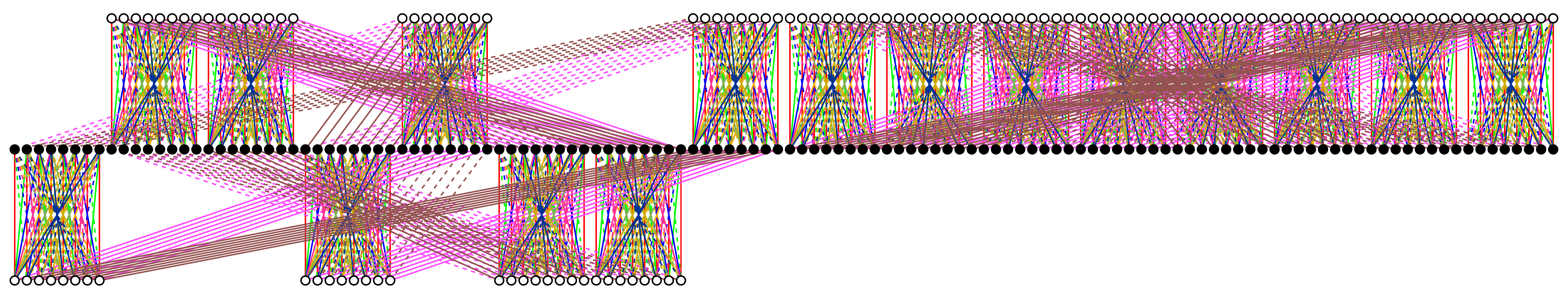}}
       \put(72,4){\footnotesize(first eight plus 11th and 12th supersymmetry only)}
       \put(81,15){$\underbrace{\rule{78mm}{0mm}}
                   _{\text{some $8{+}8$ blocks remain to be judiciously `lowered'}}$}
     \end{picture}}
\end{equation}
The second half of the $8{+}8$ node blocks will have to follow this relative positioning pattern within the right-hand half. Hiding the 11th and 12th color edges, and revealing the edges corresponding to the 13th and 14th supersymmetry determines the relative positioning of the $8{+}8$ blocks in the right-hand half as compared to the left-hand half:
\begin{equation}
 \vC{\begin{picture}(160,30)
      \put(0,0){\includegraphics[width=160mm]{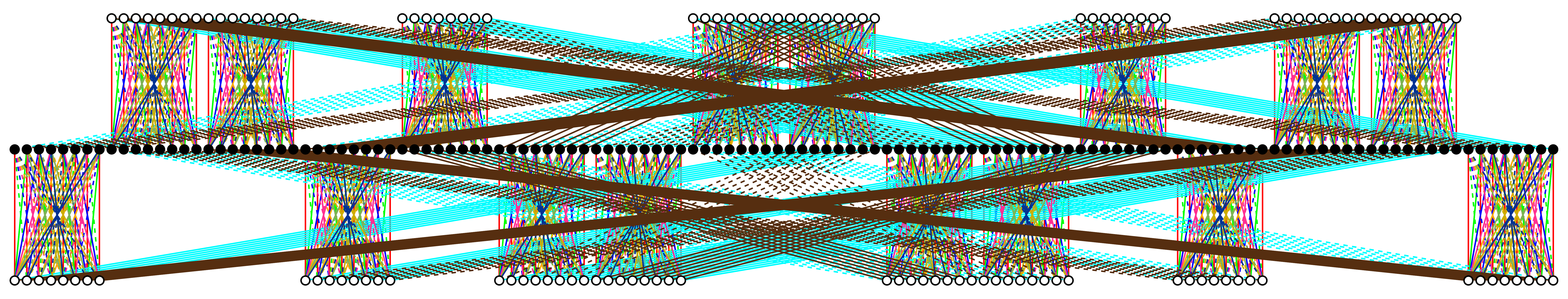}}
     \end{picture}}
\end{equation}
Finally, swapping 13th and 14th for 15th and 16th supersymmetry verifies that this arrangement produces no ambidextrous 2-colored bow-ties:
\begin{equation}
 \vC{\begin{picture}(160,30)
      \put(0,0){\includegraphics[width=160mm]{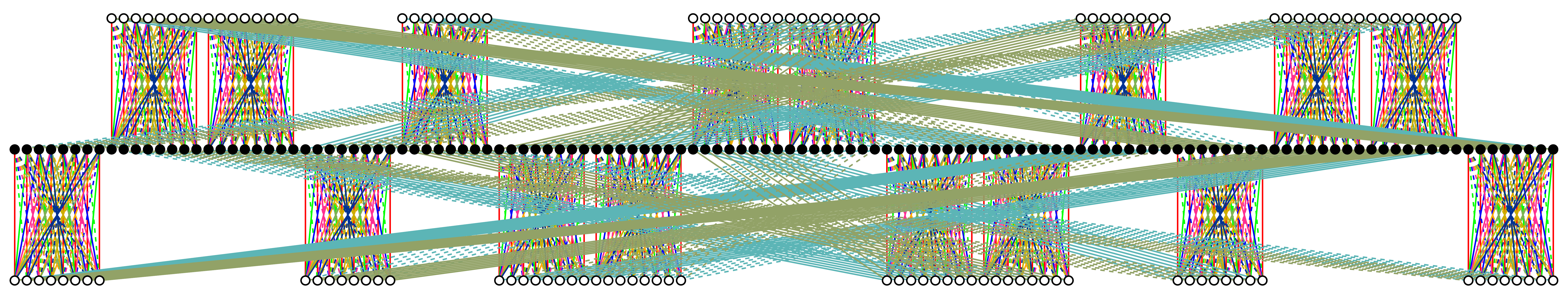}}
     \end{picture}}
\end{equation}
Thus, the edges of the first eight colors connect nodes within the $8{+}8$-node blocks and do form 2-colored bow-ties, but all correspond to, say, left-handed supercharges. In turn, edges of the latter eight colors connect nodes from a ``lowered'' $8{+}8$-node block to another that is ``higher,'' and these edges depict the action of right-handed supercharges. This then avoids forming ambidextrous 2-colored bow-ties.
 Putting this together results in:
\begin{equation}
 \vC{\begin{picture}(160,30)
      \put(0,0){\includegraphics[width=160mm]{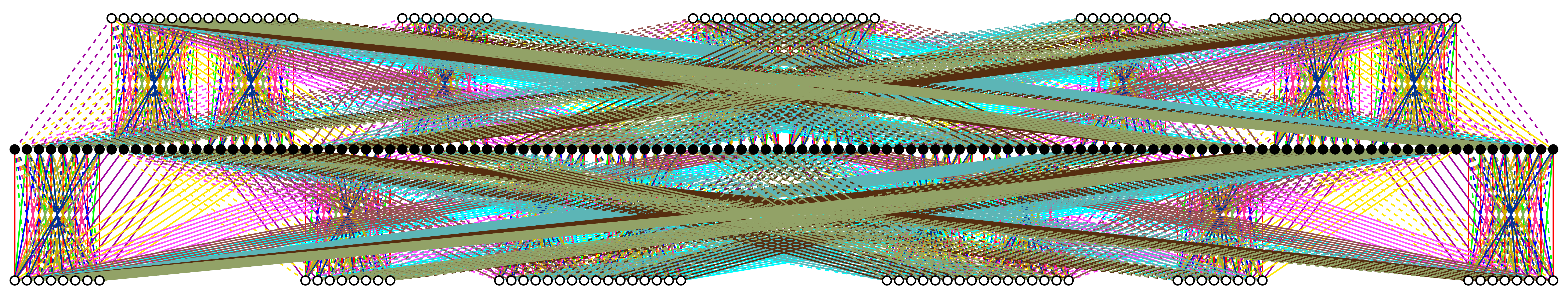}}
       \put(0,25){\footnotesize all:}
     \end{picture}} \label{e:E8xE8}
\end{equation}

In a similarly depicted fashion, we could alternatively proceed as follows:
\begin{gather}
 \vC{\begin{picture}(160,17)
      \put(0,0){\includegraphics[width=160mm]{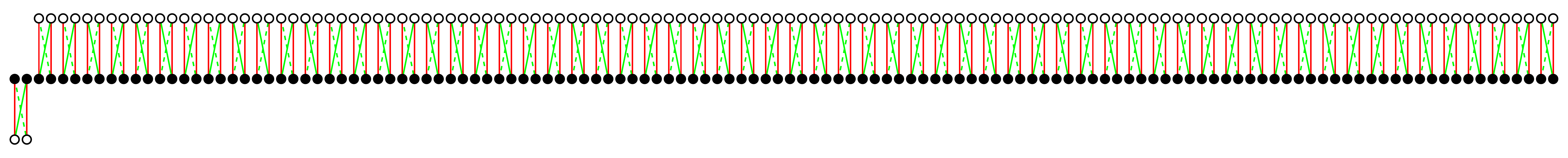}}
       \put(5,3){\footnotesize(first two supersymmetries only)}
     \end{picture}} \label{e:E160}\\
 \vC{\begin{picture}(160,17)
      \put(0,0){\includegraphics[width=160mm]{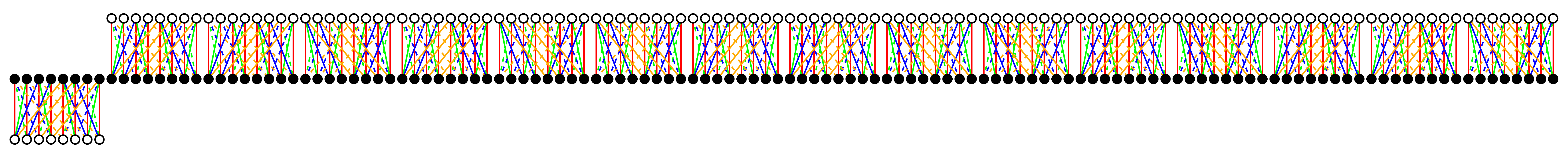}}
       \put(12,3){\footnotesize(first four supersymmetries only)}
     \end{picture}}
\end{gather}
Note the different pattern emerging as more and more edges, to be identified with $\rD_{\a+}$-action, are added
\begin{gather}
 \vC{\begin{picture}(160,20)
      \put(0,0){\includegraphics[width=160mm]{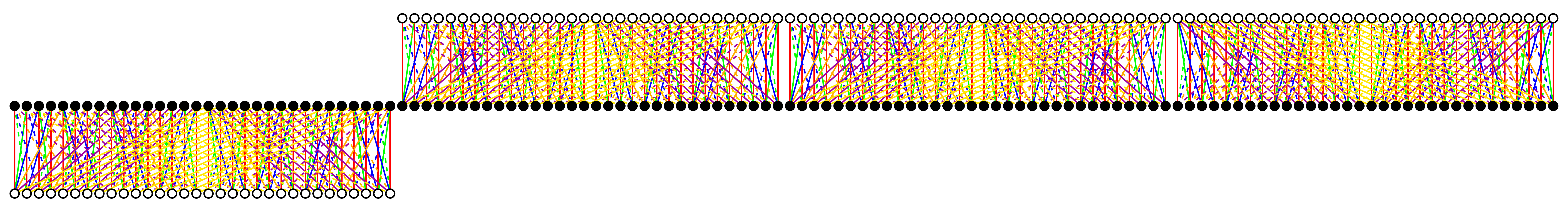}}
       \put(42,5){\footnotesize(first six supersymmetries only)}
     \end{picture}}\\
 \vC{\begin{picture}(160,32)
      \put(0,0){\includegraphics[width=160mm]{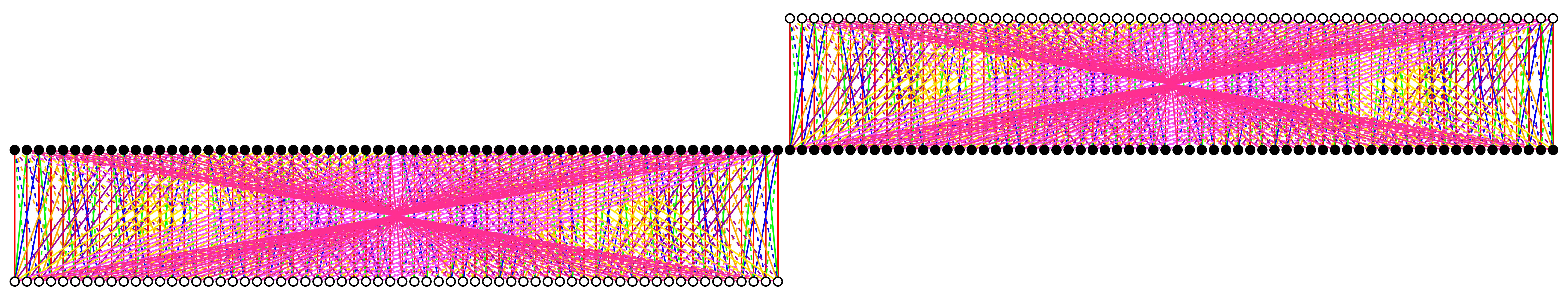}}
       \put(82,7){\footnotesize(first eight supersymmetries only)}
     \end{picture}}
\end{gather}
At this point all eight edges, to be identified with $\rD_{\a+}$-action, have been added, dividing the nodes into two two fairly uniform $32{+}32$-node Adinkras|in sharp distinction from\eq{e:1st8}. The next eight edges, to be identified with $\rD_{\ad-}$-action, are now being added so as to {\em\/not\/} form bow-ties with the previous eight:
\begin{equation}
 \vC{\begin{picture}(160,32)
      \put(0,0){\includegraphics[width=160mm]{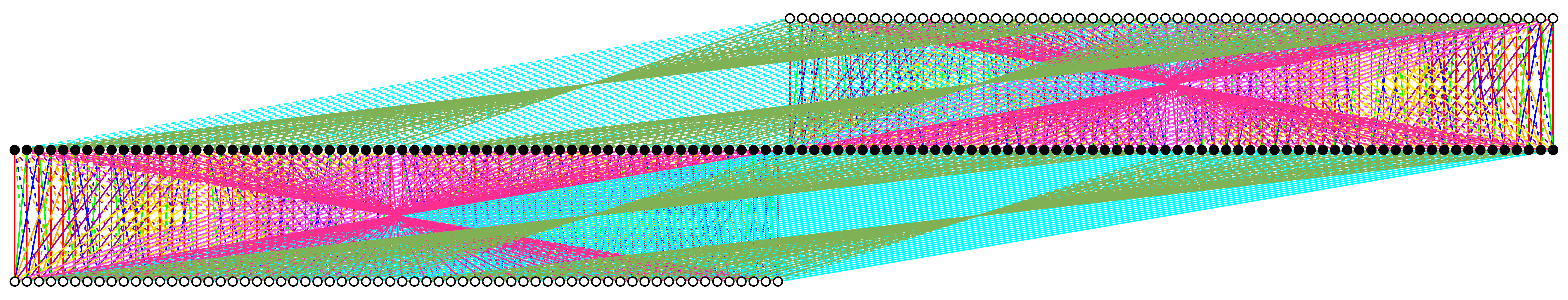}}
       \put(110,3){\footnotesize(first ten supersymmetries only)}
     \end{picture}}
\end{equation}
swapping 9th and 10th for 11th and 12th supersymmetry:
\begin{equation}
  \vC{\begin{picture}(160,32)
      \put(0,0){\includegraphics[width=160mm]{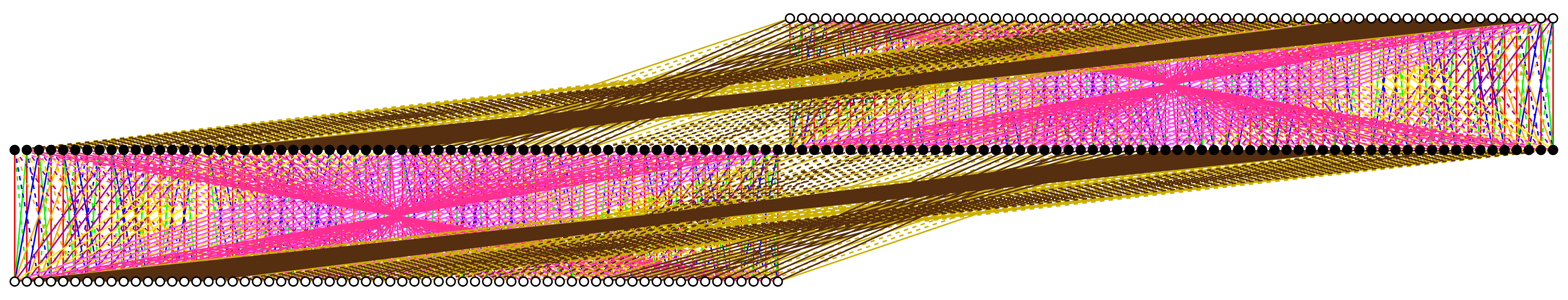}}
     \end{picture}}
\end{equation}
swapping 11th and 12th for 13th and 14th supersymmetry:
\begin{equation}
  \vC{\begin{picture}(160,30)
      \put(0,0){\includegraphics[width=160mm]{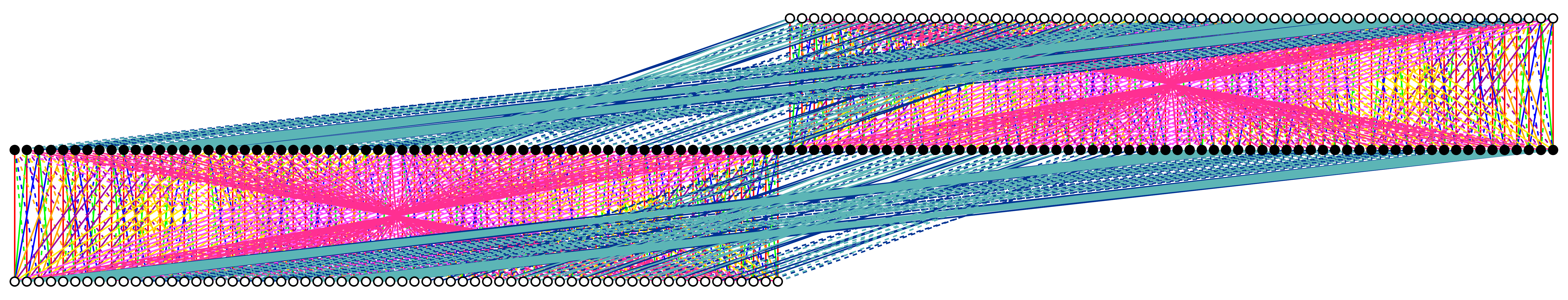}}
     \end{picture}}
\end{equation}
swapping 13th and 14th for 15th and 16th supersymmetry:
\begin{equation}
  \vC{\begin{picture}(160,32)
      \put(0,0){\includegraphics[width=160mm]{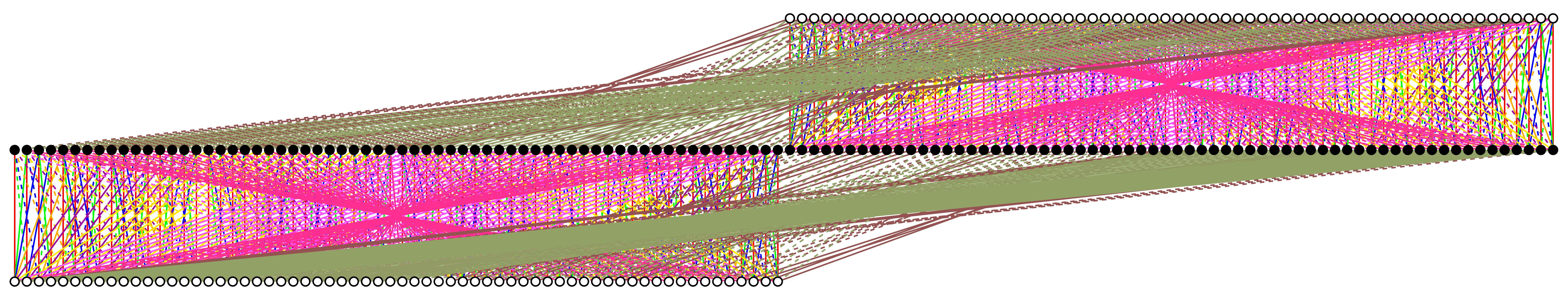}}
     \end{picture}}
\end{equation}
Putting this together results in:
\begin{equation}
  \vC{\begin{picture}(160,30)
      \put(0,0){\includegraphics[width=160mm]{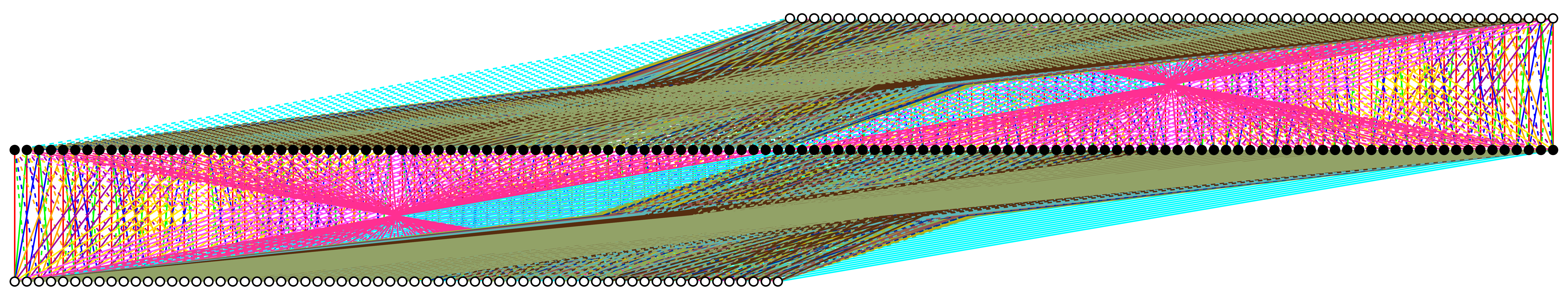}}
       \put(0,25){\footnotesize all 16 supersymmetries:}
     \end{picture}}
 \label{e:E16}
\end{equation}

Both\eq{e:E8xE8} and\eq{e:E16} are 8-fold iterated $\ZZ_2$-quotients (in the manner of\eq{e:B484>242}) of the 16-cube rearranged so that all fermions are at the same height (have the same engineering dimension), while the bosons are judiciously partitioned into the ``lower'' and the ``upper'' half, so as to exhibit a $(\ZZ_2)^8$ symmetry.
 The particular $(\ZZ_2)^8$-action that has been employed to produce\eq{e:E8xE8} from the 16-cube is encoded by the (binary) $e_8+e_8$ doubly even linear block code, whereas quotienting by a $e_{16}$-encoded $(\ZZ_2)^8$ symmetry results in\cite{r6-3,r6-3.2}.
 
As shown in the procedures\eqs{e:E8}{e:484E8}, \eqs{e:E8xE80}{e:E8xE8} and\eqs{e:E160}{e:E16}, the partitioning of the white nodes into the `lower'(propagating)/`upper'(auxiliary) ones has been unambiguously enforced by avoiding the obstruction of Theorem~\ref{T:bowT} while iteratively including the edges corresponding to all supersymmetries, for which the supersymmetry action is encoded by the binary codes $e_8$, $e_8\oplus e_8$ and $e_{16}$, respectively. We therefore conclude that the depictions\eq{e:484E8} are essentially unique for each chromotopology, \ie, each binary code encoding the supersymmetry action\cite{r6-3}. That is, all other choices depict supermultiplets that are field redefinitions of those given here---except for the {\em\/twisting\/}\cite{r6-3}, which is alternatively obtained:
\begin{enumerate}\itemsep=-3pt\vspace{-2mm}
 \item either by swapping the white$\,{\iff}\,$black node assignments, flipping the Adinkra upside-down, and then repeating the procedures\eqs{e:E8}{e:484E8}, \eqs{e:E8xE80}{e:E8xE8} and\eqs{e:E160}{e:E16},
 \item or by swapping the solid/dashed designation of the edges of a single color, corresponding to flipping the sign of a single $\rD$ superderivative.
\end{enumerate}

Now, Ref.\cite{r6-3.2} proves that|appearances to the contrary|the {\em\/valise\/} rendition of\eq{e:E8xE8} depicts a worldline supermultiplet that is in fact isomorphic|and by judicious component field redefinitions only|to the supermultiplet depicted by the {\em\/valise\/} rendition of\eq{e:E16}. In turn, the Adinkras\eq{e:E8xE8} and\eq{e:E16} admit {\em\/twisted\/} variants that are not equivalent to the un-twisted originals.
That the {\em\/valise\/} supermultiplets should be equivalent follows from the fact that the Clifford algebra $\mathfrak{Cl}(16,1)=\IR(256)\oplus\IR(256)$, thus having precisely two inequivalent $256$-dimensional representations|and one is obtained from the other by {\em\/twisting\/}.

 We conjecture that this isomorphism is not obstructed by the partitioning of the bosonic nodes into the ``upper'' and ``lower'' half as done in the process\eqs{e:E8xE80}{e:E8xE8} and\eqs{e:E160}{e:E16}, so that the $(64|128|64)$-node Adinkra\eq{e:E8xE8} and\eq{e:E16} in fact represent ``internally'' isomorphic supermultiplets: Unlike {\em\/twisting\/} that requires a redefinition of the basis of superderivatives (and supercharges), this isomorphism requires only a component field redefinition. While the chiral and the twisted-chiral supermultiplet are usefully inequivalent in that they afford writing Lagrangians that could not be written with only one or only the other kind\cite{rGHR}|if the isomorphism between the {\em\/valise\/} renditions of\eq{e:E8xE8} and\eq{e:E16} also holds for their versions depicted in\eq{e:E8xE8} and\eq{e:E16}|these Adinkras simply depict two distinct but equivalent bases. If in turn the isomorphism is obstructed by the height arrangement of\eq{e:E8xE8} and\eq{e:E16}, these then depict genuinely distinct supermultiplets, and together with their twisted variants this would imply the existence of {\em\/four\/} distinct irreducible off-shell $(64|128|64)$-component supermultiplets of worldsheet $(8,8)$-supersymmetry.

A comparison of the off-shell worldsheet $(8,8)$-supermultiplets depicted by the Adinkras\eq{e:E8xE8} and\eq{e:E16} with the constructions in the literature is beyond the scope of this note, as is the exploration of the number of usefully inequivalent variants, as indicated for the supermultiplets\eq{e:TM-I} and\eq{e:TM-II}.

\subsection{$(16,16)$-Supersymmetry}
In fact, owing to the generative similarities in the explicit construction of these minimal supermultiplets of $(2,2)$,- $(4,4)$- and $(8,8)$-supersymmetry and their twisted variants, we conjecture that the same construction produces an irreducible off-shell $(2^{14}|2^{15}|2^{14})$-component supermultiplets of worldsheet $(16,16)$-supersymmetry and its twisted variant. This time, there are actually 85 doubly even linear block codes\cite{rCP,rCPS} that can be used to project the 32-cube folded to the three levels such as\eq{e:E8xE8} and\eq{e:E16}. This results in 85 distinct Adinkras, together with a twisted variant of each.

 Now, the {\em\/valise\/} renditions of all 85 Adinkras, projected using respectively the 85 distinct doubly even codes, all depict isomorphic {\em\/worldline\/} supermultiplets, as do their twisted variants, thus merely giving varied depictions of only two distinct {\em\/valise\/} worldline supermultiplets.
 
 Again, we conjecture that raising a judicious half of the bosons to the ``upper'' level in each of these 85+85 Adinkras does not obstruct this isomorphism, which then extends to the depicted worldsheet supermultiplets.
If so, then this collection of 85 distinct 3-level Adinkras simply depicts 85 distinct bases for only one equivalence class of off-shell worldsheet $(16,16)$-supermultiplets, and the same holds for their twisted variants.

\section{Conclusions}
\label{e:coda}
In the foregoing analysis, the twin theorems~\ref{T:bowT} and~\ref{T:SSR}, and Corollary~\ref{C:ObSuSy} have been used to filter those worldline off-shell supermultiplets from the huge class of Refs.\cite{r6-3,r6-3.2,r6-1.2} that extend to off-shell supermultiplets of worldsheet $(p,q)$-supersymmetry.

In extending to worldsheet $(2,2)$-supermultiplets, only five Adinkras\eq{e:Five2.2} and their boson\,$\iff$\,fer\-mion flips (from a total of 64\cite{r6-3.2}|not counting numerous nodal permutations!) depict off-shell worldsheet $(2,2)$-supermultiplets. Of these, Adinkra {\bsf C} decomposes into a direct sum of Adinkras {\bsf D} and {\bsf E}. In turn, Adinkras {\bsf A}, {\bsf B}, {\bsf D} and {\bsf E} correspond to the well-known intact (unconstrained, ungauged, unprojected\dots), semi-chiral, chiral and twisted-chiral superfield, respectively.

The Reader may find it gratifying that no surprising off-shell $(2,2)$-supermultiplets have been uncovered by dimensionally extending supermultiplets from the collection of 64 worldline supermultiplets of $(N\,{=}\,4)$-extended supersymmetry\cite{r6-3.2}. However, this by no means guarantees that no surprises will be found for $p{+}q\geq4$, and the following remarks are in order:
\begin{enumerate}\itemsep=-3pt\vspace{-2mm}
 \item The number of inequivalent chromotopologies of Adinkras is a strongly combinatorially growing function of $N$: to date, even distributed computing efforts running on supercomputer clusters have stalled at $N=29$\cite{rRLM-Codes}|and this is without considering the combinatorial complexity involved with assigning different possible engineering dimensions to the component (super)fields in a supermultiplet! Even with a filtering based on the obstruction described in Corollary~\ref{C:ObSuSy}, it is likely that the number of off-shell supermultiplets of ambidextrous worldsheet $(p,q)$-supersymmetry (\ie, when $p,q\neq0$) is nevertheless a combinatorially growing function of $p{+}q$. In turn, for {\em\/unidextrous\/} $(N,0)$- and $(0,N)$-supersymmetry, this obstruction is always absent and the whole huge class\cite{r6-3,r6-3.2,r6-1.2} of worldline off-shell supermultiplets extends to worldsheet off-shell supermultiplets.
 \item It is always possible to construct an indefinite number of new supermultiplets by ``linear algebra'': by subjecting direct sums of tensor products of these {\em\/adinkraic\/} supermultiplets to superdifferential constraints, gauging and projection. Such constructions extend the Weyl construction from Lie algebra representation theory, the discussion of which is deferred to subsequent effort. Suffice it here to say, however, that the well-known {\em\/linear supermultiplet\/}\cite{r1001,rPW,rWB,rBK} is but a simple example of such a construction, the Adinkra of which does not occur in the line-up\eq{e:Five2.2}, but which can be constructed in terms of those.
\end{enumerate}

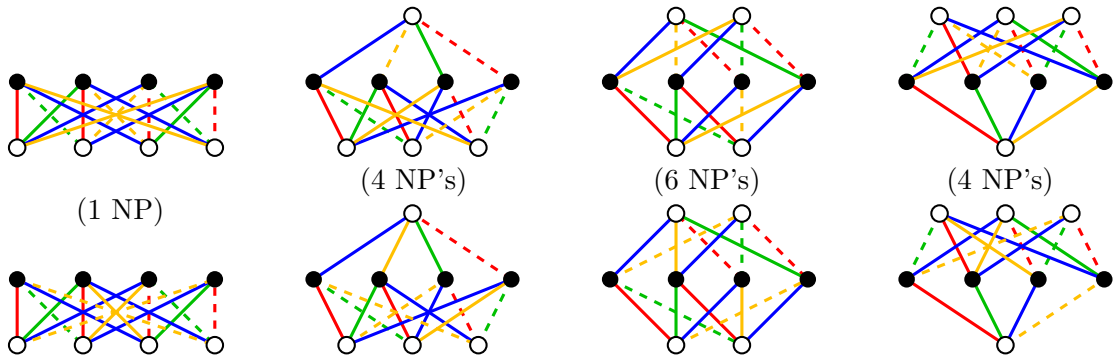
\begin{figure}[htp]
 \begin{center}
  \begin{picture}(140,45)(0,4)
   \put(0,0){\includegraphics[width=150mm]{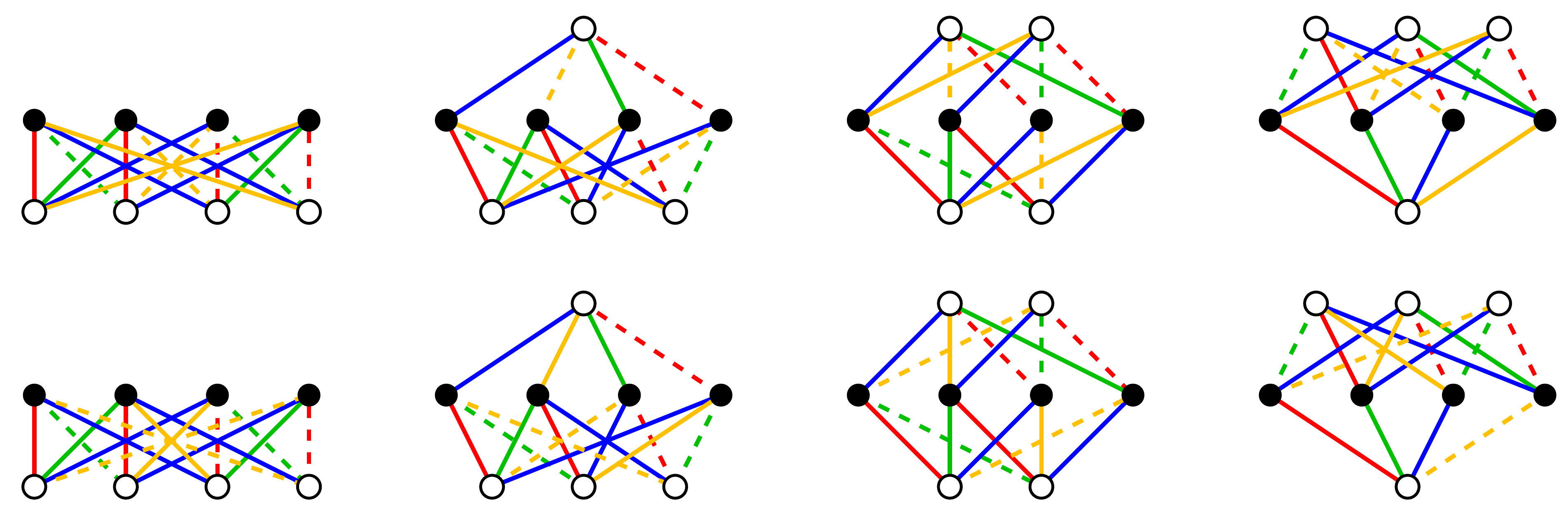}}
    \put(17,19){\small\Cx{(1 NP)}}
    \put(56,23){\small\Cx{(4 NP's)}}
    \put(95,23){\small\Cx{(6 NP's)}}
    \put(134,23){\small\Cx{(4 NP's)}}
 \end{picture}
 \end{center}
 \caption{The four distinct height configurations of the ``half-sized'' $N=4$ Adinkras, their relatively twisted variants (such as the chiral and twisted-chiral Adinkras, second from the right) stacked one above the other. The number of inequivalent nodal permutations (NP's) are shown in parentheses.}
 \label{f:N4}
\end{figure}

\paragraph{Comparisons:}
It is worthwhile comparing the present approach with that of Refs.\cite{rFIL,rFL}, where the line-up of 30 ``half-sized'' supermultiplets of the $(N\,{=}\,4)$-extended worldline supersymmetry is explicitly tested for extending {\em\/directly\/} to supermultiplets of ${\cal N}{=}1$ supersymmetry in $(3{+}1)$-dimensional spacetime.
 These supermultiplets are depicted by the various ``nodal permutations'' of the Adinkras in Figure~\ref{f:N4}.
For these 30 Adinkras, the computations of Refs.\cite{rFIL,rFL} require numerically checking a system of certain $4\times4$ matrix equations for each Adinkra, which adds up to
 $2\cdot30\cdot(2\cdot30)=3{,}600$
such $4\times4$ matrix equations.

By contrast, simple inspection reveals that of the Adinkras in Figure~\ref{f:N4} only the two shown second from the right satisfy the twin theorems~\ref{T:bowT} and~\ref{T:SSR}, and for these we must select the \C1{red}-\C2{green} pair for $\rD_{\a+}$ and the \C3{blue}-\C4{gold} pair for $\rD_{\ad-}$. Any permutation of nodes across the levels (other than a simple upside-down flip) necessarily violates these theorems and we conclude that
\begin{equation}
 \vC{\begin{picture}(140,20)
   \put(0,0){\includegraphics[width=140mm]{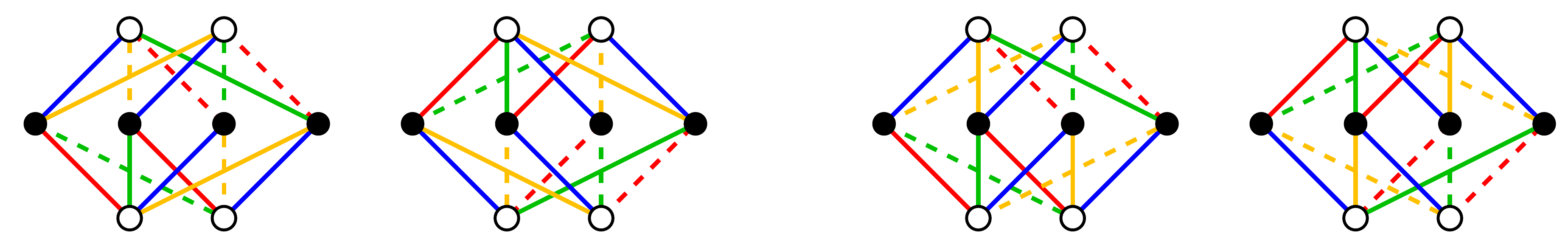}}
 \end{picture}}
 \label{e:4N4}
\end{equation}
are the only distinct ``half-sized'' $N=4$ Adinkras that extend to depict worldsheet $(2,2)$-super\-mul\-tiplets: the second one is the upside-down version of the first, and the fourth one is the upside-down version of the third; the right-hand side two are twisted variants of left-hand side two, where the solid/dashed parity is flipped for only the golden edges.
 However, on closer inspection, we note that changing the signs of the two right-hand side fermions and the two top bosons turns the second Adinkra in the line-up\eq{e:4N4} into the left-right mirror image of the first one, proving that these two depict the same supermultiplet. Whence only the two ``half-sized'' Adinkras in the line-up\eq{e:Five2.2} and their boson\,$\iff$\,fermion flips extend to depict supermultiplets of worldsheet $(2,2)$-supersymmetry.
 
As worldsheet supersymmetry is a subset of the ${\cal N}=1$ supersymmetry in $(3{+}1)$-dimensional spacetime, the filtering presented herein (twin theorems~\ref{T:bowT} and~\ref{T:SSR}, \ie, Corollary~\ref{C:ObSuSy}) provides an intermediate step that enhances the criteria of Refs.\cite{rFIL,rFL}. This effectively reduces the need for computations of Refs.\cite{rFIL}, already 30-fold for the lowest-$N$ case\ft{Supersymmetry in $d$-dimensional spacetimes has $N=2^{^{\lfloor(d{+}1)/2\rfloor}}{\cal N}$ real generators, except for $(d{-}2)=0\pmod8$ when a model can have half as many chiral supersymmetries, such as on the worldsheet.}! Being that the huge number of Adinkras depends highly combinatorially on $N$, this improvement increases dramatically with $N$; see Section~\ref{s:XMpls+} for some $(4,4)$- and $(8,8)$-supersymmetric examples.

Even in the case of $(2,2)$-supersymmetry, a simple inspection of the 64 Adinkras (most of which admit many nodal permutations) leaves only the five Adinkras in\eq{e:Five2.2} and their boson\,$\iff$\,fermion flips. Furthermore, as shown above in the illustration\eq{e:B484>242} and guaranteed by the work of Ref.\cite{r6-3.2}, the Adinkra {\bsf C} in the line-up\eq{e:Five2.2} decomposes as a direct sum $\text{\bsf D}\oplus\text{\bsf E}$, so that we only need to consider the Adinkras {\bsf A}, {\bsf B}, {\bsf D}, {\bsf E} and their boson\,$\iff$\,fermion flips|a total of eight. We leave it to the Reader to tally up the number of numerical criteria {\em\/\'a la\/} Ref.\cite{rFIL} required to check this list, and the improvement factor afforded by the reduction to\eq{e:Five2.2}.

In turn, the intuitive (see Sections~\ref{s:XMpls} and~\ref{s:XMpls+}) considerations presented herein suffice for adinkraic supermultiplets of all worldsheet $(p,q)$-supersymmetry, which are not subject to gauge equivalence or Bianchi-type self-(anti)duality conditions; these will be addressed under separate cover.

\paragraph{Summary and Outlook}
The unpublished work in\cite{rGLP} made an unexpected assertion, bringing to light evidence for the existence of a ``supersymmetry holography'' that was unknown at the time. This evidence was garnered from many works referenced here from that period and made on the basis of observation of ``Garden Algebra'' structures found universally in all unconstrained supersymmetric quantum mechanical 
systems\cite{rGR1}.  

Adinkras are graphical representation of these algebraic structures.  The correctness of the assertion implies that Adinkras are actually holograms of representations of supersymmetry in higher dimensional spacetimes. As a hologram, an Adinkra must contain all the information of the higher dimensional theory and permit a reconstruction of the higher dimensional theory only from the data solely contained in the Adinkra.  The works of\cite{rFIL,rFL} show this in some specific examples in the context of $3{+}1$-dimensional, ${\cal N} = 1$ supersymmetric representations. The current work gives a filtering procedure that can be applied to this end to $0{+}1$-dimensional systems, and is expected to produce {\em\/all\/} intact off-shell $(p,q)$-supermultiplets in $1{+}1$ dimensions.

Finally, the criterion employed herein (twin theorems~\ref{T:bowT} and~\ref{T:SSR}, \ie, the Corollary~\ref{C:ObSuSy}) is necessary for dimensional extension to higher-dimensional theories (see remark~\ref{i:3} to Corollary~\ref{C:ObSuSy}), but is evidently not sufficient: For example, Adinkras~{\bsf D} and~{\bsf E} in the line-up\eq{e:Five2.2}, the chiral and the twisted-chiral $(2,2)$-supermultiplet, cannot {\em\/both\/} simultaneously extend to $(3{+}1)$-dimensional spacetime and indeed one of them (depending on the spin-structure) fails the numerical tests of Ref.\cite{rFIL}. That is, extension from worldsheet to higher-dimensional supersymmetry does involve additional obstructions, evidently related to the fact that the Lorentz groups in all higher-dimensional spacetimes is non-abelian.

\bigskip\bigskip
\paragraph{\bfseries Acknowledgments:}
We thank Willie Merrell for insightful discussions on semi-chiral supermultiplets, and Charles Doran, Michael Faux, Kevin Iga, Greg Landweber and Robert Miller for prior extensive collaboration on the classification of worldline off-shell supermultiplets, of which the present work is a generalization..
 SJG's research was supported in part by the endowment of the John S.~Toll Professorship, the University of Maryland Center for String \& Particle Theory, National Science Foundation Grant PHY-0354401. SJG's work is also supported by U.S. Department of Energy (D.O.E.) under cooperative agreement DE-FG02-5ER-41360.  SJG offers additional gratitude to the M.~L.~K.\ Visiting Professorship and to the M.~I.~T.\ Center for Theoretical Physics for support and hospitality extended during the undertaking of this work.
 TH is grateful to the Department of Energy for the generous support through the grant DE-FG02-94ER-40854, as well as the Department of Physics, University of Central Florida, Orlando FL, and the Physics Department of the Faculty of Natural Sciences of the University of Novi Sad, Serbia, for recurring hospitality and resources.
 Some Adinkras were drawn with the help of the \Am~\copyright\,2008 by G.~Landweber.

\appendix
\section{The Semi-Chiral Supermultiplet}
The Adinkra\eq{e:B2662} defines a supermultiplet by assigning a component superfield to each node:
\begin{equation}
 \vC{\begin{picture}(80,50)
   \put(0,0){\includegraphics[height=52mm]{B2662.pdf}}
      \put(33.5,47){\cB{$\BX^+_1$}}
      \put(48.5,47){\cB{$\BX^+_2$}}
      \put(3.5,32){\cB{${\bf f}_1$}}
      \put(18,32){\cB{${\bf f}_2$}}
      \put(32,32){\cB{${\bf f}^\pp_1$}}
      \put(47,32){\cB{${\bf f}^\pp_2$}}
      \put(62,32){\cB{${\bf f}_3$}}
      \put(76.5,32){\cB{${\bf f}_4$}}
      \put(3,17){\cB{$\BJ^+_1$}}
      \put(17.75,17){\cB{$\BJ^+_2$}}
      \put(32,17){\cB{$\BJ^-_1$}}
      \put(47,17){\cB{$\BJ^-_2$}}
      \put(61.25,17){\cB{$\BJ^+_3$}}
      \put(76.5,17){\cB{$\BJ^+_4$}}
      \put(33.5,2){\cB{$\BF_1$}}
      \put(48.5,2){\cB{$\BF_2$}}
      \put(17,7){\footnotesize\C3{$\rD_{1-}$}}
      \put(59,7){\footnotesize\C4{$\rD_{2-}$}}
      \put(26,9){\footnotesize\C1{$\rD_{1+}$}}
      \put(36.5,14){\footnotesize\C2{$\rD_{2+}$}}
 \end{picture}}
 \label{e:SM2662}
\end{equation}
and read off the superdifferential relations following the dictionary in Table~\ref{t:A}:
\begin{subequations}
 \label{e:SS2662}\small
\begin{alignat}9
 \C1{\rD_{1+}}\BF_1      &= i\,\BJ^-_1,&\quad
 \C2{\rD_{2+}}\BF_1      &= i\,\BJ^-_2,&\quad
 \C3{\rD_{1-}}\BF_1      &= i\,\BJ^+_1,&\quad
 \C4{\rD_{2-}}\BF_1      &= i\,\BJ^+_3,\\
 \C1{\rD_{1+}}\BF_2      &= i\,\BJ^-_2,&\quad
 \C2{\rD_{2+}}\BF_2      &=-i\,\BJ^-_1,&\quad
 \C3{\rD_{1-}}\BF_2      &= i\,\BJ^+_2,&\quad
 \C4{\rD_{2-}}\BF_2      &= i\,\BJ^+_4,\\
 \C1{\rD_{1+}}\BJ^+_1   &=-{\bf f}_1,&\quad
 \C2{\rD_{2+}}\BJ^+_1   &=-{\bf f}_2,&\quad
 \C3{\rD_{1-}}\BJ^+_1   &= \vd_\mm\BF_1,&\quad
 \C4{\rD_{2-}}\BJ^+_1   &=-{\bf f}_{1\mm},\\
 \C1{\rD_{1+}}\BJ^+_2   &=-{\bf f}_2,&\quad
 \C2{\rD_{2+}}\BJ^+_2   &= {\bf f}_1,&\quad
 \C3{\rD_{1-}}\BJ^+_2   &= \vd_\mm\BF_2,&\quad
 \C4{\rD_{2-}}\BJ^+_2   &=-{\bf f}^\pp_2,\\
 \C1{\rD_{1+}}\BJ^-_1   &= \vd_\pp\BF_1,&\quad
 \C2{\rD_{2+}}\BJ^-_1   &=-\vd_\pp\BF_2,&\quad
 \C3{\rD_{1-}}\BJ^-_1   &= {\bf f}_1,&\quad
 \C4{\rD_{2-}}\BJ^-_1   &= {\bf f}_3,\\
 \C1{\rD_{1+}}\BJ^-_2   &= \vd_\pp\BF_2,&\quad
 \C2{\rD_{2+}}\BJ^-_2   &= \vd_\pp\BF_1,&\quad
 \C3{\rD_{1-}}\BJ^-_2   &= {\bf f}_2,&\quad
 \C4{\rD_{2-}}\BJ^-_2   &= {\bf f}_3,\\
 \C1{\rD_{1+}}\BJ^+_3   &=-{\bf f}_3,&\quad
 \C2{\rD_{2+}}\BJ^+_3   &=-{\bf f}_4,&\quad
 \C3{\rD_{1-}}\BJ^+_3   &= {\bf f}^\pp_2,&\quad
 \C4{\rD_{2-}}\BJ^+_3   &= \vd_\mm\BF_1,\\
 \C1{\rD_{1+}}\BJ^+_4   &=-{\bf f}_4,&\quad
 \C2{\rD_{2+}}\BJ^+_4   &= {\bf f}_3,&\quad
 \C3{\rD_{1-}}\BJ^+_4   &= {\bf f}^\pp_2,&\quad
 \C4{\rD_{2-}}\BJ^+_4   &= \vd_\mm\BF_2,\\
 \C1{\rD_{1+}}{\bf f}_1      &=-i\,\vd_\pp\BJ^+_1,&\quad
 \C2{\rD_{2+}}{\bf f}_1      &= i\,\vd_\mm\BJ^+_2,&\quad
 \C3{\rD_{1-}}{\bf f}_1      &= i\,\vd_\mm\BJ^-_1,&\quad
 \C4{\rD_{2-}}{\bf f}_1      &=-i\,\BX^+_1,\\
 \C1{\rD_{1+}}{\bf f}_2      &=-i\,\vd_\pp\BJ^+_2,&\quad
 \C2{\rD_{2+}}{\bf f}_2      &=-i\,\vd_\pp\BJ^+_1,&\quad
 \C3{\rD_{1-}}{\bf f}_2      &= i\,\vd_\mm\BJ^-_2,&\quad
 \C4{\rD_{2-}}{\bf f}_2      &=-i\,\BX^+_2,\\
 \C1{\rD_{1+}}{\bf f}_{1\mm} &= i\,\BX^+_1,&\quad
 \C2{\rD_{2+}}{\bf f}_{1\mm} &= i\,\BX^+_2,&\quad
 \C3{\rD_{1-}}{\bf f}_{1\mm} &= i\,\vd_\mm\BJ^+_3,&\quad
 \C4{\rD_{2-}}{\bf f}_{1\mm} &=-i\,\vd_\mm\BJ^+_1,\\
 \C1{\rD_{1+}}{\bf f}^\pp_2 &= i\,\BX^+_2,&\quad
 \C2{\rD_{2+}}{\bf f}^\pp_2 &=-i\,\BX^+_1,&\quad
 \C3{\rD_{1-}}{\bf f}^\pp_2 &= i\,\vd_\mm\BJ^+_4,&\quad
 \C4{\rD_{2-}}{\bf f}^\pp_2 &=-i\,\vd_\mm\BJ^+_2,\\
 \C1{\rD_{1+}}{\bf f}_3      &=-i\,\vd_\pp\BJ^+_3,&\quad
 \C2{\rD_{2+}}{\bf f}_3      &= i\,\vd_\mm\BJ^+_4,&\quad
 \C3{\rD_{1-}}{\bf f}_3      &= i\,\BX^+_1,&\quad
 \C4{\rD_{2-}}{\bf f}_3      &= i\,\vd_\mm\BJ^-_1,\\
 \C1{\rD_{1+}}{\bf f}_4      &=-i\,\vd_\pp\BJ^+_4,&\quad
 \C2{\rD_{2+}}{\bf f}_4      &=-i\,\vd_\pp\BJ^+_3,&\quad
 \C3{\rD_{1-}}{\bf f}_4      &= i\,\BX^+_2,&\quad
 \C4{\rD_{2-}}{\bf f}_4      &= i\,\vd_\mm\BJ^-_2,\\
 \C1{\rD_{1+}}\BX^+_1   &= \vd_\pp{\bf f}_{1\mm},&\quad
 \C2{\rD_{2+}}\BX^+_1   &=-\vd_\pp{\bf f}^\pp_2,&\quad
 \C3{\rD_{1-}}\BX^+_1   &= \vd_\mm\BF_3,&\quad
 \C4{\rD_{2-}}\BX^+_1   &=-\vd_\mm{\bf f}_1,\\
 \C1{\rD_{1+}}\BX^+_2   &= \vd_\pp{\bf f}^\pp_2,&\quad
 \C2{\rD_{2+}}\BX^+_2   &= \vd_\pp{\bf f}_{1\mm},&\quad
 \C3{\rD_{1-}}\BX^+_2   &= \vd_\mm\BF_4,&\quad
 \C4{\rD_{2-}}\BX^+_2   &=-\vd_\mm{\bf f}_2.
\end{alignat}
\end{subequations}
It is possible to complexify {\em\/simultaneously\/} the component superfields
\begin{subequations}
 \label{e:C2662}
\begin{alignat}9
  \bBF^c    &\Defl(\BF_1{+}i\BF_2),&\quad
  \bJ^c_{1-}&\Defl(\BJ^+_1{+}i\BJ^+_2),&\quad
  \bJ^c_+   &\Defl(\BJ^-_1{+}i\BJ^-_2),&\quad
  \bJ^c_{2-}&\Defl(\BJ^+_3{+}i\BJ^+_4),\\
  \Bf^c_1   &\Defl({\bf f}_1{+}i{\bf f}_2),&\quad
  \Bf^c_\mm &\Defl({\bf f}_{1\mm}{+}i{\bf f}^\pp_2),&\quad
  \Bf^c_2   &\Defl({\bf f}_3{+}i{\bf f}_4),&\quad
  \bX^c_-   &\Defl(\BX^+_1{+}i\BX^+_2),
\end{alignat}
\end{subequations}
and also the left-handed superderivatives, $\BD^c_+\Defl(\rD_{1+}+i\rD_{2+})$, leaving however the $\rD_{1-},\rD_{2-}$-action unchanged. This simplifies the Adinkra\eq{e:SM2662}:
\begin{equation}
 \vC{\begin{picture}(150,45)(0,2)\footnotesize\unitlength=.9231mm
   \put(0,0){\includegraphics[height=48mm]{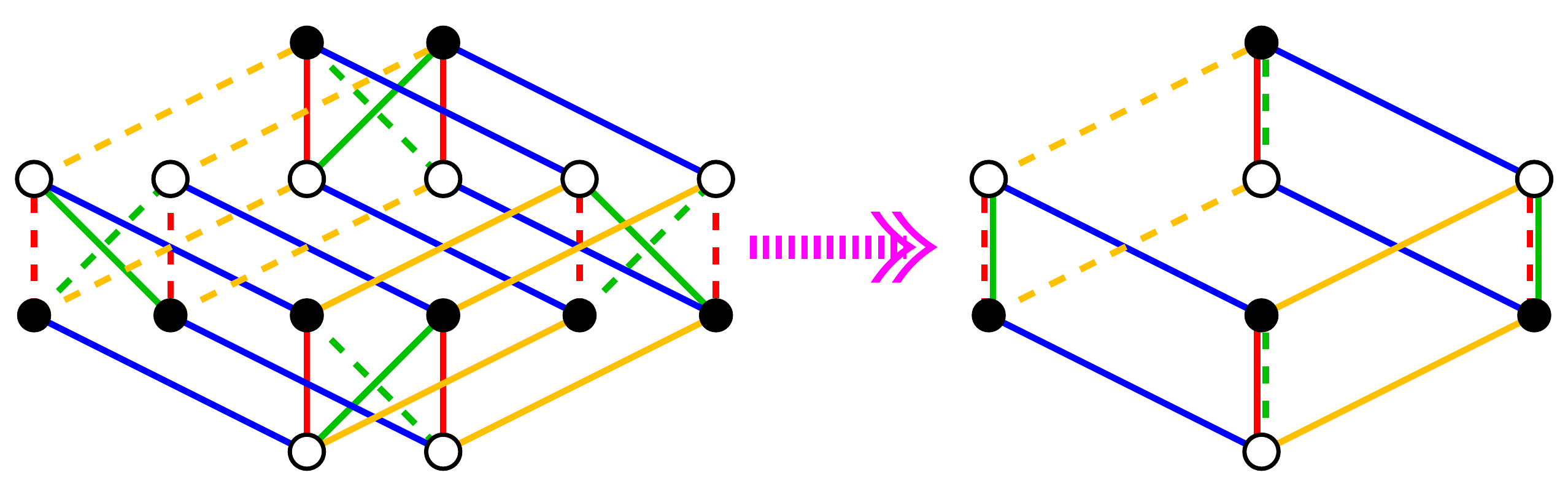}}
      \put(33.5,47){\cB{$\BX^+_1$}}
      \put(48.5,47){\cB{$\BX^+_2$}}
      \put(4,32){\cB{${\bf f}_1$}}
      \put(17.75,32){\cB{${\bf f}_2$}}
      \put(32,32){\cB{${\bf f}_{1\mm}$}}
      \put(47,32){\cB{${\bf f}^\pp_2$}}
      \put(62.25,32){\cB{${\bf f}_3$}}
      \put(76.5,32){\cB{${\bf f}_4$}}
      \put(3,17){\cB{$\BJ^+_1$}}
      \put(17.75,17){\cB{$\BJ^+_2$}}
      \put(32,17){\cB{$\BJ^-_1$}}
      \put(47,17){\cB{$\BJ^-_2$}}
      \put(61.25,17){\cB{$\BJ^+_3$}}
      \put(76.5,17){\cB{$\BJ^+_4$}}
      \put(33.5,2){\cB{$\BF_1$}}
      \put(48.5,2){\cB{$\BF_2$}}
      \put(135,47){\cB{$\bX^c_-$}}
      \put(106,32){\cB{$\Bf^c_1$}}
      \put(135,32){\cB{$\Bf^c_\mm$}}
      \put(164,32){\cB{$\Bf^c_2$}}
      \put(106,17){\cB{$\bJ^c_{1-}$}}
      \put(135,17){\cB{$\bJ^c_+$}}
      \put(164,17){\cB{$\bJ^c_{2-}$}}
      \put(135,2){\cB{$\bBF^c$}}
 \end{picture}}
 \label{e:SM2662c}
\end{equation}
The \C1{red}-\C2{green} double edge indicates the complex action of $\BD^c_+$; by contrast, the action of \C3{$\rD_{1-}$ (blue edges)} and \C4{$\rD_{2-}$ (orange edges)} is not so paired, which hints at the possibility that\eq{e:SM2662} is a particularly constrained, but initially complex supermultiplet. The graded dimension count (number of real degrees of freedom per height level) suggests that\eq{e:SM2662} is the Adinkra of the semi-chiral supermultiplets\cite{rSChSF0,rSChSF}, and we now turn to prove this.

\paragraph{Proof (that the Adinkra\eq{e:SM2662} depicts the semi-chiral superfield):}
The semi-chiral supermultiplets is defined as a {\em\/complex\/} intact $(2,2)$-superfield subject to the single {\em\/complex\/} superdifferential constraint\cite{rSChSF0,rSChSF}:
\begin{align}
  \bar\BD_-^c\BS^c&= 0.
 \label{e:SCh}
\intertext{Rewriting $\BD_-^c=\rD_{1-}-i\rD_{2-}$ and $\BS^c=\bS_1+i\bS_2$, this is seen to consist of {\em\/two\/} real superdifferential constraints:}
  \qquad\qquad\bar\BD_-^c\BS^c&=0\quad\Iff\quad
  \Big\{\begin{array}{rl}
  \rD_{1-}\bS_1&=-\rD_{2-}\bS_2,\\
  \rD_{1-}\bS_2&=\rD_{2-}\bS_1.
  \end{array}
 \label{e:SChR}
\end{align}
We use the definitions of real component fields of $\bS$:
\begin{subequations}
\begin{gather}
  \cS_i\Defl\frc14[\rD_{1+},\rD_{2+}][\rD_{1-},\rD_{2-}]\bS_i|,\\
  \S^-_{i\ad}\Defl\frc12[\rD_{1+},\rD_{2+}]\rD_{\ad-}\bS_i|,\qquad
  \S^+_{i\a}\Defl\frc12[\rD_{1-},\rD_{2-}]\rD_{\a+}\bS_i|,\\
  S^\mm_i\Defl\frc{i}2[\rD_{1+},\rD_{2+}]\bS_i|,\qquad
  S_{i\a\ad}\Defl\frc{i}2[\rD_{\a+},\rD_{\ad-}]\bS_i|,\qquad
  S^\pp_i\Defl\frc{i}2[\rD_{1-},\rD_{2-}]\bS_i|,\\
  \s^+_{i\ad}\Defl i\,\rD_{\ad-}\bS_i|,\qquad
  \s^-_{i\a}\Defl i\,\rD_{\a+}\bS_i|,\\
  s_i\Defl i\,\rD_{\ad-}\bS_i|,\qquad i=1,2.
\end{gather}
\end{subequations}
and project the components of the superdifferential constraint by applying the tesseract of superderivatives\eq{e:Ds} on the equation and projecting to the worldsheet.
 Evaluation of the components (with convenient constant pre-factors) of\eq{e:SCh} produces, in turn:
\begin{subequations}
 \label{e:S1+S2}
\begin{alignat}9
 i\bar\BD_-^c\BS^c|
 &=0:&\quad
 \s^+_{22}&=-\s^+_{11},&\quad
 \s^+_{21}&=\s^+_{12};\label{e:s22}\\
 i\rD_{1+}\bar\BD_-^c\BS^c|
 &=0:&\quad
 S_{212}&=-S_{111},&\quad
 S_{211}&=S_{112};\label{e:S212}\\
 i\rD_{2+}\bar\BD_-^c\BS^c|
 &=0:&\quad
 S_{222}&=-S_{121},&\quad
 S_{221}&=S_{122};\label{e:S222}\\
 i\rD_{1-}\bar\BD_-^c\BS^c|
 &=0:&\quad
 S^\pp_2&=(\vd_\mm s_1),&\quad
 S^\pp_1&=-(\vd_\mm s_2);\\
 i\rD_{2-}\bar\BD_-^c\BS^c|
 &=0:&\quad
 &\Cx{\quad~ditto,}&\quad
 &\Cx{\quad~ditto;}\\
 \inv2[\rD_{1+},\rD_{2+}]\bar\BD_-^c\BS^c|
 &=0:&\quad
 \S^-_{22}&=-\S^-_{11},&\quad
 \S^-_{21}&=\S^-_{12};\label{e:S22}\\
 \inv2[\rD_{1+},\rD_{1-}]\bar\BD_-^c\BS^c|
 &=0:&\quad
 \S^+_{21}&=-(\vd_\mm\s^-_{11}),&\quad
 \S^+_{11}&=(\vd_\mm\s^-_{21});\\
 \inv2[\rD_{1+},\rD_{2-}]\bar\BD_-^c\BS^c|
 &=0:&\quad
 &\Cx{\quad~ditto,}&\quad
 &\Cx{\quad~ditto;}\\
 \inv2[\rD_{2+},\rD_{1-}]\bar\BD_-^c\BS^c|
 &=0:&\quad
 \S^+_{22}&=-(\vd_\mm\s^-_{12}),&\quad
 \S^+_{12}&=(\vd_\mm\s^-_{22});\\
 \inv2[\rD_{2+},\rD_{2-}]\bar\BD_-^c\BS^c|
 &=0:&\quad
 &\Cx{\quad~ditto,}&\quad
 &\Cx{\quad~ditto;}\\
 \inv2[\rD_{1-},\rD_{2-}]\bar\BD_-^c\BS^c|
 &=0:&\quad
 (\vd_\mm\s^+_{21})&=(\vd_\mm\s^+_{12}),&\quad
 (\vd_\mm\s^+_{22})&=-(\vd_\mm\s^+_{11});\\[-2mm]
 &&&\Lx{\footnotesize~~[these are implied by\eq{e:s22}]}\nn\\
 \inv2[\rD_{1+},\rD_{2+}]\rD_{1-}\bar\BD_-^c\BS^c|
 &=0:&\quad
 \cS_2&=-(\vd_\mm S^\mm_1),&\quad
 \cS_1&=(\vd_\mm S^\mm_2);\\
 \inv2[\rD_{1+},\rD_{2+}]\rD_{2-}\bar\BD_-^c\BS^c|
 &=0:&\quad
 &\Cx{\quad~ditto,}&\quad
 &\Cx{\quad~ditto;}\\
 \inv2[\rD_{1-},\rD_{2-}]\rD_{1+}\bar\BD_-^c\BS^c|
 &=0:&\quad
 (\vd_\mm S_{211})&=(\vd_\mm S_{112}),&\quad
 (\vd_\mm S_{212})&=-(\vd_\mm S_{111});\\[-2mm]
 &&&\Lx{\footnotesize~~[these are implied by\eq{e:S212}]}\nn\\
 \inv2[\rD_{1-},\rD_{2-}]\rD_{2+}\bar\BD_-^c\BS^c|
 &=0:&\quad
 (\vd_\mm S_{221})&=(\vd_\mm S_{122}),&\quad
 (\vd_\mm S_{222})&=-(\vd_\mm S_{121});\\[-2mm]
 &&&\Lx{\footnotesize~~[these are implied by\eq{e:S222}]}\nn\\
 \inv4[\rD_{1+},\rD_{2+}][\rD_{1-},\rD_{2-}]\bar\BD_-^c\BS^c|
 &=0:&\quad
 (\vd_\mm\S^-_{22})&=-(\vd_\mm\S^-_{11}),&\quad
 (\vd_\mm\S^-_{21})&=(\vd_\mm\S^-_{12}).\\[-2mm]
 &&&\Lx{\footnotesize~~[these are implied by\eq{e:S22}]}\nn
\end{alignat}
\end{subequations}
The identifications noted as implied by earlier identifications or being a copy (``ditto'') of a previous identification may be dropped.
In deriving these, the following operatorial identities were useful:
\begin{subequations}
\begin{alignat}9
 [\rD_{1+},\rD_{1-}]\rD_{1-}
 &=2\rD_{1+}\rD_{1-}\rD_{1-}
  =2i\vd_\mm\rD_{1+},\\
 [\rD_{1+},\rD_{1-}]\rD_{2-}
 &=2\rD_{1+}\rD_{1-}\rD_{2-}
  =[\rD_{1-},\rD_{2-}]\rD_{1+},\\
 [\rD_{1+},\rD_{2-}]\rD_{1-}
 &=2\rD_{1+}\rD_{2-}\rD_{1-}
  =-[\rD_{1-},\rD_{2-}]\rD_{1+},\\
 [\rD_{1+},\rD_{2-}]\rD_{2-}
 &=2\rD_{1+}\rD_{2-}\rD_{2-}
  =2i\vd_\mm\rD_{1+},\\
 [\rD_{2+},\rD_{1-}]\rD_{1-}
 &=2\rD_{2+}\rD_{1-}\rD_{1-}=2i\vd_\mm\rD_{2+},\\
 [\rD_{2+},\rD_{1-}]\rD_{2-}
 &=2\rD_{2+}\rD_{1-}\rD_{2-}=[\rD_{1-},\rD_{2-}]\rD_{2+},\\
 [\rD_{2+},\rD_{2-}]\rD_{1-}
 &=2\rD_{2+}\rD_{2-}\rD_{1-}=-[\rD_{1-},\rD_{2-}]\rD_{2+},\\
 [\rD_{2+},\rD_{2-}]\rD_{2-}
 &=2\rD_{2+}\rD_{2-}\rD_{2-}=2i\vd_\mm\rD_{2+},\\
 [\rD_{1-},\rD_{2-}]\rD_{1-}
 &=-2\rD_{2-}\rD_{1-}\rD_{1-}=-2i\vd_\mm\rD_{2-},\\ 
 [\rD_{1-},\rD_{2-}]\rD_{2-}
 &=2\rD_{1-}\rD_{2-}\rD_{2-}=2i\vd_\mm\rD_{1-}.
\end{alignat}
\end{subequations}

These component field level identifications\eq{e:S1+S2} imply the corresponding identifications of the nodes in the Adinkras $\cA(\bS_1)$ and $\cA(\bS_2)$, and ``fuse'' them by identifying each node from $\cA(\bS_1)$ that corresponds to a component field appearing in the identifications\eq{e:S1+S2} with precisely one node from $\cA(\bS_2)$.
 In the resulting fusion, $\cA(\bS_1)\#\cA(\bS_2)$:
\begin{enumerate}\itemsep=-3pt\vspace{-2mm}
  \item There are $(2|6|6|2)$ nodes per height. For example, the top (fifth level) nodes have been identified: $\cS_1=(\vd_\mm S^\mm_2)$ and $\cS_2=-(\vd_\mm S^\mm_1)$, so that all edges that used to lead to the top (fifth) level, now lead to the $S^\mm_1$ and $S^\mm_2$ nodes in the previously middle (third) level.
  \item Each remaining node again has precisely one edge of each kind adjacent to it and so belongs to a proper Adinkra, which by the classification of Refs.\cite{r6-1,r6-3} filtered by theorems~\ref{T:bowT} and~\ref{T:SSR} must be the one in\eq{e:SM2662}.
\end{enumerate}
This above identification process is depicted below:
\begin{equation}
 \vC{\begin{picture}(160,35)
   \put(0,0){\includegraphics[width=160mm]{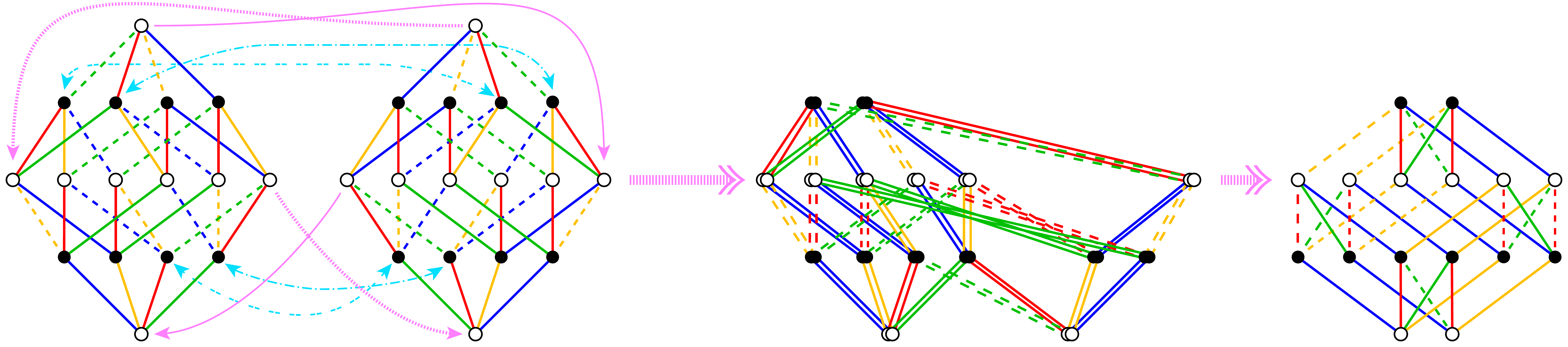}}
    \put(0,2){\footnotesize$\cA(\bS_1)$}
    \put(53,2){\footnotesize$\cA(\bS_2)$}
    \put(77,30){\parbox{55mm}{\footnotesize\baselineskip=10pt\centering
                              Edges being identified must match
                              in their sold/dashed quality}}
 \end{picture}}
\end{equation}
where the curved arrows spanning between $\cA(\bS_1)$ and $\cA(\bS_2)$ indicate a few of the identifications\eq{e:S1+S2}; note that these ``real part'' and ``imaginary part'' Adinkras were drawn in mirror image to each other. The illustration in the middle depicts the result of the identifications, where the transported vertices were brought close, but not precisely to their destination so as to show the perfect overlay|upon sign-changes in a few component fields consistent with\eq{e:S1+S2}. Upon some horizontal repositioning of the nodes, the result is the right-hand side Adinkra, which is identical with\eq{e:SM2662}, thus proving that this is indeed the Adinkra of the semi-chiral supermultiplet.\QED

\section{Dirac Algebra and Related Conventions}
For the discussion around equations\eqs{e:TM-I}{e:TM-Ia}, the following set of conventions for the $\g$-matrices was followed:
\begin{gather}
 \eta_{ a b} = \text{diag}(1,-1),\quad
 \ve_{ab}\ve^{cd} = - \d_{[a}{}^c\,\d_{b]}{}^d,\quad
 \ve^{01} = +1,\\[1mm]
  (\g^a)_{\sss A}{}^{\sss C} (\g^b)_{\sss C}{}^{\sss B}
  =\h^{ab}\,\d_{\sss A}{}^{\sss B} - \ve^{ab}(\g^3)_{\sss A}{}^{\sss B}. 
\end{gather}
The last one of these relations implies
\begin{equation}
 \g^a \g_a = 2 \,\Ione,\quad
 \g^3 \g^a = -\ve^{ab}\g_b. 
\end{equation}
Moreover, it follows that
\begin{gather}
(\g^3)_{\sss A}{}^{\sss B} (\g^a)_{\sss B}{}^{\sss A} = 0,\\
(\g^3)_{\sss A}{}^{\sss B} (\g^3)_{\sss B}{}^{\sss D} =
 \inv2 \ve_{ab} (\g^3)_{\sss A}{}^{\sss B}
                 (\g^a)_{\sss B}{}^{\sss C} (\g^b)_{\sss C}{}^{\sss D} =  
 \inv2 (\g_b)_{\sss A}{}^{\sss C} (\g^b)_{\sss C}{}^{\sss D} =\d_{\sss A}{}^{\sss D}~.
\end{gather}

Denoting the spinorial metric $C_{\sss AB}$, some useful Fierz identities are:
\begin{align}
   C_{\sss AB} C^{\sss CD} &= \d_{\sss[A}{}^{\sss C}\,\d_{\sss B]}{}^{\sss D},\\
   (\g^a)_{\sss AB} (\g_a)^{\sss CD} + (\g^3)_{\sss AB} (\g^3)^{\sss CD}
 &= -\d_{\sss (A}{}^{\sss C}\,\d_{\sss B)}{}^{\sss D},\\
     (\g^a)_{\sss(A}{}^{\sss C} (\g_a)_{\sss B)}{}^{\sss D}
     +(\g^3)_{\sss(A}{}^{\sss C} (\g^3)_{\sss B)}{}^{\sss D}
 &= \d_{\sss(A}{}^{\sss C}\,\d_{\sss B)}{}^{\sss D}, \\
    (\g^a)_{\sss(A}{}^{\sss C} (\g_a)_{\sss B)}{}^{\sss D}
 &= -2 (\g^3)_{\sss AB} (\g^3)^{\sss CD}, \\
   2(\g^a)_{\sss AB} (\g_a)^{\sss CD}
    +(\g^3)_{\sss(A}{}^{\sss C} (\g^3)_{\sss B)}{}^{\sss D}
 &= -\d_{\sss(A}{}^{\sss C}\,\d_{\sss B)}{}^{\sss D},\\
     (\g_a)_{\sss A}{}^{\sss D}\d_{\sss B} {}^{\sss C}
     +(\g^3 \g_a)_{\sss A}{}^{\sss C} (\g^3)_{\sss B}{}^{\sss D}
 &=(\g^3 \g_a)_{\sss AB} (\g^3)^{\sss CD}, \\
   (\g^3 \g_a\g^3)_{\sss A}{}^{\sss D}
 &= -(\g_a)_{\sss A}{}^{\sss D}.\\
    (\g^c)_{\sss A}{}^{\sss B} (\g_c)_{\sss C}{}^{\sss D}
    +(\g^3)_{\sss A}{}^{\sss B} (\g^3)_{\sss C}{}^{\sss D}
 &=\d_{\sss C}{}^{\sss B}\,\d_{\sss A}{}^{\sss D}
    - C_{\sss AC}\,C^{\sss BD}, \\
   (\g^c)_{\sss A}{}^{\sss C} (\g^3\g_c)_{\sss B}{}^{\sss D}
 &= C_{\sss AB}(\g^3)^{\sss CD} + (\g^3)_{\sss AB}\,C^{\sss CD} 
\end{align}

As an explicit representation, we may define the $1{+}1$-dimensional $\g$-matrices
in terms of the usual Pauli matrices according to
\begin{equation}
 (\g^0)_{\sss A}{}^{\sss B} \equiv (\s^2)_{\sss A}{}^{\sss B},\quad
 (\g^1)_{\sss A}{}^{\sss B} \equiv - i (\s^1)_{\sss A}{}^{\sss B},\quad
 (\g^3)_{\sss A}{}^{\sss B} \equiv (\s^3)_{\sss A}{}^{\sss B}.  
\end{equation}
The spinor metric $C_{\sss AB}$ and its inverse $C^{\sss AB}$ may then be chosen as
\begin{equation}
 C_{\sss AB} \equiv (\s^2)_{\sss AB},\quad
 C^{\sss AB} \equiv -(\s^2)^{\sss AB}.
\end{equation}
Using this explicit representation, it is easy to show the following
symmetry properties
\begin{alignat}9
(\g^a)_{\sss AB} &= (\g^a)_{\sss BA},&\qquad
(\g^3)_{\sss AB} &= (\g^3)_{\sss BA},&\qquad
     C_{\sss AB} &= - C_{\sss BA}, \\
(\g^a)^{\sss AB} &= (\g^a)^{\sss BA},&\qquad
(\g^3)^{\sss AB} &= (\g^3)^{\sss BA},&\qquad
     C^{\sss AB} &= - C^{\sss BA}. 
\end{alignat}
In a similar manner the following complex conjugation properties
can be derived
\begin{equation}
[(\g^a)_{\sss A}{}^{\sss B}]^* = -(\g^a)_{\sss A}{}^{\sss B},\qquad
[(\g^3)_{\sss A}{}^{\sss B}]^* = +(\g^3)_{\sss A}{}^{\sss B},
\end{equation}
\begin{alignat}9
[ (\g^a)_{\sss AB}]^* &= (\g^a)_{\sss AB},&\qquad
[ (\g^3)_{\sss AB}]^* &= -(\g^3)_{\sss AB},&\qquad
[ C_{\sss AB}]^*      &= -C_{\sss AB}, \\
[ (\g^a)^{\sss AB}]^* &= (\g^a)^{\sss AB},&\qquad
[ (\g^3)^{\sss AB}]^* &= -(\g^3)^{\sss AB},&\qquad
[ C^{\sss A B}]^*     &= -C^{\sss AB}. 
\end{alignat}

Due to the relation $[(\g^a)_{\sss A}{}^{\sss B}]^* = -(\g^a)_{\sss A}{}^{\sss B}$, we see that this choice of gamma matrices is in a Majorana representation and thus the simplest spinors such as $\j^{\sss A}(x)$ may be chosen to be real, \ie,
\begin{equation}
 [\, \psi^{\sss A}(x) \, \, ]^* = \psi^{\sss A}(x),
\end{equation}
and we can raise and lower spinor indices according to
\begin{equation}
  \j^{\sss A}(x) ~=~ C{}^{\sss AB} \,  \psi_{\sss B}(x),\qquad
  \j_{\sss A}(x) ~=~  \psi^{\sss B}(x)  \,  C{}_{\sss BA}.
\end{equation}
It then follows that
\begin{equation}
 [\, \j_{\sss A}(x) \, \, ]^* =  - \j_{\sss A}(x).
\end{equation}
Of course, it is always possible to introduce complex spinors also.

The extraction of light-cone coordinates begins with the observation that
$\g^3$ in this set of conventions is diagonal.  This means that our 2-component
Majorana spinors can be written as
\begin{equation}
  \j_{\sss A}(x) = \Bm{~\j_+(x)~\\~\j_-(x)~}.
\end{equation}
By defining chiral projection operators via the equations
\begin{equation}
 \hP{\pm} \equiv \inv2\,\big[\,\Ione ~\pm ~ \g^3 \,\big],
\end{equation}
it follows that
\begin{equation}
  (\hP+)_{\sss A}{}^{\sss B}\,\j_{\sss B}(x)
  =\Bm{~\j_+(x)~\\~0~ },\qquad
  (\hP-)_{\sss A}{}^{\sss B}\,\j_{\sss B}(x)
  =\Bm{~0~\\~\j_-(x)~ },
\end{equation}
in terms of the one-component Majorana spinors $\psi_{+}(x) $ and $\psi_{-}(x) $.

The projection operators $\hP\pm$ satisfy the following relations:
\begin{alignat}9
 \hP{\pm}\,\g^3\,\hP{\pm} &= \pm\,\hP{\pm},\qquad
 \hP{\pm}\,\g^3\,\hP{\mp} &= 0,\\
 \hP{\pm}\,\g^a\,\hP{\pm} &= 0,\qquad
 \hP{\pm}\,\g^3\,\g^a\,\hP{\pm} &=0,
\end{alignat}
and project the following worldsheet derivatives:
\begin{alignat}9
 \hP+\,\g^a\,\hP-\,\vd_a
 &=\Bm{~0 & -i(\vd_\t+\vd_\s)~\\
       ~0 & ~~~~0~}&
 &=\Bm{~~0 & -i\vd_{\pp}~\\
       ~~0 & ~~0~},\\[1mm]
 \hP-\,\g^a\,\hP+\,\vd_a
 &=\Bm{~0 & ~0~\\
       ~i(\vd_\t-\vd_\s)~ & ~0~}&
 &=\Bm{~0 & ~~0~~\\
       ~i\vd_{\mm}  & ~~0~~},\\
 \hP+\,\g^3\,\g^a\,\hP-\,\vd_a
 &=\Bm{~0 & -i(\vd_\t+\vd_\s)~\\
       ~0 & ~~~~0~}&
 &=\Bm{~~0 & -i\vd_{\pp}~\\
       ~~0 & ~~0~},\\
 \hP-\,\g^3\,\g^a\,\hP+\,\vd_a
 &=\Bm{~~~0 & ~0~\\
       -i(\vd_\t-\vd_\s) & ~0~}&
 &=\Bm{~0 & ~0~~\\
       -i\vd_{\mm} & ~0~~}.
\end{alignat}

Finally, our rules for manipulating the $\SU(2)$ indices are very similar to the ones
used for the $\Spin(1,1)\approx\textsl{SL}(2,\IR)$ spinor indices.  The $\SU(2)$ metric
$C_{i \, j}$ and its inverse $C^{i \, j}$ can be identified as
\begin{equation}
 C_{ij} \equiv (\s^2)_{ij},\qquad
 C^{ij} \equiv -(\s^2 )^{ij},
\end{equation}  
so that
\begin{equation}
 C_{ij} = - C_{ji},\qquad
 C^{ij} = - C^{ji},\qquad
 C_{ij}\,C^{kl} = \d{}_i{}^k\,\d{}_j{}^l - \d{}_i{}^l\,\d{}_j{}^k.
\end{equation}
We raise and lower $\SU(2)$ indices according to
\begin{equation}
  \j^i(x) = C^{ij}\,\j_j(x),\qquad
  \j_i(x) = \j^j(x)\,C_{ji},
\end{equation}
that are directly the analogs for raising and lowering indices on $\textsl{SL}(2,\IR)$ tensors.
Note also that
\begin{equation}
 (C^{ij})^*=C_{ij},\qquad\text{and}\qquad (C_{ij})^*=C^{ij}~.
\end{equation}

\clearpage
\bibliographystyle{elsart-numX}
\bibliography{Refs}
\end{document}